\documentclass[lettersize,journal]{IEEEtran}
\usepackage{amsmath,amsfonts,amssymb}
\usepackage{array}
\usepackage[caption=false,font=footnotesize,labelfont=rm,textfont=rm]{subfig}
\usepackage{textcomp}
\usepackage{stfloats}
\usepackage{url}
\usepackage{verbatim}
\usepackage{graphicx}
\usepackage{cite}
\usepackage[table]{xcolor}
 \usepackage{threeparttable}
\usepackage{makecell}
\usepackage{diagbox}
\usepackage{hyperref}
\usepackage{amssymb}
\usepackage{multirow}
\usepackage{stmaryrd}
\usepackage{booktabs}
\usepackage{epstopdf}
\usepackage{wasysym}
\usepackage{pifont}
\usepackage[ruled,vlined]{algorithm2e}
\usepackage{algorithmic}
\usepackage{amsthm}
\usepackage{indentfirst}
\usepackage{soul}
\usepackage{fancyhdr}
\usepackage{xcolor}
\definecolor{blue}{rgb}{0,0,0}  

\newtheorem{assumption}{Assumption}
\newtheorem{thm}{\bf Theorem}
\newtheorem{corollary}{Corollary}

\usepackage{color, xcolor}
\usepackage{fancyhdr}

\begin{document}

\title{DFPL:  Decentralized  Federated Prototype Learning Across Heterogeneous Data Distributions}
\author{Hongliang Zhang, Fenghua Xu, Zhongyuan Yu, Shanchen Pang, Chunqiang Hu, and Jiguo Yu,~\IEEEmembership{Fellow,~IEEE},

\thanks{This work was supported by the National Natural Science Foundation of China under Grants 62272256 and 62202250, the Major Program of Shandong Provincial Natural Science Foundation for the Fundamental Research under Grant ZR2022ZD03, the National Science Foundation of Shandong Province under Grant ZR2021QF079, the Colleges and Universities 20 Terms Foundation of Jinan City under Grant 202228093, and the Shandong Province Youth Innovation Team Project under Grant 2024KJH032. ($\textit{Corresponding author}$: $\textit{Jiguo Yu}$.)}
\thanks{H. Zhang is with the Key Laboratory of Computing Power Network and Information Security, Ministry of Education, Shandong Computer Science Center, Qilu University of Technology (Shandong Academy of Sciences), Jinan, 250353, China, Email: b1043123004@stu.qlu.edu.cn.}
\thanks{F. Xu is with the Cyber Security Institute, University of Science and Technology of China,  Hefei, 230026, China, Email: nstlxfh@gmail.com.}
\thanks{Z. Yu and S. Pang  are with the College of computer science and technology, China University of Petroleum,  Qingdao, 266580, China, Email:  yuzhy24601@gmail.com, pangsc@upc.edu.cn.}
\thanks{C. Hu is with the School of Big Data and Software Engineering, Chongqing University,  Chongqing, 400044, China, Email:  chu@cqu.edu.cn.}
\thanks{J. Yu is with  School of Computer Science and Engineering, University of Electronic Science and Technology of China, Chengdu, 611731, China, and also with the Big Data Institute, Qilu University of Technology, Jinan, 250353, China, Email: jiguoyu@sina.com; jiguoyu17@uestc.edu.cn.}
}

\markboth{}%
{Shell \MakeLowercase{\textit{et al.}}: A Sample Article Using IEEEtran.cls for IEEE Journals}



\maketitle

\thispagestyle{fancy}
\begin{abstract}
Federated learning  is a distributed machine learning paradigm  through centralized model aggregation.
However, standard federated learning relies on a centralized server, making it vulnerable to server failures.
While  existing solutions utilize blockchain technology to implement Decentralized Federated Learning (DFL), the statistical heterogeneity of data distributions among clients severely degrades the performance of DFL.
Driven by this issue, this paper proposes a decentralized federated prototype learning framework, named DFPL, which  significantly improves the performance of   DFL  under heterogeneous data distributions.
Specifically,  DFPL  introduces prototype learning into DFL to mitigate   the impact of statistical heterogeneity and reduces the amount of parameters exchanged between clients.
Additionally, blockchain is embedded    into our framework, enabling  the training and mining processes to be executed locally on  each client.
From a theoretical perspective, we analyze the  convergence   of DFPL by modeling the required computational resources during  both training and mining.
The experiment results highlight  the superiority of  DFPL  in both model performance and communication efficiency  across  four benchmark datasets with heterogeneous data distributions.
\end{abstract}


\section{Introduction}\label{sec1}

 Federated Learning (FL) \cite{mcmahan2017communication}, a distributed machine learning paradigm, allows multiple clients to  collaboratively  train a global  model by centralized aggregation without sharing  raw  data.
However, the centralized  aggregation suffers from serious security threats.  (\textit{i}) The global model is inaccurate if the server is attacked (e.g., data tampering attack \cite{9945975}). (\textit{ii}) FL is interrupted if the server fails due  to physical damage \cite{9403374}\cite{zhang2024survey}.
To solve these  threats,   Blockchain-assisted FL (BFL) frameworks  \cite{9079513,9272656,9551794,9714771,9399813}   are proposed, replacing the central server with a third-party blockchain network.
Nevertheless, the aggregation stage  in BFL frameworks is susceptible  to pooling and collusion attacks  from  third-party blockchain networks \cite{10884002,9833437,9969914}.
The risk  of  pooling or collusion attacks undermines the decentralization of blockchain,    compromising its security and degrading  FL performance.
To mitigate  these   threats, Decentralized  Federated Learning (DFL) solutions \cite{chen2018machine, 9664296, 9945975, 10177803,10439977,wang2025decentralized} have emerged.
The DFL framework   {\color{blue}integrates}  training and mining into the clients,  allowing them  to communicate directly  with  neighboring nodes   without relying on any  third-party.
\begin{figure}[t]
  \centering
  \subfloat[MNIST]
  {
      \label{6710051}  \includegraphics[width=0.46\linewidth]{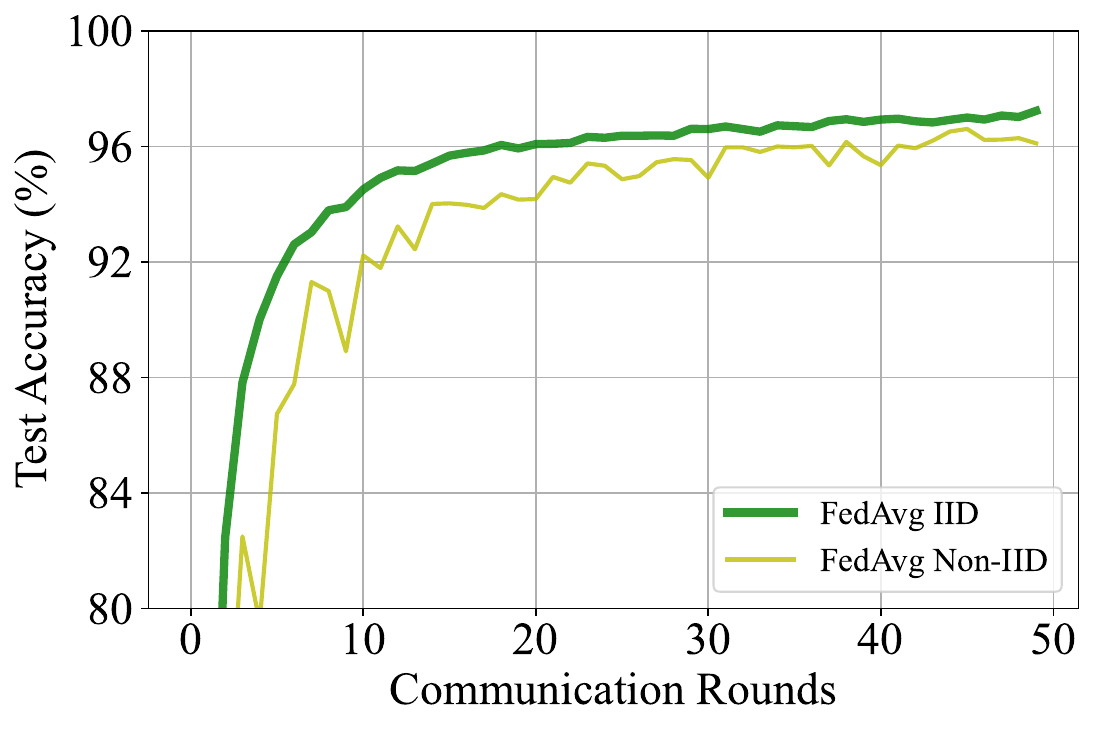}
  }
  \subfloat[CIFAR10]
  {
      \label{6710051}  \includegraphics[width=0.456\linewidth]{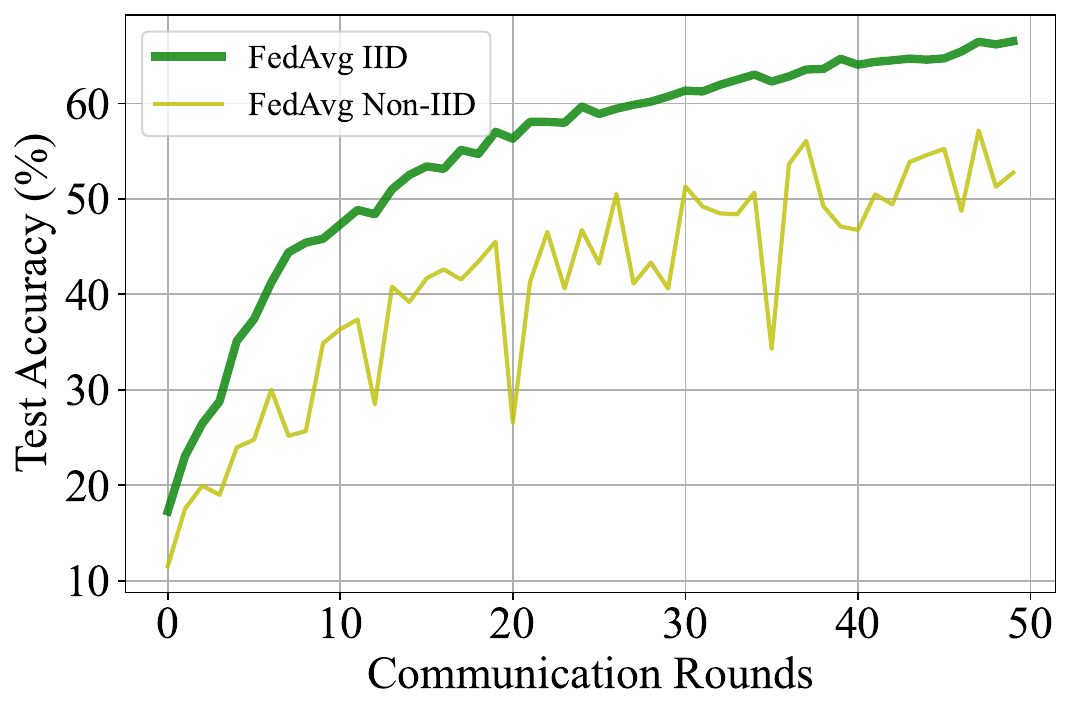}
  }
\caption{The performance of DFL using the FedAvg \cite{mcmahan2017communication} aggregation strategy  in IID and Non-IID settings, where the Non-IID data is simulated using a Dirichlet distribution \cite{lin2016dirichlet}    with a concentration parameter of 0.1.}
\label{6710051}
\end{figure}

However, in practical scenarios, the performance of DFL rapidly degrades when the data distribution is statistically heterogeneous \cite{9835537}\cite{10468591}.
Specifically,  the local data stored on  clients is  usually non-independent and identically distributed (Non-IID), which leads to data heterogeneity.
The Non-IID data   causes divergent  local optimal solutions among clients, resulting  in performance degradation of DFL, as shown in Fig. \ref{6710051}.
 {\color{blue}This is because   each client's model tends to adapt to its own data distribution during local training, thereby converging to a local optimum suitable for the local distribution rather than the global data distribution \cite{li2020federated}. 
 When these local optima are aggregated, the resulting global model deviates from the true global optimum, which leads to suboptimal FL performance.}
Researchers  have  been devoted to addressing the performance degradation of FL  {\color{blue}caused by}  Non-IID  {\color{blue}data  \cite{li2021model,sun2023fedspeed,10066639,zhang2025swim}.}
They focus on aligning local models {\color{blue}with} global models  {\color{blue}to  improve the consistency of local models}.
However, these works \cite{li2021model,sun2023fedspeed,10066639,zhang2025swim} rely on a fixed central server to perform model  aggregation   and global optimization.
This design contradicts the decentralized nature of DFL, where no individual client can consistently act as a server  throughout  federation training process.
Therefore, how to improve the performance of DFL under heterogeneous data distributions is a current  challenge.

Fortunately, prototype learning has gradually attracted  attention in FL \cite{tan2022fedproto}\cite{mu2023fedproc}. These works leverage   prototype learning  to build global representations by   {\color{blue}aggregating} prototypes submitted by clients.
The so-called ``prototype''  can be viewed  as an abstract representation of a class \cite{Yang2018CVPR}.
For instance, when recognizing a ``cat'', each  client  has its unique ``imaginary picture'' or ``prototype'' to capture  the feature of ``cat''.
By exchanging these prototypes, clients can share  the abstract representations of classes like ``cat''.
This prototype-based knowledge integration is independent  of local class distributions, mitigating  the negative impacts caused by heterogeneous data distributions.
Consequently, a critical question is: \textit{can we leverage prototype learning to address the statistical heterogeneity in decentralized federated learning  and achieve high-performance federated training?}

To answer this question, we propose a  decentralized  federated prototype learning framework, named DFPL, which enhances  the performance of distributed learning under heterogeneous data distributions while reducing the amount of parameters transmitted among clients.
The key novelty of our work lies in: 
(\textbf{i}) DFPL integrates blockchain into federated prototype learning,    enabling each client to perform both  training and  mining locally, which facilitates direct communication  with  neighboring nodes   without relying on any  third-party.
(\textbf{ii})  A novel local optimization function is designed  leveraging prototype learning, where clients exchange only prototypes instead of model parameters or gradients, which mitigates the impact of statistical heterogeneity in decentralized federated learning.
(\textbf{iii})  We  model the computational resources required by DFPL for  {\color{blue}both} training and mining, and further derive the relationship between its convergence and computational resource requirements.



Our main contributions can be highlighted as follows:
\begin{itemize}
\item We  introduce prototype learning into DFL and propose a decentralized federated prototype learning framework (DFPL), in which each client aligns its  prototypes with those of the other clients while minimizing its own classification loss, which effectively  address the challenges caused  by Non-IID data in DFL.

\item
{\color{blue}We provide theoretical convergence analyses of DFPL} {\color{blue}by incorporating} the computational resource allocation between training and mining processes, providing theoretical guidance for deploying DFPL in resource-constrained environments.

\item We experimentally  {\color{blue}examine} the impact of the trade-off between training and mining on model performance under limited computational resource settings.
    In addition, the experiments show  that the DFPL  achieves  communication efficiency and enhanced performance across four benchmark datasets with different Non-IID settings.
\end{itemize}

The rest of the paper is organized as follows. Section \ref{sec3} discusses recent works  related to blockchain in FL and prototype learning.
A detailed presentation of DFPL is provided in Section \ref{sec4}.
Section \ref{sec5} provides the  theoretical analysis about DFPL.
Section \ref{sec6} reviews the experimental results. Finally, Section \ref{sec7}  concludes this paper.
\section{Related Work}\label{sec3}
This section reviews research on federated learning  and prototype learning schemes.

\subsection{FL with Blockchain}

Due to  {\color{blue}privacy and security concerns,} accessing data from distributed clients  is challenging for the server.
To address this concerns, Google proposed FL, where each {\color{blue}client}  submits its local model update  to a central server.
 {\color{blue}The server then aggregates} these local updates to obtain a global model update, thereby effectively {\color{blue}preserving}  data privacy.
However, centralized aggregation is vulnerable to central server attacks and server failures.
To mitigate the potential threats posed by centralized  aggregation,  existing  works \cite{9079513,9272656,9551794,9714771,9399813} {\color{blue}have focused} on BFL frameworks, where clients upload their local model updates to distributed servers for global aggregation.
These frameworks utilize the advantages of blockchain, i.e., anonymity, being tamper-proof, and traceability, to introduce third-party blockchain networks into the FL system to replace the central server.
For instance, the work in \cite{9551794} {\color{blue}proposed} a BFL framework that uses blockchain technology for model aggregation, {\color{blue}thereby} avoiding the single point of failure caused  by a centralized server and ensuring  security for  cross-domain clients.
Similarly, the study   in \cite{9399813} {\color{blue}proposed} a blockchain-enabled asynchronous federated learning architecture,  in which  each client uploads its local model to   distributed servers whenever global aggregation is required.

However, the distributed servers that comprise third-party blockchain networks can launch  pooling and collusion attacks against the federated learning system.
Specifically, miners controlling  the network's mining hash rate  may  manipulate  model aggregation by rejecting  legitimate blocks and generating  biased ones.
To avoid the concern, DFL \cite{chen2018machine, 9664296, 9945975, 10177803,10439977,wang2025decentralized} frameworks have been proposed,  which integrate  blockchain technology to {\color{blue}achieve} a fully decentralized FL system.
Unlike   third-party blockchain networks,   DFL frameworks  embed both training and mining functionality within  clients.
For example, the works \cite{chen2018machine} and \cite{9664296} apply blockchain with  Proof of Work (PoW) consensus into DFL, and {\color{blue}assign}    training and mining tasks  to clients, where the former exchanges gradients among clients, and the latter exchanges model parameters, thereby effectively addressing {\color{blue}the} dependency on third-party networks.
 This design makes DFL  suitable  for practical applications, such as  healthcare \cite{9754271}, Industry 4.0 \cite{10457953}, and mobile services \cite{9783194}.

Later, other studies have explored DFL in depth from  various  {\color{blue}perspectives}  \cite{9716792,9713700,10323597,10056968}.
Specifically, the work in \cite{9716792} {\color{blue}designed} a robust decentralized stochastic gradient descent approach to solve unreliable communication {\color{blue}protocols} in DFL.
The recent work in \cite{9713700} proposed a  DFL framework that balances communication efficiency and model performance  by periodically implementing local updates and communication.
The {\color{blue}study in} \cite{10323597} proposed an incentive mechanism  to encourage clients to {\color{blue}faithfully}  follow the protocols of DFL.
The work in \cite{10056968} proposed  a privacy-preserving and reliable DFL {\color{blue}framework that aims to enable}  batch joining and leaving of clients while minimizing  delay and achieving {\color{blue}a}  high  accuracy model.
Collectively, these above works demonstrate  that  DFL has a very popular research domain from various  perspectives.
 
\subsection{FL with Non-IID Data}
 {\color{blue}
To mitigate the performance degradation of federated learning under Non-IID data, common strategies include improving the model aggregation process and introducing additional constraints during local training.
Specifically, several studies have proposed alternative model aggregation schemes on the server side.
For example, the work in \cite{hsu2019measuring} incorporates a global momentum term at the server to guide the direction of global model updates.
In addition, the work in \cite{NEURIPS202012} normalizes the model updates submitted by each client before aggregation.
 Similarly, FedBN \cite{li2021fedbn} aggregates layer-wise normalized model updates to obtain the global model.
In addition, the work  in \cite{Wang_2024_CVPR} trains the global model on the server using condensed  data and soft labels received from clients.
Different from the above model aggregation-based works, many studies alleviate the impact of data heterogeneity by adding constraints to local training.
Specifically, FedProx \cite{li2020federated} introduces a prox-term into local training to address client drift caused by local training overfitting, and FedDyn \cite{durmus2021federated} incorporates a dynamic regularization term into local training so that each client's loss better aligns with the global loss.
Distinct from the above works \cite{li2020federated}\cite{durmus2021federated}, MOON \cite{li2021model} enhances the similarity between the global and local models by adding contrastive regularization to the local optimization function.
Inspired by MOON, the works in \cite{10066639}   \cite{zhang2025swim} have proposed several variants of contrastive regularization.
Although the above works can effectively improve the performance of federated learning under heterogeneous data distributions, they  rely on the centralized federated learning paradigm and  cannot be directly applied in DFL.
}

\subsection{Prototype Learning}
The concept of a prototype,  defined  as the average of multiple feature representations, has been explored in  various machine learning tasks.
In image classification, a prototype is the representation of a class, computed as the average of the feature vectors within that class \cite{NIPS2017cb8da676}.
Additionally, representing a sentence by averaging its word embeddings can achieve competitive performance on natural language processing tasks \cite{Babenko2015ICCV}.
Since prototypes  represent     {\color{blue}abstract  knowledge},  several studies have employed them to  deal with statistical heterogeneity of data distributions in federated learning.
For instance, the work in \cite{michieli2021prototype}  leverages prototypes to  compute client deviations,  {\color{blue}which are then used to guide} federated optimization.
 Similarly, FedProc \cite{mu2023fedproc}  is designed to align  local prototypes with global prototypes during model training.
Their experiments demonstrate that  exchanging prototypes  offers   significant advantages in handling  statistical heterogeneity.

Prototype learning has been integrated with FL to effectively address the statistical heterogeneity issue, and blockchain has been applied to achieve DFL.
However, the interaction  between prototype learning and DFL  remains  unexplored.
 {\color{blue}Our work differs from existing  based-prototype studies \cite{tan2022fedproto}\cite{mu2023fedproc} \cite{michieli2021prototype} in the following aspects:
(\textit{i}) Our work introduces prototype learning into DFL  to tackle  the statistical heterogeneity  through  exchanging prototypes instead of model parameters or gradients.
(\textit{ii}) Our  work integrates training and mining at the client side and  records the verified global prototypes on the blockchain.
(\textit{iii}) We provide detailed convergence analysis of DPFL in conjunction with the allocation of computational resources between training and mining.}
\section{Design Details of DFPL}\label{sec4}
This section  outlines    our design goals,  presents the  DFPL framework,  provides a detailed explanation of local model training, and models     the required computational resources  for both training and mining within  DFPL.
\subsection{Design Goals}
Our study utilizes prototype learning to design {\color{blue}a}  distributed federated prototype learning framework  that overcomes  statistical heterogeneity of  data distributions  and achieve high-performance federated training.
In addition, we  consider communication efficiency   for real-world deployments.
Therefore, DFPL {\color{blue}is designed to meet} the following design goals:
\begin{itemize}
\item \textbf{Accuracy}:
DFPL should {\color{blue}achieve high performance}  across  various data distributions and exhibit  superior generalization capability on diverse datasets.
\item \textbf{Efficiency}:
DFPL should reduce the number of communication parameters transmitted among clients compared to other FL algorithms.
\end{itemize}
\subsection{Framework of DFPL} \label{need}
\begin{figure*}[!t]
\centering
\centerline{\includegraphics[width=0.9\linewidth]{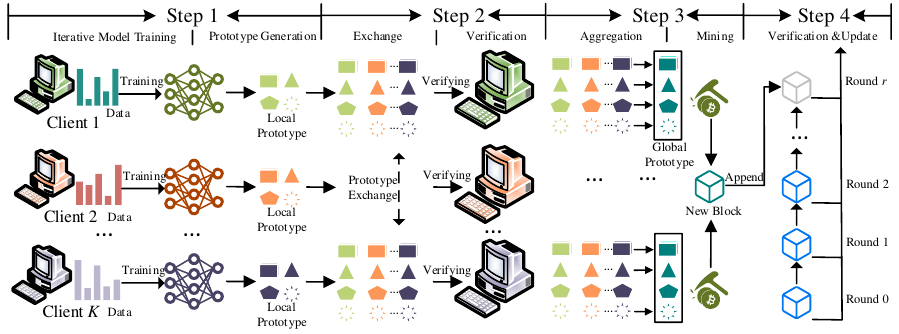}}
\caption{The four steps of the DFPL framework during the $r$-th communication round.}
\label{010804}
\end{figure*}

In this subsection, we  present the  DFPL framework.
Let $\mathcal{K} = \{1,\cdots, K\}$ denote the  set of clients.
Our DFPL consists of $K$ clients, where each client $k$ has its own local dataset $\mathcal{D}_k =\{(\textit{\textbf{x}}_{k,(d)},y_{k,(d)})\}_{d = 1}^{|\mathcal{D}_k|}$, and $|\mathcal{D}_k|$ represents the amount of  {\color{blue}samples} at the   client $k$,  and $\textit{\textbf{x}}_{k,(d)}/y_{k,(d)}$ denotes the feature$/$label of sample $d$ at the   client $k$.
The  DFPL framework  is illustrated  in Fig. \ref{010804}, where  each client integrates both the training and mining processes.
Specifically, for an  arbitrary communication round, each client performs  the   following steps:
\begin{itemize}
\item \textit{Step 1 (Local model training)}.
At the beginning  of {\color{blue}the} $r$-th communication round, each client iteratively  trains its  local model using its dataset $\mathcal{D}_k$.
After  $E$ local iterations, each client generates  its local prototypes for {\color{blue}exchange}.
The detailed training procedure  is  described in Section \ref{11252031}.

\item \textit{Step 2 (Exchange and verification)}.
Each client signs its local prototypes using  digital signature technology, and then exchanges them  with  neighboring   clients.
Subsequently, all  clients  verify the  signatures of the {\color{blue}received  prototypes}, and store the verified  prototypes locally.

\item \textit{Step 3 (Aggregation and mining)}.
In the step, each client computes global prototypes by aggregating  the verified local prototypes,  and then appends  these global prototypes  to its candidate block.
Then, the client applies PoW \cite{gervais2016security} consensus  to  adjust  the nonce  until the resulting hash value  is lower than the target threshold  defined by the block generation difficulty.
The first client to find a valid nonce becomes the mining winner and is authorized  to add its candidate block to the blockchain.

\item \textit{Step 4 (Validation  and updating)}.
The mining winner distributes the new block across the blockchain  network.
Each client    verifies the new block by comparing  global prototypes  contained in the block with its computed global prototypes in \textit{Step 3}.
If a majority of  clients verify the new block, the new block is appended  to the blockchain.
Finally, each client updates its locally stored global prototypes for the next communication round.
\end{itemize}
The above  {\color{blue}four steps are iteratively executed} until the predefined  number of communication rounds are reached.

\subsection{Prototype Calculation  and   Objective of DFPL}

{\color{blue}This section discusses} the  prototype calculation and optimization function employed  by each client in DFPL.
\subsubsection{Prototype Calculation}
For  classification tasks in DFPL, we assume that   each client's local model has the same architecture, consisting of a feature extractor and a decision classifier.
The computation of  prototypes depends on the  feature extractor.
Formally, let $f(\textit{\textbf{r}}_k;\cdot)$ and $g(\textit{\textbf{z}}_k;\cdot)$ denote the feature extractor and decision classifier, respectively, where  $\textit{\textbf{r}}_k$ and $\textit{\textbf{z}}_k$ are the parameters of feature extractor and  decision classifier at client $k$.
Consequently,  the local model of  client $k$  is denoted as $\mathcal{F}((\textit{\textbf{r}}_k,\textit{\textbf{z}}_k); \cdot)=g(\textit{\textbf{z}}_k; \cdot)\circ f(\textit{\textbf{r}}_k; \cdot)$, where $\circ$ denotes  the composition operation.
For simplicity, let $\textit{\textbf{w}}_{k}$  denote $(\textit{\textbf{r}}_{k},\textit{\textbf{z}}_{k})$, resulting in $\mathcal{F}((\textit{\textbf{r}}_{k},\textit{\textbf{z}}_{k});\cdot)= \mathcal{F}(\textit{\textbf{w}}_{k};\cdot)$.
Next, we utilize  the above  symbolic representations to describe  the computation of local prototypes.
Let $\mathcal{I}$ denote the  set of classes,  where $i \in \mathcal{I}$ denotes a class.
Let $\textit{\textbf{p}}^{i}_{k}$   denote the local prototype of class $i$  at  client $k$, calculated  as:
\begin{equation}
\begin{aligned}
\textit{\textbf{p}}^{i}_{k}= \frac{1}{|\mathcal{D}^i_k|}\sum_{d =1}^{|\mathcal{D}^i_k|} f(\textit{\textbf{r}}_{k};\textit{\textbf{x}}_{k,(d)}), \label{9261}
\end{aligned}
\end{equation}
where $|\mathcal{D}^i_k|$ denotes the number of samples with class $i$ at  client $k$.
Thus, $\textit{\textbf{p}}^{i}_{k}$ is considered as the feature representation of class $i$ at  client $k$.
The client $k$ can get  local prototypes for each  class $i \in\mathcal{I}$, denoted $\{\textit{\textbf{p}}^{i}_{k}\}_{i \in \mathcal{I}}$.

{\color{blue}Furthermore,  to calculate the  global prototype, it is obtained by  averaging all local prototypes belonging to the same class, formulated as: 
\begin{equation}\scalebox{1}{$
\begin{aligned}
\textit{\textbf{P}}^i= \frac{1}{K}\sum_{k =1}^{K} \textit{\textbf{p}}^i_k, \forall i \in \mathcal{I}, \label{11101120}
\end{aligned}$}
\end{equation}
where $\textit{\textbf{P}}^i$ denotes the global prototype of class $i$.}
 {\color{blue}The process is similar to traditional model parameter aggregation \cite{9664296}, where local results from different clients are averaged to obtain a global update.
The key difference is that model parameter aggregation are more sensitive to local data distribution, leading to significant deviations among different clients. 
In contrast, global prototypes are aggregated based on categorical features, which alleviates deviations introduced by data heterogeneity.}

\subsubsection{Optimization Function}
During local model training, our design objective  for each client is  to align its local prototype with those from other clients,  while minimizing the loss of its local learning task.
To this end, we design a novel local optimization function with two components:  a classification loss term and an auxiliary loss term,   defined as:
\begin{equation}
\begin{aligned}
\mathcal{L}(\textit{\textbf{w}}_k;  \mathcal{D}_k, &\textit{\textbf{p}}_k^i, \textit{\textbf{P}}^i)=\mathcal{L}_S(\mathcal{F}_k(\textit{\textbf{w}}_k;\textit{\textbf{x}}_{k,(d)}),y_{k,(d)})\\
&+\lambda\cdot\mathcal{L}_R\left(\textit{\textbf{p}}_k^i,\textit{\textbf{P}}^i\right),
\forall d\in\{1,\cdots,|\mathcal{D}_k|\}
,\label{120201}
\end{aligned}
\end{equation}
where  $\mathcal{L}_S(\cdot)$ denotes the classification loss term (e.g., the cross-entropy loss),  $\lambda$ is the importance  weight of the auxiliary loss term, and $\mathcal{L}_R(\cdot)$ defines  the auxiliary  loss term  as:
\begin{equation}
\mathcal{L}_R\left(\textit{\textbf{p}}_k^i,\textit{\textbf{P}}^i\right)=\frac{1}{|\mathcal{I}|}\sum_{i \in \mathcal{I}}\|\textit{\textbf{p}}_k^{i}-\textit{\textbf{P}}^{i}\|_2,
\end{equation}
where  $\Vert\cdot\Vert_2$ is the $\ell_2$-norm  used to measure  the distance between the local prototype $\textit{\textbf{p}}_k^{i}$ and the global  prototype $\textit{\textbf{P}}^{i}$.
Each client $k$  performs local  model training  based on the optimization function.

The DFPL aims to  tackle   the federated optimization problem by minimizing  the  sum of loss across all  clients.
Thus, the global optimization function of  DFPL  is formulated as:
\begin{equation}
\begin{aligned}
\mathop{\arg\min}_{\textit{\textbf{w}}_{1},\textit{\textbf{w}}_{2},\cdots,\textit{\textbf{w}}_{k}}\{\sum_{k\in \mathcal{K}}\frac{|\mathcal{D}_k|}{|\mathcal{D}|}&\mathcal{L}_{S}(\mathcal{F}_{k}(\textit{\textbf{w}}_k;\textit{\textbf{x}}_{k,(d)})),y_{k,(d)})\\
&+ \lambda\cdot\sum_{k\in \mathcal{K}}\frac{|\mathcal{D}_k|}{|\mathcal{D}|}\mathcal{L}_{R}\left(\textit{\textbf{p}}_k^i,\textit{\textbf{P}}^i\right)\},
\end{aligned}
\end{equation}
where $|\mathcal{D}|$ denotes the total number of samples from   all clients.
After introducing the  objective of DFPL, the following outlines the detailed process of local model training.

\subsection{Detailed Implementation in Local Model Training} \label{11252031}
The local model training consists of two stages:  iterative model training and   prototype generation.
 The complete  procedure  is outlined  in $\textbf {Algorithm \ref{algorithm:test1}}$.

\subsubsection{Iterative Model Training}
During  the $r$-th communication round, each  client $k$ utilizes its own local dataset $ \mathcal{D}_k$ to iteratively train its local model.
To clearly denote  the state of variables during the iterative process, the superscript $(e)$ represents the current local iteration, and the subscript  $r$ denotes the current communication round.
 Firstly, each client $k$  uses the local model parameters  $\textit{\textbf{w}}_{k,r-1}^{(E)}$ from the  previous   round   as the starting point   for  local model training in the current $r$-th round.
 Subsequently, the detailed iterative process is described  as follows.
\begin{itemize}
  \item  In the $e$-th local  iteration, the client $k$  randomly selects  a subset  $\mathcal{D}_{k,r}^{(e)}$ from its local dataset $\mathcal{D}_{k}$.
  \item The client $k$ calculates the local prototypes for all classes  using  formula (\ref{9261}) to get the set $\{\textit{\textbf{p}}^{i,(e)}_{k, r}\}_{i \in \mathcal{I}}$.
  \item  The empirical loss for client $k$ is calculated using the optimization function $\mathcal{L}(\textit{\textbf{w}}_{k,r}^{(e)};  \mathcal{D}_{k,r}^{(e)},\textit{\textbf{p}}^{i,(e)}_{k, r},\textit{\textbf{P}}^i_r)$.
      The gradient   is computed  as:
\begin{equation}
\textit{\textbf{g}}_{k,r}^{(e)}  = \nabla\mathcal{L}(\textit{\textbf{w}}_{k,r}^{(e)};  \mathcal{D}_{k,r}^{(e)}, \textit{\textbf{p}}^{i,(e)}_{k, r},\textit{\textbf{P}}^i_r),\label{925-2}
\end{equation}
where $\nabla$ denotes derivation operation, $\textit{\textbf{P}}^i_r$ denotes the global prototype of class $i$ in $r$-th communication round.
  \item The model parameters  $\textit{\textbf{w}}_{k,r}^{(e)}$ are  updated according to the following formula:
\begin{equation}
\textit{\textbf{w}}_{k,r}^{(e+1)}  = \textit{\textbf{w}}_{k,r}^{(e )} - \eta \textit{\textbf{g}}_{k,r}^{(e)},\label{925-3}
\end{equation}
where $\eta$ denotes the local learning rate.
\end{itemize}
After  {\color{blue}the}  above process is repeated for $E$ {\color{blue}iterations}, each  client $k$ obtains the local model parameters $\textit{\textbf{w}}_{k,r}^{(E)}$.

\begin{algorithm}[t]
\begin{small}
    \caption{$\textbf {Local\_Model\_Training}$}
    \label{algorithm:test1}
        \LinesNumbered
    \KwIn {$\textit{\textbf{w}}_{k,r-1}^{(E)}$,  $ \mathcal{D}_k$, $\eta$,  $E$,  $\{\textit{\textbf{P}}^{i}_{r}\}_{i\in \mathcal{I}}$}
    \KwOut {$\{\textit{\textbf{p}}_{k,r}^{i,(E)}\}_{i\in \mathcal{I}}$}
$\textit{\textbf{w}}_{k,r}^{(0)} = \textit{\textbf{w}}_{k,r-1}^{(E)}$;\\
    \For{each $e$ $\in$ $\{0,1,\cdots,E-1\}$}{
	    	Randomly samples  $\mathcal{D}_{k}^{(e)}\subset \mathcal{D}_k$;\\
    \For{each $i$ $\in$ $\mathcal{I}$}{
Compute $\textit{\textbf{p}}^{i,(e)}_{k,r}$  using  formula (\ref{9261}) ;\\}
Compute	 $\mathcal{L}(\textit{\textbf{w}}_{k,r}^{(e)};  \mathcal{D}_{k,r}^{(e)}, \textit{\textbf{p}}^{i,(e)}_{k, r},\textit{\textbf{P}}^i_r)$;\\
Calculate  gradient	$\textit{\textbf{g}}_{k,r}^{(e)} = \nabla\mathcal{L}(\cdot)$;\\
Update model  parameters 	$\textit{\textbf{w}}_{k,r}^{(e+1)} = \textit{\textbf{w}}_{k,r}^{(e)} - \eta \textit{\textbf{g}}_{k,r}^{(e)}$;\\
    }
        \For{each $i$ $\in$ $ \mathcal{I}$}{
		$\textit{\textbf{p}}^{i,(E)}_{k,r}= \frac{1}{| \mathcal{D}^i_k|}\sum_{d= 1}^{| \mathcal{D}^i_k|} f_k(\textit{\textbf{r}}_{k,r}^{(E)};x_{k,(d)})$;\\
    }
    \Return{$\{\textit{\textbf{p}}_{k,r}^{i,(E)}\}_{i\in \mathcal{I}}$}
    \end{small}
\end{algorithm}
\subsubsection{Prototype Generation}
During the iterative model training stage, each client $k$ generates its  local prototypes based on  {\color{blue}its} current model parameters and dataset to calculate the auxiliary loss.
Consequently, the local prototypes evolve dynamically throughout the iterative training process.
To exchange more representative prototypes, the  clients  perform  the prototype generation stage.
Specifically, after completing the iterative model training, the client $k$ obtains  the local model  parameters $\textit{\textbf{w}}_{k,r}^{(E)}$.
 {\color{blue}Using} these parameters,   {\color{blue}each client feeds}  their own local dataset $\mathcal{D}_k$ into  feature extractor $f(\textit{\textbf{r}}_{k,r}^{(E)};\cdot)$ to generate   local prototypes $\{\textit{\textbf{p}}_{k,r}^{i,(E)}\}_{i\in \mathcal{I}}$ for  {\color{blue}exchange}, as shown in line 10 of Algorithm \ref{algorithm:test1}.

\subsection{Computation Time Modeling}\label{11221248}
In this subsection, we model the  time  required for training and mining to explore their relationship with the convergence of DFPL.

\textit{Model Iteration Rate}.
The model iteration rate  refers to the  time required to complete one  local iteration.
The training time for each local iteration at the  client $k$ is calculated as:
\begin{equation}
\alpha_k =   \frac{|\mathcal{D}_k^{(e)}|\rho}{f},
\end{equation}
where $|\mathcal{D}_k^{(e)}|$ denotes  the number of samples in $e$-th local iteration for client $k$, and $\rho$ is the number of CPU cycles required to train single  sample,  and $f$ denotes the CPU cycles per second for each client.
We assume  that all clients have identical hardware resources, and process  the same number of local samples per  local iteration.
Consequently, the training time is uniform across clients, i.e.,   $\alpha = \alpha_k, \forall_k$.

\textit{Block Generation Rate}.
The block generation rate refers to the  time required to mine a single block during one  communication   round, which is  determined by the computational complexity of  the hash function and the total computational power of  blockchain network.
In PoW,  the average CPU cycles required to generate a block    {\color{blue}are} defined as $ \mathbb{E}[\text{PoW}] = \mu\tau$, where $\tau$ denotes the average {\color{blue}total number} of CPU cycles required to generate a block, $\mu$ represents the mining difficulty.
 Thus, the average mining time for generating  one block is defined as:
\begin{equation}
\beta =   \frac{\mathbb{E}[\text{PoW}]}{K f} = \frac{\mu\tau}{K f},
\end{equation}
where  $K$ denotes the number of clients in DFPL.

Consider that a typical FL learning task is completed within a fixed time duration of  $t_{sum}$.
Under the same hardware conditions, each client performs   $ft_{sum}$ CPU cycles  within the time $t_{sum}$, which is expressed as:
\begin{equation}
ft_{sum} = (fE\alpha + f\beta)R+ \gamma, \label{1130}
\end{equation}
where $fE\alpha$  {\color{blue}represents}  the CPU cycles required for  local model training  by  client $k$, and $f\beta$  represents   the   CPU cycles required for mining in one communication round, and $R$ is the total  number of communication rounds.
Additionally, $\gamma$ denotes  the CPU cycles consumed  by  extra computations.
Thus,  the following condition holds: $ft_{sum} \geq (fE\alpha + f\beta)R$.
Since the  CPU cycles required for  extra computations are negligible compared to those required for one communication round, i.e., $\gamma \ll (fE\alpha + f\beta)$,  the formula (\ref{1130}) can be simplified as:
\begin{equation}
E = \left \lfloor \frac{1}{\alpha}\left( \frac{t_{sum}}{R}-\beta\right) \right\rfloor, \label{113001}
\end{equation}
where $\left\lfloor \cdot \right\rfloor$ denotes the floor function.
Therefore, we ignore the CPU cycles consumed by the extra computation $\gamma$, and use $t_{sum} = (E\alpha + \beta)R$ in  convergence analysis.

\section{Analysis}\label{sec5}
This section presents the analysis of the convergence, complexity, and privacy of DFPL.
\subsection{Convergence Analysis}
We make the following assumptions about the local optimization  function (i.e., formula (\ref{120201})), which  are consistent with  those commonly adopted  {\color{blue}by}  existing  FL frameworks \cite{li2020federated,NEURIPS202012,tan2022fedproto}. 
These assumptions are essential for {\color{blue}the} convergence analysis  of  gradient-based method \cite{bottou2018optimization}.
In the following assumptions, we introduce  subscripts $(rE+e)$ and $k$ to   the optimization function $\mathcal{L}_{k, rE+e}$ to indicate the state at the $(rE+e)$-th iteration for client $k$.

\begin{assumption}
(Lipschitz Smooth)  The local optimization function of each client $k$ is $L_1$-Lipschitz smooth, which means that the gradient of the local optimization function is $L_1$-Lipschitz continuous,
\begin{equation}
\begin{gathered}
\left\|\nabla \mathcal{L}_{k, rE+e_1}-\nabla \mathcal{L}_{k, rE+e_2}\right\|_2 \leq L_1\left\|\textit{\textbf{w}}_{k,r}^{(e_1)}-\textit{\textbf{w}}_{k,r}^{(e_2)}\right\|_2.
\end{gathered}
\end{equation}
This also implies the following quadratic bound,
\begin{equation}\scalebox{0.9}{$
\begin{aligned} 
\mathcal{L}_{k, rE+e_1}-\mathcal{L}_{k, rE+e_2} \leq&\left\langle\nabla \mathcal{L}_{k, rE+e_2},\left(\textit{\textbf{w}}_{k,r}^{(e_1)}-\textit{\textbf{w}}_{k,r}^{(e_2)}\right)\right\rangle\\
&+\frac{L_1}{2}\left\|\textit{\textbf{w}}_{k,r}^{(e_1)}-\textit{\textbf{w}}_{k,r}^{(e_2)}\right\|_2^2.
\end{aligned}$}
\end{equation}
\end{assumption}

\begin{assumption}
(Unbiased Gradient and Bounded Variance)
The stochastic gradient $\textit{\textbf{g}}_{k, r}^{(e)}=\nabla \mathcal{L}\left(\textit{\textbf{w}}^{(e)}_{k,r}, \mathcal{D}_k^{(e)}\right)$ is an unbiased estimator of the local gradient for each client $k$.
Suppose its expectation:
\begin{equation}
\mathbb{E}_{\mathcal{D}_k^{(e)} \sim
 \mathcal{D}_k}\left[\textit{\textbf{g}}_{k, r}^{(e)}\right]=\nabla \mathcal{L}\left(\textit{\textbf{w}}_{k, r}^{(e)}\right),
\end{equation}
and its variance is bounded by $\sigma^2$:
\begin{equation}
\mathbb{E}[\|\textit{\textbf{g}}_{k,r}^{(e)}-\nabla\mathcal{L}(\textit{\textbf{w}}^{(e)}_{k,r})\|_2^2]\leq\sigma^2, \quad\sigma^2\geq0.
\end{equation}
\end{assumption}

\begin{assumption}
(Bounded Expectation of Euclidean norm of Stochastic Gradients). The expectation of the stochastic gradient is bounded by G:
\begin{equation}
\mathbb{E}[\|\textit{\textbf{g}}_{k,r}^{(e)}\|_2]\leq G.
\end{equation}
\end{assumption}

\begin{assumption}
(Lipschitz Continuity). Each feature extractor is $L_2$-Lipschitz continuous, that is,
\begin{equation}
\begin{gathered}
\|f_k(\textit{\textbf{r}}_{k,r}^{(e_1)})-f_k(\textit{\textbf{r}}_{k,r}^{(e_2)})\|_2\leq L_2\|\textit{\textbf{r}}_{k,r}^{(e_1)}-\textit{\textbf{r}}_{k,r}^{(e_2)}\|_2.
\end{gathered}
\end{equation}
\end{assumption}

Notably, we introduce  ``$\frac{1}{2}$'' to the process of  local iteration, denoted as $\{\frac{1}{2}, 1, \cdots, E-1\}$, where $rE$ represents the time step before prototype aggregation in the $r$-th round, and $rE + 1/2$ represents the time step between prototype aggregation and the first local iteration of the $(r+1)$-th round.

\begin{thm} (Upper Bound on Variation)
Let Assumption 1 to 4 hold, for   an arbitrary client $k$, after each communication round, the variation of the local optimization  function  of DFPL  is bounded as follows:
\begin{equation}
\begin{aligned}
\mathbb{E}&\left[\mathcal{L}_{k,(r+1) E+\frac{1}{2}}\right] -\mathcal{L}_{k,r E+\frac{1}{2}} \leq j(\lambda, \eta, \alpha, \beta,R, t_{sum}),\nonumber
\end{aligned}
\end{equation}
where
\begin{equation}
\begin{aligned}
&j(\lambda, \eta, \alpha, \beta,R,  t_{sum}) = \\&\left(\frac{L_1 \eta^2}{2}-\eta\right) Q+ \left(\frac{L_1\eta^2\sigma^2}{2}+ L_2\eta G\lambda\right)\left(\frac{t_{sum}-\beta R}{\alpha R}\right),\nonumber
\end{aligned}
\end{equation}
and $Q =\sum_{e=\frac{1}{2}}^{E-1}\left\|\nabla \mathcal{L}_{k,rE+e}\right\|_2^2$.
\end{thm}
 {\color{blue}Theorem 1   provides  the variation bound of the optimization function for each client in DFPL under computational resource constraints.
When all clients are equipped with identical hardware, this theorem is applicable to any blockchain-assisted federated learning based on prototype aggregation.}
The upper bound indicates  that the learning performance depends on the total number of  rounds
$R$, the  training time $\alpha$ per local iteration, the average mining time $\beta$ per block, the learning rate $\eta$, and the total computing time $ t_{sum}$.

\begin{proof}
Let Assumption 1 and 2 hold, for  an arbitrary client $k$,  we can obtain the following formula:
\begin{equation}
\begin{aligned}
&\mathcal{L}_{k, rE+1}-\mathcal{L}_{k, rE+\frac{1}{2}} \leq \\& \left\langle\nabla \mathcal{L}_{k, rE+\frac{1}{2}},(\underbrace{\textit{\textbf{w}}_{k,r}^{(1)}-\textit{\textbf{w}}_{k,r}^{(\frac{1}{2})}}_{A_1})\right\rangle+\frac{L_1}{2}\|\underbrace{\textit{\textbf{w}}_{k,r}^{(1)}-\textit{\textbf{w}}_{k,r}^{(\frac{1}{2})}}_{A_1}\|_2^2.\label{5301557}
\end{aligned}
\end{equation}
Since     $\textit{\textbf{w}}_{k,r}^{(1)} = \textit{\textbf{w}}_{k,r}^{(\frac{1}{2})} - \eta\textit{\textbf{g}}_{k,r}^{(\frac{1}{2})}$, we can get $A_1 = - \eta\textit{\textbf{g}}_{k,r}^{(\frac{1}{2})}$, so the formula (\ref{5301557}) can be rewritten  as:
\begin{equation}\scalebox{0.9}{$
\begin{aligned}
&\mathcal{L}_{k, rE+1}-\mathcal{L}_{k, rE+\frac{1}{2}} \leq- \eta\left\langle\nabla \mathcal{L}_{k, rE+\frac{1}{2}},\textit{\textbf{g}}_{k,r}^{(\frac{1}{2})}\right\rangle+\frac{L_1\eta^2}{2}\left\|\textit{\textbf{g}}_{k,r}^{(\frac{1}{2})}\right\|_2^2.\nonumber\label{5301609}
\end{aligned}$}
\end{equation}
Taking expectation over both sides of the above formula on the  dataset $\mathcal{D}_k^{(e)}$, we can obtain the following:
\begin{equation}\scalebox{0.9}{$
\begin{aligned}
&\mathbb{E}\left[\mathcal{L}_{k, rE+1}\right] -\mathcal{L}_{k, rE+\frac{1}{2}}\\&\leq-\eta\mathbb{E}\left[\left\langle\nabla \mathcal{L}_{k, rE+\frac{1}{2}},\textit{\textbf{g}}_{k,r}^{(\frac{1}{2})}\right\rangle\right] +\frac{L_1\eta^2}{2}\underbrace{\mathbb{E}[\|\textit{\textbf{g}}_{k,r}^{(\frac{1}{2})}\|_2^2]}_{A_2}\\
&\overset{\text{(a)}}{=} \frac{L_1\eta^2}{2}(\|\nabla \mathcal{L}_{k, rE+\frac{1}{2}}\|_2^2+Var(\textit{\textbf{g}}_{k,r}^{(\frac{1}{2})}))-\eta\mathbb{E}[\langle\nabla \mathcal{L}_{k, rE+\frac{1}{2}},\textit{\textbf{g}}_{k,r}^{(\frac{1}{2})}\rangle]\\
&\overset{\text{(b)}}{=}-\eta \|\nabla \mathcal{L}_{k, rE+\frac{1}{2}}\|_2^2+\frac{L_1\eta^2}{2}\left(\|\nabla \mathcal{L}_{k, rE+\frac{1}{2}}\|_2^2+Var(\textit{\textbf{g}}_{k,r}^{(\frac{1}{2})})\right)\\
&=\left(\frac{L_1\eta^2}{2}-\eta\right)\|\nabla \mathcal{L}_{k, rE+\frac{1}{2}}\|_2^2+\frac{L_1\eta^2}{2}Var\left(\textit{\textbf{g}}_{k,r}^{(\frac{1}{2})}\right)\\
&\overset{\text{(c)}}{\leq} \left(\frac{L_1\eta^2}{2}-\eta\right)\|\nabla \mathcal{L}_{k, rE+\frac{1}{2}}\|_2^2+\frac{L_1\eta^2}{2}\sigma^2, \nonumber
\end{aligned}$}
\end{equation}
where (a) follows from $Var(x) =\mathbb{E}[x^2]-(\mathbb{E}[x])^2$, we can get:
\begin{equation}\scalebox{0.9}{$
\begin{aligned}
\underbrace{\mathbb{E}[\|\textit{\textbf{g}}_{k,r}^{(\frac{1}{2})}\|^2_2]}_{A_2}& = Var(\|\textit{\textbf{g}}_{k,r}^{(\frac{1}{2})}\|_2)+(\mathbb{E}[\|\textit{\textbf{g}}_{k,r}^{(\frac{1}{2})}\|_2] )^2\\&\overset{\text{(d)}}{=}Var(\|\textit{\textbf{g}}_{k,r}^{(\frac{1}{2})}\|_2)+\|\nabla \mathcal{L}_{k, rE+\frac{1}{2}}\|_2^2, \nonumber
\end{aligned}$}
\end{equation}
where (d) follows from Assumption 2.
The (b) and (c) follow from Assumption 2.
Then, after $E$ local  iterations, we can get:
\begin{equation}\scalebox{0.9}{$
\begin{aligned}
&\mathbb{E}\left[\mathcal{L}_{k, r(E+1)}\right] -\mathcal{L}_{k, rE+\frac{1}{2}}\leq\\&\left(\frac{L_1\eta^2}{2}-\eta\right)\sum_{e = \frac{1}{2}}^{E-1}\|\nabla \mathcal{L}_{k, rE+\frac{1}{2}}\|_2^2+\frac{L_1E\eta^2}{2}\sigma^2.\\ \label{5301501}
\end{aligned}$}
\end{equation}
 {\color{blue}Additionally, let Assumption 3 and 4 hold.
The optimization function is updated after the client receives the latest global prototype,  which satisfies the following relation:}
\begin{equation}\scalebox{0.9}{$
\begin{aligned}
&\mathcal{L}_{k, (r+1) E+\frac{1}{2}}-\mathcal{L}_{k, (r+1) E} \\
&= \lambda\frac{1}{|\mathcal{I}|}\sum_{i \in \mathcal{I}}\|\textit{\textbf{p}}_{k,r+1}^{i,(E)}-\textit{\textbf{P}}^{i}_{r+2}\|_2 -\lambda\frac{1}{|\mathcal{I}|}\sum_{i \in \mathcal{I}}\|\textit{\textbf{p}}_{k,r+1}^{i,(E)}-\textit{\textbf{P}}^{i}_{r+1}\|_2 \\
&\overset{\text{(e)}}{\leq}\lambda\frac{1}{|\mathcal{I}|}\sum_{i \in \mathcal{I}}\|\textit{\textbf{P}}^{i}_{r+2}-\textit{\textbf{P}}^{i}_{r+1}\|_2\\&\overset{\text{(f)}}{\leq}\lambda\|\textit{\textbf{P}}^{i^\prime}_{r+2}-\textit{\textbf{P}}^{i^\prime}_{r+1}\|_2\overset{\text{(g)}}{=}\lambda\|\frac{1}{K} \sum_{k \in \mathcal{K}} \textit{\textbf{p}}_{k,r+1}^{i^\prime,(E)}-\frac{1}{K} \sum_{k \in \mathcal{K}} \textit{\textbf{p}}_{k,r }^{i^\prime,(E)}\|_2\\
&\overset{\text{(h)}}{=}\lambda \|\frac{1}{K} \sum_{k \in \mathcal{K}}\frac{1}{|\mathcal{D}^{i^\prime}_k|}\sum_{d = 1} ^{|\mathcal{D}^{i^\prime}_k|}(f_k(\textit{\textbf{r}}_{k,r+1}^{(E)};\textit{\textbf{x}}_{k,(d)})-f_k(\textit{\textbf{r}}_{k,r}^{(E)};\textit{\textbf{x}}_{k,(d)}))\|_2\\
&\overset{\text{(i)}}{\leq}\lambda \frac{1}{K} \sum_{k \in \mathcal{K}}\frac{1}{|\mathcal{D}^{i^\prime}_k|}\sum_{d = 1} ^{|\mathcal{D}^{i^\prime}_k|}\|f_k(\textit{\textbf{r}}_{k,r+1}^{(E)};\textit{\textbf{x}}_{k,(d)})-f_k(\textit{\textbf{r}}_{k,r}^{(E)};\textit{\textbf{x}}_{k,(d)})\|_2\\
&\overset{\text{(j)}}{\leq}\lambda L_2 \frac{1}{K}\sum_{k \in \mathcal{K}} \|\textit{\textbf{r}}_{k,r+1}^{(E)}-\textit{\textbf{r}}_{k,r}^{(E)} \|_2 \\& \overset{\text{(k)}}{\leq}\lambda L_2\frac{1}{K} \sum_{k \in \mathcal{K}} \|\textit{\textbf{w}}_{k,r+1}^{(E)}-\textit{\textbf{w}}_{k,r}^{(E)} \|_2\\
&\overset{\text{(l)}}{=}\lambda L_2\eta\frac{1}{K} \sum_{k \in \mathcal{K}} \|\sum_{e = \frac{1}{2}}^{E-1}\textit{\textbf{g}}_{k,r}^{(e)}\|_2\leq\lambda L_2\eta \frac{1}{K} \sum_{k \in \mathcal{K}} \sum_{e = \frac{1}{2}}^{E-1}\|\textit{\textbf{g}}_{k,r}^{(e)}\|_2. \nonumber
\end{aligned}$}
\end{equation}
Taking expectation  {\color{blue}over} both sides  {\color{blue}with respect to} random dataset $\mathcal{D}_k^{(e)}$, we can get the following:
\begin{equation}\scalebox{0.9}{$
\begin{aligned}
\mathbb{E}[\mathcal{L}_{k, (r+1) E+\frac{1}{2}}]-\mathcal{L}_{k, (r+1) E} &\leq\lambda L_2 \eta \frac{1}{K} \sum_{k \in \mathcal{K}} \sum_{e = \frac{1}{2}}^{E-1}\mathbb{E}[\|\textit{\textbf{g}}_{k,r}^{(e)}\|_2]\\
&\overset{\text{(m)}}\leq \lambda L_2 \eta EG, \label{5311621}
\end{aligned}$}
\end{equation}
where (e) follows from $\|a-b\|_2-\|a-c\|_2 \leq\|b-c\|_2$,  (f)   follows from class $i^\prime \in \mathcal{I}$  to maximize $\|\textit{\textbf{P}}^{i}_{r+2}-\textit{\textbf{P}}^{i}_{r+1}\|_2$, (g) follows from  the calculation of  global  prototype,  (h) follows from the calculation of  local  prototype, (i) and (m) follows from $\|\sum a_i\|_2 \leq \sum \|a_i\|_2$, (j) follows from $L_2$-Lipschitz continuity in Assumption 4, (k) follows from the fact that $\textit{\textbf{r}}_{k,r+1}^{(E)}$ is a subset of $\textit{\textbf{w}}_{k,r+1}^{(E)}$, (l) follows from parameters update, (m) follows from Assumption 3.

After the $(r+1)$-th communication round, the variation of the optimization  function  is obtained by {\color{blue}substituting} formula (\ref{5301501}) into formula (\ref{5311621}), and is bounded as follows:
\begin{equation}\scalebox{0.9}{$
\begin{aligned}
\mathbb{E}[\mathcal{L}_{k, (r+1) E+\frac{1}{2}}]-\mathcal{L}_{k, rE+\frac{1}{2}}  \leq&(\frac{L_1\eta^2}{2}-\eta)\sum_{e = \frac{1}{2}}^{E-1}\|\nabla \mathcal{L}_{k, rE+e}\|_2^2\\&+\frac{L_1E\eta^2}{2}\sigma^2 +\lambda L_2 \eta EG. \label{611255}
\end{aligned}$}
\end{equation}
Since each client  {\color{blue}performs} both training and mining in DFPL,   the relationship between  the training time $\alpha$ per round  and mining time $\beta$ per round  is denoted as $t_{sum} = (E\alpha + \beta)R$, as described in Section \ref{11221248}.  {\color{blue}Substituting} $E= \frac{t_{sum}-\beta R}{\alpha R}$ into formula (\ref{611255}), it yields
\begin{equation}\scalebox{0.9}{$
\begin{aligned}
&\mathbb{E}\left[\mathcal{L}_{k,(r+1) E+\frac{1}{2}}\right] -\mathcal{L}_{k,r E+\frac{1}{2}} \leq j(\lambda, \eta, \alpha, \beta, R, t_{sum})\\
 &= \left(\frac{L_1 \eta^2}{2}-\eta\right) Q+ \left(\frac{L_1\eta^2\sigma^2}{2}+ \lambda L_2 \eta  G\right)\left(\frac{t_{sum}-\beta R}{\alpha R}\right),
 \label{113003}\nonumber
\end{aligned}$}
\end{equation}
where  {\color{blue}the relation} $Q = \sum_{e=\frac{1}{2}}^{E-1}\left\|\nabla \mathcal{L}_{k,r E+e}\right\|_2^2$  {\color{blue}holds}.  Thus, Theorem 1 is proved.
\end{proof}

\begin{corollary}
 (DFPL Convergence).
 {\color{blue}Given fixed} $\alpha, \beta, R,$ and $t_{sum}$, for an arbitrary  client $k$ in DFPL, the optimization function  monotonically decreases in communication round  when
\begin{equation}
\eta^{(e^{\prime})}_{k,r}<\frac{2(\alpha R\sum_{e=\frac{1}{2}}^{e=e^{\prime}}\left\|\nabla \mathcal{L}_{k,r E+e}\right\|_2^2-(L_{2}t_{sum}+ L_{2}\beta R)\lambda G}{L_{1}(\alpha R\sum_{e=\frac{1}{2}}^{e=e^{\prime}}\left\|\nabla \mathcal{L}_{k,r E+e}\right\|_2^2+t_{sum}\sigma^{2}-\sigma^{2}\beta R)},\label{01013005}\nonumber
\end{equation}
and
\begin{equation}
\lambda_{k,r}<\frac{\alpha R\left \|\nabla \mathcal{L}_{k,r E+\frac{1}{2}}\right\|_2^2}{L_{2}G(t_{sum}-\beta R)},\label{010301}\nonumber
\end{equation}
where  $\eta^{(e^{\prime})}_{k,r}$ denotes the local  learning rate of client $k$ at $e$-th local iteration  ($e^{\prime} \in \{\frac{1}{2},1, \cdots,  E-1 \}$), $\lambda_{k,r}$ denotes the importance weight  for the auxiliary  loss item of client $k$ in the $r$-th communication round.
\end{corollary}
Corollary 1 guarantees  that the variation bound is negative, ensuring  convergence of the optimization function. It provides guidance  the choice of appropriate values for $\lambda, \eta, \alpha, \beta,R$, and $t_{sum}$ to achieve  convergence.
\begin{proof}
For  an arbitrary client $k$, we need to ensure that the  optimization function  decreases    after each communication round, so we have:  $$j(\lambda, \eta, \alpha, \beta, R, t_{sum}) \textless 0.$$
From the relation, we can derive:
\begin{equation}\scalebox{0.9}{$
 \left(\frac{L_1 \eta^2}{2}-\eta\right)Q+ \left(\frac{L_1\eta^2\sigma^2}{2}+ L_2\eta G\lambda\right)\left(\frac{t_{sum}-\beta R}{\alpha R}\right)\textless 0. \label{0113005}$}
\end{equation}
Simplifying formula (\ref{0113005}), we obtain:
\begin{equation}
\eta<\frac{2(\alpha RQ-\lambda L_{2}t_{sum}G+\lambda L_{2}G\beta R)}{L_{1}(\alpha RQ+t_{sum}\sigma^{2}-\sigma^{2}\beta R)}.\label{0113007}
\end{equation}
To guarantee  $\eta >0$, the following relationship must hold:
$$
\frac{2(\alpha RQ-\lambda L_{2}t_{sum}G+\lambda L_{2}G\beta R)}{L_{1}(\alpha RQ+t_{sum}\sigma^{2}-\sigma^{2}\beta R)} >0.
$$
Since  $t_{sum} > \beta R$,  the  denominator of above  formula is always  positive. Thus, we  need to make sure that the numerator is also positive, so we can get
\begin{equation}
\lambda<\frac{\alpha RQ}{L_{2}G(t_{sum}-\beta R)}.\label{0113008}
\end{equation}
Additionally, we need to make sure that each client's optimization function decreases  {\color{blue}at every local iteration}. Formally, $\eta^{(e^{\prime})}_{k,r}$ and $\lambda_{k,r}$  {\color{blue}have to} satisfy the formulas (\ref{0113007}) and (\ref{0113008}), where $\eta^{(e^{\prime})}_{k,r}$  denotes the local learning rate of cliant $k$ at the $e^{\prime}$-th local iteration ($e^{\prime} \in \{\frac{1}{2},1, \cdots, E-1\}$),  $\lambda_{k,r}$ denotes the importance weight  for auxiliary  loss  {\color{blue}term} of client $k$ in $r$-th communication round.
Thus, we can easily  get:
\begin{equation}\scalebox{0.8}{$
\eta^{(e^{\prime})}_{k,r}<\underbrace{\frac{2(\alpha R\sum_{e=\frac{1}{2}}^{e=e^{\prime}}\left\|\nabla \mathcal{L}_{k,r E+e}\right\|_2^2-(L_{2}t_{sum}+ L_{2}\beta R)\lambda G}{L_{1}(\alpha R\sum_{e=\frac{1}{2}}^{e=e^{\prime}}\left\|\nabla \mathcal{L}_{k,r E+e}\right\|_2^2+t_{sum}\sigma^{2}-\sigma^{2}\beta R)}}_{A_3}.\label{0101306}$}
\end{equation}
When $e^{\prime} = \frac{1}{2}$, $A_3$ obtains the minimum value.
Thus, according to formula (\ref{0113008})$, \lambda_{k,r}$  needs to satisfy the following:
\begin{equation}
\lambda_{k,r}<\frac{\alpha R\left\|\nabla \mathcal{L}_{k,r E+\frac{1}{2}}\right\|_2^2}{L_{2}G(t_{sum}-\beta R)} \label{611748}
\end{equation}
Therefore, the optimization function converges if the local iteration of DFPL satisfies the formulas (\ref{0101306}) and (\ref{611748}).
\end{proof}

\begin{corollary}
(Convergence rate of DFPL).
Let Assumptions 1 to 4 hold, and   {\color{blue}define}  $\triangle = \mathcal{L}_0 - \mathcal{L}_\star $, where $\mathcal{L}_\star$ denotes the optimal   {\color{blue}objective value}.
For an arbitrary client $k$, given any $\chi >0$,  {\color{blue}if the number of communication rounds $R$ satisfies}
$$
 R > \frac{1}{\beta}\left(t_{sum}-\frac{2 \triangle \alpha}{\left(2\eta-L_1 \eta^2)\chi-\eta(L_1\eta\sigma^2+2\lambda L_2G\right)}\right),
$$
we have
$$
\frac{1}{RE}\sum_{r=0}^{R-1}\sum_{e=\frac{1}{2}}^{E-1}\mathbb{E}[\|\nabla \mathcal{L}_{k,r E+e}\|_2^2] <\chi,
$$
when the local learning rate $\eta$ satisfies $\eta<\frac{2(\chi-\lambda L_{2}G)}{L_{1}(\chi+\sigma^{2})} $, and $\lambda$ satisfies $\lambda<\frac{\chi}{L_{2}G}$.
\end{corollary}
The corollary   provides DFPL's convergence rate. The $\ell_2$-norm of the gradient can be confined to any bound $\chi$ with appropriate $R$, $\eta$, and $\lambda$.
Moreover, we observe that the smaller $\chi$ is, the larger $R$ becomes, which means that the tighter the constraint, the more communication rounds are required.
\begin{proof}
Considering the time step from $e = \frac{1}{2}$ to $e = E-1$ in each communication round and the number of communication rounds from $r = 0$ to $r = R -1$, we calculate the sum of the  bounds on the variation of the optimization function  on the basis of formula (\ref{611255}), so we can get
\begin{equation}\scalebox{0.8}{$
\begin{aligned}
&\sum_{r= 0}^{R-1}(\mathbb{E}[\mathcal{L}_{k, (r+1) E+\frac{1}{2}}]-\mathcal{L}_{k, rE+\frac{1}{2}}) \leq\\
&\sum_{r= 0}^{R-1}\left((\frac{L_1\eta^2}{2}-\eta)\sum_{e = \frac{1}{2}}^{E-1}\|\nabla \mathcal{L}_{k, rE+e}\|_2^2\right)+\frac{L_1E\eta^2}{2}\sigma^2R +\lambda L_2 \eta EGR. \label{612057}\nonumber
\end{aligned}$}
\end{equation}
After simplifying the formula (\ref{612057}), we can  obtain the  following:
\begin{equation}\scalebox{0.8}{$
\begin{aligned}
&\frac{1}{RE}\sum_{r=0}^{R-1}\sum_{e=\frac{1}{2}}^{E-1}\mathbb{E}[\|\nabla \mathcal{L}_{k,r E+e}\|_2^2] \leq\\
&\underbrace{\frac{2\sum_{r= 0}^{R-1}\left(\mathcal{L}_{k, rE+\frac{1}{2}}-\mathbb{E}[\mathcal{L}_{k, r(E+1)+\frac{1}{2}}]\right)+(L_1\sigma^2\eta^2+2\lambda L_2G\eta) ER}{(2\eta-L_1\eta^2)ER}}_{A_4}, \label{612137}\nonumber
\end{aligned}$}
\end{equation}
Given any $\chi>0$, let
\begin{equation}
A_4  < \chi,\label{621021}
\end{equation}
and let $\triangle = \mathcal{L}_{k,0} - \mathcal{L}_\star $, where $\mathcal{L}_\star$ denotes the optimal  function, we can get $\sum_{r=0}^{R-1}(\mathcal{L}_{k,rE+\frac{1}{2}}-\mathbb{E}[\mathcal{L}_{k,(r+1)E+\frac{1}{2}}]) <\triangle$.
The formula (\ref{621021}) holds when
\begin{equation}
\begin{aligned}
\frac{2\triangle+L_1\sigma^2\eta^2ER+2\lambda L_2G\eta ER}{(2\eta-L_1\eta^2)ER}  < \chi.\label{621022}
\end{aligned}
\end{equation}
After simplifying the formula (\ref{621022}), we can  obtain:
$$
\frac{1}{RE}\sum_{r=0}^{R-1}\sum_{e=\frac{1}{2}}^{E-1}\mathbb{E}[\|\nabla \mathcal{L}_{k,r E+e}\|_2^2] <\chi,
$$
when $R$ satisfies
\begin{equation}
\begin{aligned}
R > \frac{2\triangle}{E\chi(2\eta-L_1\eta^2)-E\eta(L_1\eta\sigma^2+2\lambda L_2G)}.\label{621041}
\end{aligned}
\end{equation}
Notably, for an arbitrary client $k\in \mathcal{K}$ of DFPL, we can get the relationship $t_{sum} = (E\alpha + \beta)R$.
Thus, we  rewrite the formula (\ref{621041}) to get:
$$
 R > \frac{1}{\beta}\left(t_{sum}-\frac{2 \alpha\triangle }{\left(2\eta-L_1 \eta^2)\chi-\eta(L_1\eta\sigma^2+2\lambda L_2G\right)}\right).
$$
To ensure that the above inequality holds, we need to guarantee
$$(2\eta-L_1 \eta^2)\chi-\eta(L_1\eta\sigma^2+2\lambda L_2G) \textgreater 0,$$
so we get $\eta<\frac{2(\chi-\lambda L_{2}G)}{L_{1}(\chi+\sigma^{2})} $ and $\lambda<\frac{\chi}{L_{2}G}$. Therefore, Corollary 2 is proved.
\end{proof}
\subsection{Time Complexity Analysis}
To evaluate the time  complexity of DFPL, we analyze the computation overhead on each client.
Since   {\color{blue}the relation}  $\gamma \ll (fE\alpha + f\beta)$  {\color{blue}holds}, the main computational costs are concentrated in  the local model training step and the mining step of  DFPL. Thus,  we  analyze the computational complexity of these two steps separately.
Specifically, the local model training step involves  iterative model training and prototype generation.
Let $T_{\textit{tm}}$ be  time overload of one local iteration at client, the time complexity of the iterative model training is $\mathcal{O}(ET_{\textit{tm}})$ in one communication round.
Let $T_{\textit{gc}}$ be the  time overload for generating the prototype from a sample, so the  complexity of prototype generation can be denoted
$\mathcal{O}(|\mathcal{D}_k|T_{\textit{gc}})$.
Thus, for any client $k$,  the time complexity of local model training is $\mathcal{O}(ET_{\textit{tm}})+\mathcal{O}(|\mathcal{D}_k|T_{\textit{gc}})$.
Notably, due to $T_{\textit{tm}}\gg T_{\textit{gc}}$, we can get the relationship $\mathcal{O}(ET_{\textit{tm}})\gg\mathcal{O}(|\mathcal{D}_k|T_{\textit{gc}})$.
For the mining step, the time complexity is $\mathcal{O}(\mu T_{\textit{m}})$, where  $T_{\textit{m}}$ denotes  time overload per unit mining difficulty and $\mu$ denotes mining difficulty.
Therefore, the time  complexity of each client   can be expressed as $\mathcal{O}(ET_{\textit{tm}}) + \mathcal{O}(\mu T_{\textit{m}})$.
This analysis   provides the theoretical foundation  for the allocation of computational resources in practical deployments of DFPL.

\subsection{Privacy Analysis}
Our DFPL integrates blockchain technology into clients participating in federated training, without relying on any third-party blockchain network, which eliminates the threat of third-party dependence.
 {\color{blue}Moreover, existing works \cite{10420449}\cite{10795238} has observed that  FL schemes based on the exchange of model parameters and gradients are vulnerable to privacy attacks, where adversaries can steal sensitive information from other clients through the transmitted parameters and gradients.}
In contrast, DFPL exchanges prototypes instead of model parameters among clients, which enhances   privacy preserving in DFL.
This is because prototypes are low-dimensional vectors generated by averaging feature representations of samples from the  same class, which is an irreversible process.
Therefore, since the adversary does not have access to the local models, it cannot reconstruct the original training samples from the prototype.

\section{EXPERIMENTS}\label{sec6}
In this section, we provide detailed evaluations of  DFPL  under different heterogeneous data distributions.
\subsection{Experimental Settings}
\subsubsection{Datasets and model}
As with previous FL works \cite{10018261,10138067,10535170}, we adopt four publicly available datasets, MNIST, FMNIST, CIFAR10, and SVHN to evaluate our DFPL.
These datasets are  standard benchmarks in the FL research.

To simulate large-scale network settings in practice, we use ResNet18 \cite{szegedy2017inception} pre-trained on ImageNet as the local model on CIFAR10 and SVHN datasets.
ResNet18 is a large-scale deep neural network with more than $2.00\times10^8$ parameters that is widely used in  FL settings.
For MNIST and FMNIST datasets, we  adopt the convolutional neural network (CNN) containing multiple convolutional  and fully connected layers  as the local  model.
 {\color{blue}Both architectures are commonly employed in FL research.}

\begin{enumerate}
\item \textbf{MNIST} \cite{lecun2002gradient}: This  is a grayscale image dataset of handwritten numbers 0 to 9. It has a training set of 60,000 samples and a testing set of 10,000 samples, and each sample has a pixel value of 28 $\times$ 28.
\item \textbf{FMNIST} \cite{xiao2017fashion}: This dataset is a grayscale image dataset about clothing that contains a training set of 60,000 samples in 10 classes and a testing set of 10,000 samples each with a pixel value of 28$\times$28.
\item \textbf{CIFAR10} \cite{krizhevsky2009learning}: The CIFAR10 dataset consists of 60,000 RGB images $(32 \times 32)$ divided into 10 classes, with each class containing 6,000 samples equally distributed.
\item \textbf{SVHN} \cite{netzer2011reading}:
    The SVHN dataset  is a digit classification benchmark dataset that contains 600,000 $32 \times 32$ RGB images of printed digits (from 0 to 9) cropped from pictures of house number plates.

\end{enumerate}

\subsubsection{FL Settings}
In our evaluation, we adopt the cross-silo setting with the fully connected topology, and all clients participate in training  during each communication round.
We set the number of  {\color{blue}clients} is set to $K$=20, which is a reasonable configuration in cross-silo federated learning.
To simplify the description of parameter settings, we define some default hyper-parameters  for our experiments.
Specifically, we use SGD {\color{blue}optimizer} to minimize the loss function and set the learning rate $\eta$=0.1.
The batch size is set to 32.
 The importance  weight $\lambda$ is set to $1$.
If the above hyper-parameters are not described in later experiments, they are default values.

\subsubsection{Settings of Non-IID Data}
\begin{figure}[!t]
  \centering
  \subfloat[\textit{Avg}=3, \textit{Std}=1]
  {
      \label{110911}  \includegraphics[width=0.45\linewidth]{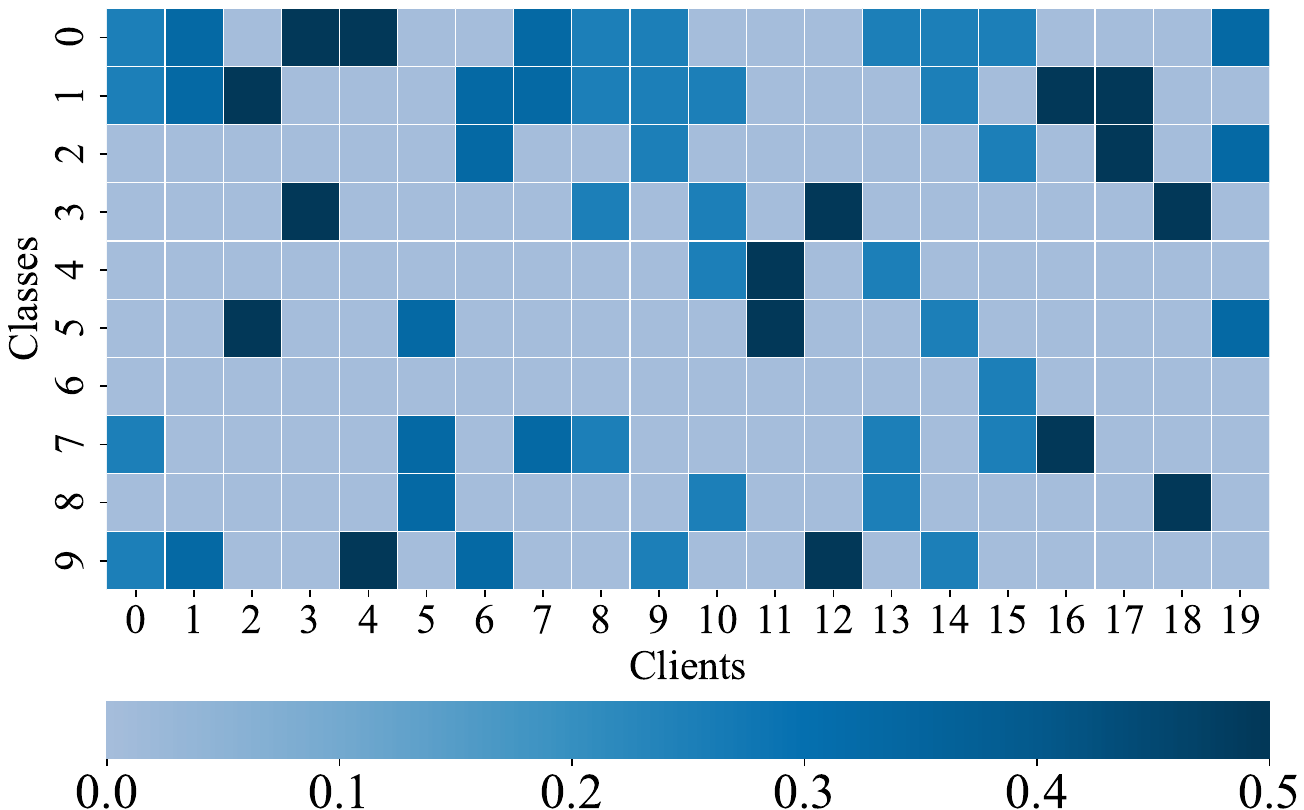}
  }
  \subfloat[\textit{Avg}=4, \textit{Std}=1]
  {
      \label{110912}  \includegraphics[width=0.45\linewidth]{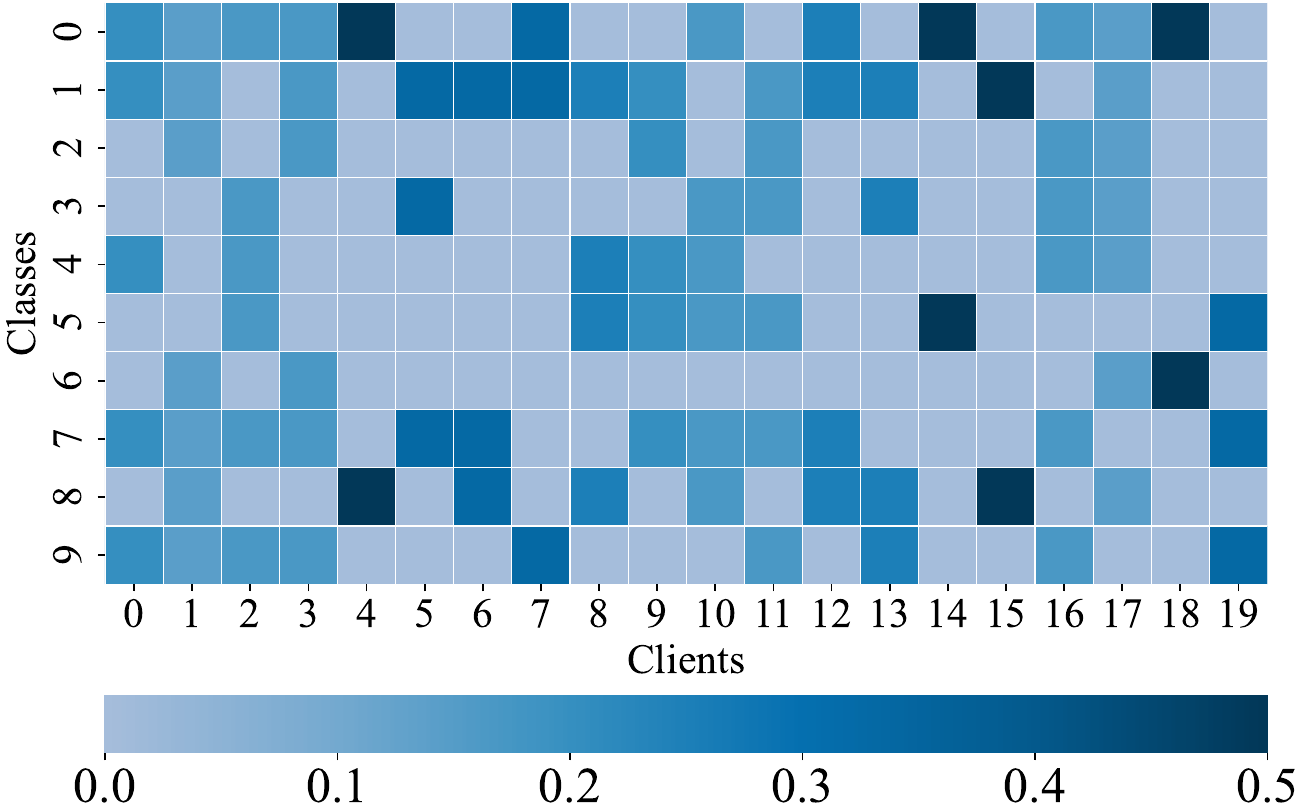}
  }
\caption{The heat map about heterogeneous data distributions of CIFAR10 for each client.}
\label{671005}
\end{figure}

In traditional federated learning tasks, most works use the Dirichlet distribution \cite{lin2016dirichlet} to simulate Non-IID data.
Although this method can control the degree of data heterogeneity by tuning the concentration parameter,  it cannot control the number of classes assigned to each client.
Since DFL lacks a central server to coordinate the training process, each client cannot be guaranteed to have a consistent class space.
However, the Dirichlet distribution  only adjusts the class ratio at the probability level, and cannot control the class range  between clients.
Therefore, DFL is not suitable for using Dirichlet distribution to simulate Non-IID data.
To better  simulate the Non-IID data in DFL, we introduce  heterogeneity in  class space among clients.
We consider the union of all clients' data classes to form {\color{blue}the} complete classification task in the federated setting.
 Similar to existing works \cite{tan2022fedproto}\cite{10398600}, we use \textit{Avg} to denote  the average number of data classes per client, and \textit{Std} to denote  the standard deviation of the number of data classes across all clients.
In our experiments, we fix \textit{Avg} to be 3, 4 or 5, and fix \textit{Std} to be 1, 2 or 3,  aiming to create heterogeneity in  class spaces.
To visualize the statistical  heterogeneity of data distribution, we plot a heat map of  data distribution among clients, as shown in Fig. \ref{671005}.
Darker colors indicate more samples of that class in clients.

\subsubsection{Baselines}
 {\color{blue}To demonstrate the effectiveness of DFPL}, we use  the  state-of-the-art FL schemes as comparison.
 
\begin{itemize}
\item \textbf{BLADE-FL}\cite{9664296}:
 {\color{blue}This is a DFL framework using the FedAvg algorithm. Each client leverages blockchain technology to exchange its local model parameters with neighboring clients and then aggregates the received local model parameters.}
\item \textbf{MOON} \cite{li2021model}: It integrates contrastive learning into federated learning by incorporating a regularization term into the local loss function, aiming to reduce the discrepancy between the feature representations of the local and global models.
\item \textbf{FedIntR} \cite{10066639}:  It adds an auxiliary term of intermediate layers contrast to the local optimization  function that encourages the local model to be close to the global model.
\item \textbf{FLPD} \cite{10495206}:  It introduces a knowledge distillation term to  fine-tune the local representation space with global prototype similarity, which facilitates local models to fit both the local data distribution and the global representation space.
\item \textbf{SWIM} \cite{zhang2025swim}: It introduces a sliding-window model contrast method into the local loss function, which collects multiple historical local model representations and distinguishes them as positive or negative samples based on cosine similarity.
\end{itemize}

 {\color{blue}Notably, MOON, FedIntR, FLPD, and SWIM are centralized solutions proposed to address the Non-IID problem in FL.
We embed their both local training  and aggregation methods into BLADE-FL.}

\subsubsection{ Evaluation Measure}
To evaluate the performance of DFPL under heterogeneous data distributions, we use the Test Average Accuracy (TAA) and Test Average Loss (TAL) as evaluation metrics  across different datasets.
Specifically, TAA refers to the average test accuracy of all clients on their local testing sets, while TAL denotes the average test loss of all clients on their local testing sets.

\subsection{Performance of DFPL in Non-IID settings}

\begin{figure}[t]
  \centering
  \subfloat[FMNIST, \textit{Avg}=3, \textit{Std}=2]
  {
      \label{112414113}  \includegraphics[width=0.48\linewidth]{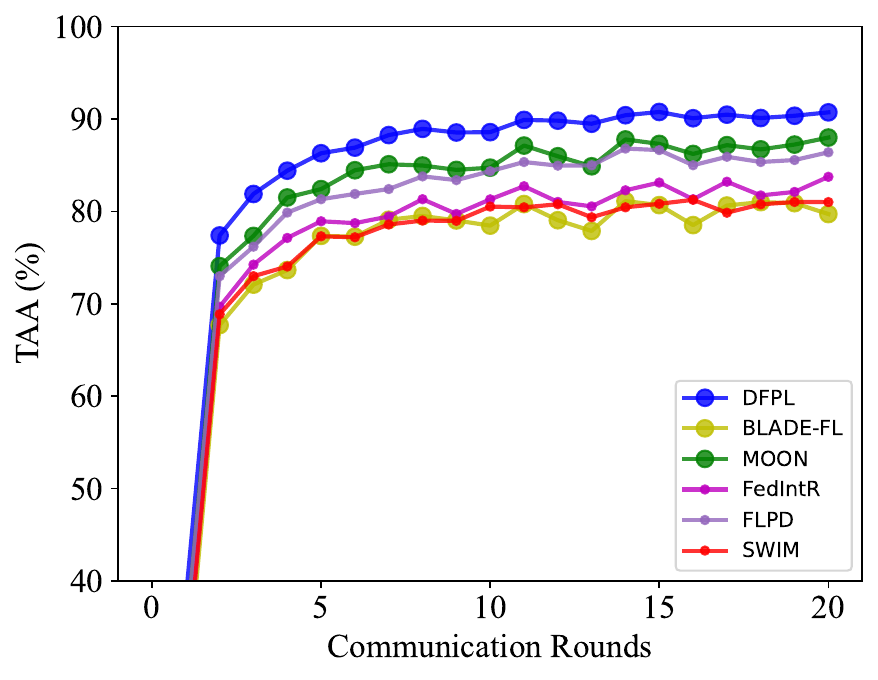}
  }
  \subfloat[FMNIST, \textit{Avg}=4, \textit{Std}=2]
  {
      \label{112414114}  \includegraphics[width=0.48\linewidth]{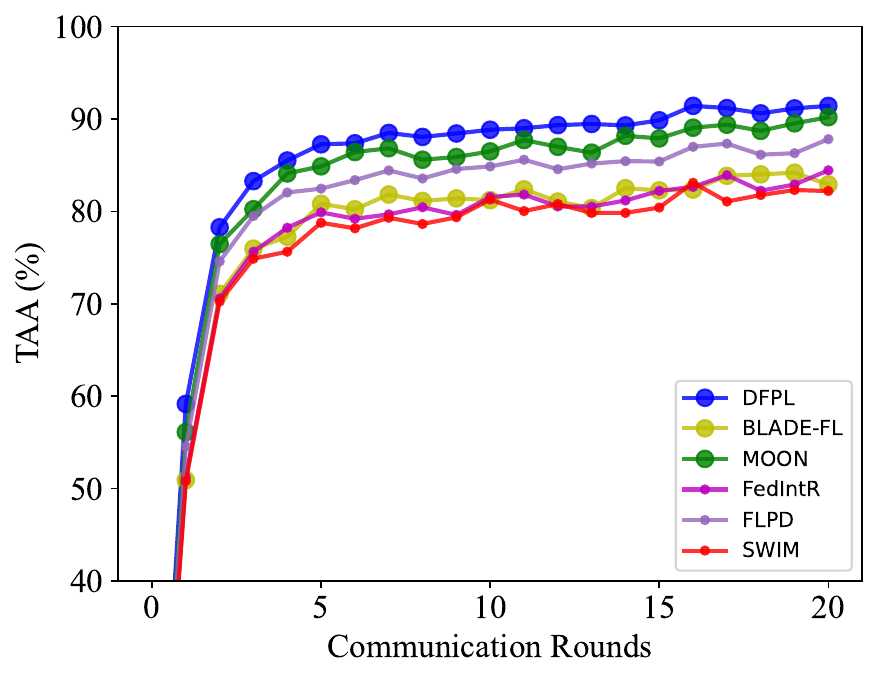}
  }

    \subfloat[CIFAR10, \textit{Avg}=3, \textit{Std}=2]
  {
      \label{112414115}  \includegraphics[width=0.48\linewidth]{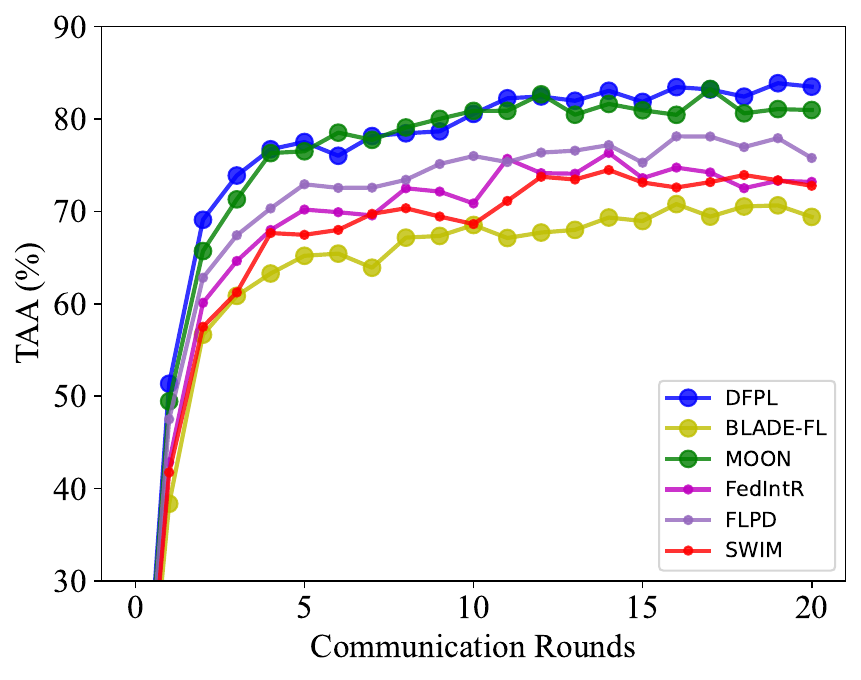}
  }
  \subfloat[CIFAR10, \textit{Avg}=4, \textit{Std}=2]
  {
      \label{112414116}  \includegraphics[width=0.48\linewidth]{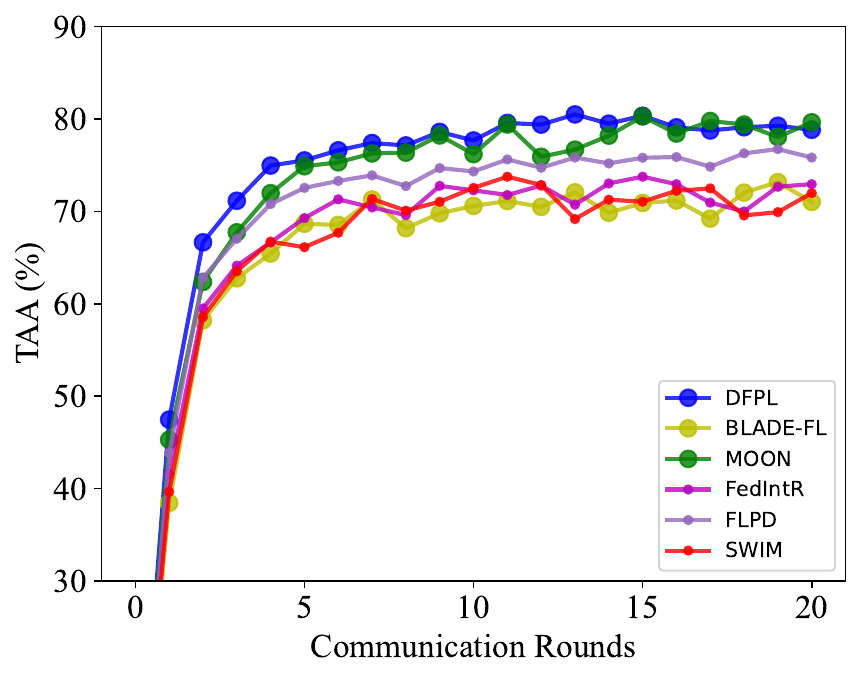}
  }
\caption{Convergence of the proposed DFPL   on the FMNIST, and CIFAR10.}
\label{11241411}
\end{figure}

\begin{table}[t]
\centering
  \caption{Test average accuracy   with diffierent distributions of Non-IID data. The best results are in bold.}
    \scalebox{0.70}{
\tabcolsep=0.20cm
\begin{tabular}{ccc|ccc|ccc}
\hline\hline
\multirow{3}{*}{\textbf{Dataset}} & \multirow{3}{*}{\textbf{Method}} & \multirow{3}{*}{\textit{Std}} & \multicolumn{6}{c}{\textbf{Test Average Accuracy (\%)}}                            \\
                      &                                              &                     & \multicolumn{3}{c|}{$R$ = 6   }  & \multicolumn{3}{c}{$R$ = 10   }                                          \\
                      &                                              &                     & \multicolumn{1}{c}{\textit{Avg = }3   } & \multicolumn{1}{c}{\textit{Avg} = 4 } &\textit{ Avg }= 5 &\textit{ Avg }= 3 &\textit{ Avg }= 4 & \textit{ Avg }=5                                               \\\cmidrule{1-9}
\multirow{18}{*}{\textbf{MNIST}}     & \multirow{3}{*}{ BLADE-FL}           &   1                   & \multicolumn{1}{c}{95.74} & \multicolumn{1}{c}{96.18} &97.32  & 96.99&97.27&97.42             \\
                   &                                               &    2                  & \multicolumn{1}{c}{97.08} & \multicolumn{1}{c}{96.60} &  \textbf{97.59}     &97.81&97.08&  \textbf{98.17}          \\
		&                                               &    3                  & \multicolumn{1}{c}{96.71} & \multicolumn{1}{c}{\textbf{97.84}} & 98.10   &97.08&\textbf{97.65}&    98.05           \\ \cmidrule{2-9}
& \multirow{3}{*}{MOON}              &    1                  & \multicolumn{1}{c}{97.91} & \multicolumn{1}{c}{96.48} &96.26  &97.92  &96.75&    96.87                    \\
                      &                                               &             2         & \multicolumn{1}{c}{97.49} & \multicolumn{1}{c}{96.35} & 95.38 &  96.83 & 97.09& 96.40                    \\
       		  &                                               &             3        & \multicolumn{1}{c}{97.15} & \multicolumn{1}{c}{95.75} & 96.13 & 96.72   &96.60&      96.74        \\\cmidrule{2-9}
 		      & \multirow{3}{*}{FedIntR}           &    1                  & \multicolumn{1}{c}{95.08} & \multicolumn{1}{c}{96.01} &\textbf{97.44}     &96.04&97.06&    \textbf{97.95}                   \\
                      &                                               &             2         & \multicolumn{1}{c}{97.15} & \multicolumn{1}{c}{96.62} & 97.48 &98.00&\textbf{97.54}&      97.74              \\
                      &                            &      3               & \multicolumn{1}{c}{96.82} & \multicolumn{1}{c}{97.57} &\textbf{98.34}               &97.35&97.64&     \textbf{98.74}    \\\cmidrule{2-9}
 & \multirow{3}{*}{FLPD}           &    1                  & \multicolumn{1}{c}{98.04} & \multicolumn{1}{c}{96.89} &96.68   &98.27&97.14&    97.46                   \\
                      &                                               &             2         & \multicolumn{1}{c}{97.43} & \multicolumn{1}{c}{97.16} & 96.92 &97.59&97.50&      97.38             \\
                      &                            &      3               & \multicolumn{1}{c}{97.10} & \multicolumn{1}{c}{96.82} & 97.04 &97.54&97.49&     97.66   \\\cmidrule{2-9}             
& \multirow{3}{*}{SWIM}              &    1                  & \multicolumn{1}{c}{94.25} & \multicolumn{1}{c}{95.05} &96.88     &95.72&97.04&    97.20                   \\
                      &                                               &             2         & \multicolumn{1}{c}{96.86} & \multicolumn{1}{c}{95.87} & 96.49 &97.80&97.41&      97.71              \\
                      &                            &      3               & \multicolumn{1}{c}{95.83} & \multicolumn{1}{c}{96.77} &98.17              &97.20&97.51&     98.34   \\\cmidrule{2-9}
                      & \multirow{3}{*}{DFPL}          &    1\cellcolor[rgb]{ .816,  .808,  .808}                  & \multicolumn{1}{c}{\textbf{98.16}\cellcolor[rgb]{ .816,  .808,  .808}} & \multicolumn{1}{c}{\textbf{97.31}\cellcolor[rgb]{ .816,  .808,  .808}} &96.72\cellcolor[rgb]{ .816,  .808,  .808}  &\textbf{98.73}\cellcolor[rgb]{ .816,  .808,  .808}  &\textbf{97.62}\cellcolor[rgb]{ .816,  .808,  .808}&    97.41\cellcolor[rgb]{ .816,  .808,  .808}                    \\
                      &                                               &             2\cellcolor[rgb]{ .816,  .808,  .808}         & \multicolumn{1}{c}{\textbf{97.80}\cellcolor[rgb]{ .816,  .808,  .808}} & \multicolumn{1}{c}{\textbf{97.33}\cellcolor[rgb]{ .816,  .808,  .808}} & 96.34\cellcolor[rgb]{ .816,  .808,  .808} &  \textbf{97.62}\cellcolor[rgb]{ .816,  .808,  .808} & 97.48\cellcolor[rgb]{ .816,  .808,  .808}& 96.80\cellcolor[rgb]{ .816,  .808,  .808}                    \\
       		  &                                               &             3\cellcolor[rgb]{ .816,  .808,  .808}        & \multicolumn{1}{c}{\textbf{97.15}\cellcolor[rgb]{ .816,  .808,  .808}} & \multicolumn{1}{c}{96.50\cellcolor[rgb]{ .816,  .808,  .808}}& 96.86\cellcolor[rgb]{ .816,  .808,  .808} & \textbf{97.67}\cellcolor[rgb]{ .816,  .808,  .808}   &97.47\cellcolor[rgb]{ .816,  .808,  .808}&      97.46\cellcolor[rgb]{ .816,  .808,  .808}              \\  \hline
\multirow{18}{*}{\makecell{\textbf{FMNIST}}}     & \multirow{3}{*}{BLADE-FL}          &                     1 & \multicolumn{1}{c}{86.87} & \multicolumn{1}{c}{79.70} & 80.68 &   88.82&81.99&    84.03                   \\
                      &                                               &           2           & \multicolumn{1}{c}{77.59} & \multicolumn{1}{c}{80.78} &  84.15&79.26 &81.75&    85.81                    \\
                      &                            &    3                  & \multicolumn{1}{c}{79.29} & \multicolumn{1}{c}{81.62} & 83.65 &   81.96   &84.58&  85.42                   \\\cmidrule{2-9}
  & \multirow{3}{*}{MOON}          &     1                & \multicolumn{1}{c}{87.31} & \multicolumn{1}{c}{87.03} &    84.13 & 87.92& 87.39&     84.88                \\
                      &                                               &                 2     & \multicolumn{1}{c}{84.20} & \multicolumn{1}{c}{86.21} &84.11  &  86.22& 87.88&    86.35             \\
                      &                             &      3                & \multicolumn{1}{c}{84.42} & \multicolumn{1}{c}{85.08} &82.28  &     86.67 & 85.23&    85.42                 \\\cmidrule{2-9}
 		      & \multirow{3}{*}{FedIntR}         &    1                  & \multicolumn{1}{c}{82.72} & \multicolumn{1}{c}{78.43} &77.88  &88.59   &79.81&    82.06                      \\
                      &                                               &             2         & \multicolumn{1}{c}{76.41} & \multicolumn{1}{c}{79.40} & 84.44 &  81.20   &81.13&    85.94                \\
                      &                             &       3               & \multicolumn{1}{c}{77.36} & \multicolumn{1}{c}{82.51} &82.03  & 79.18  &86.25&            86.36                \\\cmidrule{2-9}
 & \multirow{3}{*}{FLPD}           &    1                  & \multicolumn{1}{c}{85.75} & \multicolumn{1}{c}{85.26} &82.66   &85.97&85.78&    83.10                   \\
                      &                                               &             2         & \multicolumn{1}{c}{82.46} & \multicolumn{1}{c}{84.86} & 83.17 &83.10&84.14&      83.52             \\
                      &                            &      3               & \multicolumn{1}{c}{82.13} & \multicolumn{1}{c}{83.30} & 82.66 &83.41&83.69&     82.95   \\\cmidrule{2-9}
& \multirow{3}{*}{SWIM}         &    1                  & \multicolumn{1}{c}{82.17} & \multicolumn{1}{c}{77.46} &77.25  &87.71   &78.88&    81.12                      \\
                      &                                               &             2         & \multicolumn{1}{c}{76.23} & \multicolumn{1}{c}{79.37} & 84.36 &  80.79  &80.52&    85.50                \\
                      &                             &       3               & \multicolumn{1}{c}{76.91} & \multicolumn{1}{c}{81.87} &81.92  & 78.56  &85.33&            85.25                \\\cmidrule{2-9}
                      & \multirow{3}{*}{DFPL}          &     1\cellcolor[rgb]{ .816,  .808,  .808}                & \multicolumn{1}{c}{\textbf{92.51}\cellcolor[rgb]{ .816,  .808,  .808}} & \multicolumn{1}{c}{\textbf{89.62}\cellcolor[rgb]{ .816,  .808,  .808}} &    \textbf{86.46}\cellcolor[rgb]{ .816,  .808,  .808} & \textbf{92.85}\cellcolor[rgb]{ .816,  .808,  .808}& \textbf{90.40}\cellcolor[rgb]{ .816,  .808,  .808}&     \textbf{87.73}\cellcolor[rgb]{ .816,  .808,  .808}                \\
                      &                                               &                 2\cellcolor[rgb]{ .816,  .808,  .808}     & \multicolumn{1}{c}{\textbf{86.72}\cellcolor[rgb]{ .816,  .808,  .808}} & \multicolumn{1}{c}{\textbf{89.01}\cellcolor[rgb]{ .816,  .808,  .808}} &\textbf{86.35}\cellcolor[rgb]{ .816,  .808,  .808}  &   \textbf{89.01}\cellcolor[rgb]{ .816,  .808,  .808}& \textbf{90.00}\cellcolor[rgb]{ .816,  .808,  .808}&     \textbf{89.46}\cellcolor[rgb]{ .816,  .808,  .808}             \\
                      &                             &      3\cellcolor[rgb]{ .816,  .808,  .808}        & \multicolumn{1}{c}{\textbf{87.54}\cellcolor[rgb]{ .816,  .808,  .808}} & \multicolumn{1}{c}{\textbf{87.53}\cellcolor[rgb]{ .816,  .808,  .808}} &\textbf{85.27}\cellcolor[rgb]{ .816,  .808,  .808} &     \textbf{90.11}\cellcolor[rgb]{ .816,  .808,  .808}& \textbf{88.95}\cellcolor[rgb]{ .816,  .808,  .808}&      \textbf{ 87.42 }\cellcolor[rgb]{ .816,  .808,  .808}                   \\ \hline
\multirow{18}{*}{\textbf{CIFAR10}}     & \multirow{3}{*}{ BLADE-FL}           &   1                   & \multicolumn{1}{c}{67.85} & \multicolumn{1}{c}{70.37} & \textbf{76.41}  & 69.64&69.05& \textbf{76.24}                     \\
                   &                                               &    2                  & \multicolumn{1}{c}{62.59} & \multicolumn{1}{c}{72.45} &   \textbf{76.77}    &66.38&72.25&  \textbf{76.87}       \\
		&                                               &    3                  & \multicolumn{1}{c}{73.64} & \multicolumn{1}{c}{71.63} & 75.54   &74.95&73.52&    75.52          \\\cmidrule{2-9}
  & \multirow{3}{*}{MOON}          &    1                  & \multicolumn{1}{c}{80.76} & \multicolumn{1}{c}{77.23} &75.72  &82.12  &76.14&   74.46                   \\
                      &                                               &             2         & \multicolumn{1}{c}{78.15} & \multicolumn{1}{c}{77.22} & 75.88 &  80.39& 79.55& 75.58          \\
       		  &                                               &             3        & \multicolumn{1}{c}{75.95} & \multicolumn{1}{c}{74.40} & 72.66& 80.08   & 74.43&      75.45        \\\cmidrule{2-9}
 		      & \multirow{3}{*}{FedIntR}           &    1                  & \multicolumn{1}{c}{70.67} & \multicolumn{1}{c}{75.17} &76.32   &74.30&75.25&    75.14                   \\
                      &                                               &             2         & \multicolumn{1}{c}{70.55} & \multicolumn{1}{c}{71.63} & 76.41 &71.23&72.86&      74.71             \\
                      &                            &      3               & \multicolumn{1}{c}{72.74} & \multicolumn{1}{c}{74.13} & \textbf{75.48} &73.76&74.93&     76.59   \\\cmidrule{2-9}
 & \multirow{3}{*}{FLPD}           &    1                  & \multicolumn{1}{c}{73.10} & \multicolumn{1}{c}{76.23} &75.72   &74.38&76.44&    75.06                   \\
                      &                                               &             2         & \multicolumn{1}{c}{73.84} & \multicolumn{1}{c}{73.51} & 75.06 &74.09&73.53&      75.14             \\
                      &                            &      3               & \multicolumn{1}{c}{72.15} & \multicolumn{1}{c}{72.40} & 73.14 &72.34&72.18&     73.57   \\\cmidrule{2-9}
 & \multirow{3}{*}{SWIM}           &    1                  & \multicolumn{1}{c}{70.15} & \multicolumn{1}{c}{74.52} &75.27   &73.42&74.73&    74.27                   \\
                      &                                               &             2         & \multicolumn{1}{c}{70.22} & \multicolumn{1}{c}{70.99} & 75.77 &70.52&72.34&      73.78             \\
                      &                            &      3               & \multicolumn{1}{c}{71.88} & \multicolumn{1}{c}{73.91} & 73.80 &73.55&74.24&     76.42   \\\cmidrule{2-9}
                      & \multirow{3}{*}{DFPL}          &    1      \cellcolor[rgb]{ .816,  .808,  .808}            & \multicolumn{1}{c}{\textbf{82.96}\cellcolor[rgb]{ .816,  .808,  .808}} & \multicolumn{1}{c}{\textbf{77.99}\cellcolor[rgb]{ .816,  .808,  .808}} &75.74\cellcolor[rgb]{ .816,  .808,  .808}  &\textbf{83.18}\cellcolor[rgb]{ .816,  .808,  .808}  &\textbf{77.07}\cellcolor[rgb]{ .816,  .808,  .808}&   75.25\cellcolor[rgb]{ .816,  .808,  .808}                   \\
                      &                                               &             2    \cellcolor[rgb]{ .816,  .808,  .808}     & \multicolumn{1}{c}{\textbf{78.66}\cellcolor[rgb]{ .816,  .808,  .808}} & \multicolumn{1}{c}{\textbf{78.70}\cellcolor[rgb]{ .816,  .808,  .808}} & 76.43\cellcolor[rgb]{ .816,  .808,  .808} &  \textbf{80.72}\cellcolor[rgb]{ .816,  .808,  .808} & \textbf{80.17}\cellcolor[rgb]{ .816,  .808,  .808}& 76.47\cellcolor[rgb]{ .816,  .808,  .808}         \\
       		  &                                               &             3 \cellcolor[rgb]{ .816,  .808,  .808}       & \multicolumn{1}{c}{\textbf{76.92}\cellcolor[rgb]{ .816,  .808,  .808}} & \multicolumn{1}{c}{\textbf{74.85}\cellcolor[rgb]{ .816,  .808,  .808}} & 73.48\cellcolor[rgb]{ .816,  .808,  .808} & \textbf{80.22}\cellcolor[rgb]{ .816,  .808,  .808}& \textbf{74.98}\cellcolor[rgb]{ .816,  .808,  .808}&       \textbf{76.41}\cellcolor[rgb]{ .816,  .808,  .808}       \\  \hline
\multirow{18}{*}{\textbf{SVHN}}     & \multirow{3}{*}{ BLADE-FL}           &   1                   & \multicolumn{1}{c}{75.09} & \multicolumn{1}{c}{76.06} & 78.81 &   75.82&76.41&    77.56                 \\
                   &                                               &    2                  & \multicolumn{1}{c}{73.73} & \multicolumn{1}{c}{79.63} &  79.04&73.87 &78.46&    78.98     \\
		&                                               &    3                  & \multicolumn{1}{c}{77.48} & \multicolumn{1}{c}{79.86} & 79.89 &   77.02  &78.23&  78.91          \\\cmidrule{2-9}
                      & \multirow{3}{*}{MOON}          &    1                  & \multicolumn{1}{c}{81.06} & \multicolumn{1}{c}{79.48} &    77.73 & 83.52& 80.47&  81.25                     \\
                      &                                               &             2         & \multicolumn{1}{c}{81.44} & \multicolumn{1}{c}{79.90} &78.72  &  81.20& 80.81&    79.67         \\
       		  &                                               &             3        & \multicolumn{1}{c}{79.35} & \multicolumn{1}{c}{80.27} &79.20 &   81.76 & 81.57&   79.89      \\\cmidrule{2-9}
 		      & \multirow{3}{*}{FedIntR}           &    1                  & \multicolumn{1}{c}{76.45} & \multicolumn{1}{c}{76.66} &77.31  &77.57  &77.63&   78.73                  \\
                      &                                               &             2         & \multicolumn{1}{c}{82.41} & \multicolumn{1}{c}{77.87} & 78.05 &  77.20   &78.39&    78.60             \\
                      &                            &      3               & \multicolumn{1}{c}{76.61} & \multicolumn{1}{c}{77.34} &77.04  & 78.42  &78.01&            78.65    \\\cmidrule{2-9}
 		      & \multirow{3}{*}{FLPD}           &    1                  & \multicolumn{1}{c}{81.24} & \multicolumn{1}{c}{78.06} &76.69  &81.50  &78.29&   76.53                  \\
                      &                                               &             2         & \multicolumn{1}{c}{80.55} & \multicolumn{1}{c}{80.11} & 77.28 &  80.16   &80.27&    77.04            \\
                      &                            &      3               & \multicolumn{1}{c}{79.10} & \multicolumn{1}{c}{78.27} &76.45  & 78.65  &78.29&            76.62    \\\cmidrule{2-9}
 		      & \multirow{3}{*}{SWIM}           &    1                  & \multicolumn{1}{c}{75.66} & \multicolumn{1}{c}{76.45} &76.11  &76.85  &76.48&   78.51                  \\
                      &                                               &             2         & \multicolumn{1}{c}{81.25} & \multicolumn{1}{c}{76.45} & 77.30 &  75.35   &76.78&    78.01            \\
                      &                            &      3               & \multicolumn{1}{c}{76.33} & \multicolumn{1}{c}{76.63} &75.78  & 76.72  &76.19&            76.16    \\\cmidrule{2-9}
                      & \multirow{3}{*}{DFPL}         &    1             \cellcolor[rgb]{ .816,  .808,  .808}     & \multicolumn{1}{c}{\textbf{83.86}\cellcolor[rgb]{ .816,  .808,  .808}} & \multicolumn{1}{c}{\textbf{80.41}\cellcolor[rgb]{ .816,  .808,  .808}}&    \textbf{79.87}\cellcolor[rgb]{ .816,  .808,  .808}& \textbf{84.81}\cellcolor[rgb]{ .816,  .808,  .808}& \textbf{82.12}\cellcolor[rgb]{ .816,  .808,  .808}&     \textbf{83.07}\cellcolor[rgb]{ .816,  .808,  .808}                   \\
                      &                                             &             2 \cellcolor[rgb]{ .816,  .808,  .808}        & \multicolumn{1}{c}{\textbf{83.75}\cellcolor[rgb]{ .816,  .808,  .808}} & \multicolumn{1}{c}{\textbf{81.63}\cellcolor[rgb]{ .816,  .808,  .808}} &\textbf{79.85}\cellcolor[rgb]{ .816,  .808,  .808} &   \textbf{82.88}\cellcolor[rgb]{ .816,  .808,  .808}& \textbf{82.70}\cellcolor[rgb]{ .816,  .808,  .808}&     \textbf{81.54}\cellcolor[rgb]{ .816,  .808,  .808}     \\
       		  &                                        &             3    \cellcolor[rgb]{ .816,  .808,  .808}    & \multicolumn{1}{c}{\textbf{80.64}\cellcolor[rgb]{ .816,  .808,  .808}} & \multicolumn{1}{c}{\textbf{81.99}\cellcolor[rgb]{ .816,  .808,  .808}} &\textbf{79.70}\cellcolor[rgb]{ .816,  .808,  .808} &     \textbf{83.59}\cellcolor[rgb]{ .816,  .808,  .808} & \textbf{83.34}\cellcolor[rgb]{ .816,  .808,  .808}&      \textbf{81.20}\cellcolor[rgb]{ .816,  .808,  .808}     \\  \hline\hline
\end{tabular}}
\label{012102}
\end{table}
To evaluate the performance of DFPL,  {\color{blue}we examine the variation of its TAA during training using FMNIST  and CIFAR10 datasets under two heterogeneous  data distributions.}
 {\color{blue}In addition, we compare it with baselines.}
In the experiment, the number of local iterations $E$ is set to 20, and the results are shown in Fig. \ref{11241411}.
We observe  that the TAA of DFPL increases sharply at the beginning and then  gradually stabilizes, indicating that DFPL has reached convergence.
This is attributed to   the prototype exchange among clients, which effectively mitigates the inconsistency in model updates caused by the statistical heterogeneity of data distributions, thereby promoting the convergence of DFPL.
Additionally, we observe  that the performance of DFPL stabilizes  after 10 communication rounds in CIFAR10.
In other words, the computational resources consumed in subsequent communication rounds bring negligible benefits to DFPL.
This is because increased local iterations reduce the number of communication rounds required for convergence.
 Continuing training would only lead to over-fitting of the local models without improving the performance on the testing set.
 {\color{blue}In addition, we notice that FedIntR and SWIM exhibit large fluctuations in TAA during training.
This is because heterogeneous data distributions undermine the convergence of federated learning.
Moreover, BLADE-FL attains the lowest TAA on the CIFAR10 dataset, because it relies on simple parameter averaging, which severely deviates from the optimal solution of the global model parameters.
MOON and FLPD are also affected by data heterogeneity to varying degrees, resulting in degraded performance compared with DFPL.
}

To further evaluate the performance of DFPL,  we compare  it with BLADE-FL, MOON,  FedIntR, FLPD, and  SWIM  under  different heterogeneous  data distributions.
In the experiments, we set the number of local iterations to 20, and the number of communication rounds to 6 and 10, respectively.
Table  \ref {012102} reports  the  average test accuracy of each scheme in the last communication round.
We notice that our DFPL  slightly outperforms  the baselines, especially in the CIFAR10 and SVHN datasets.
Since MNIST is relatively  simple, the performance of DFPL is comparable to that of other schemes.
 In contrast, on CIFAR10 and SVHN, DFPL demonstrates superior  accuracy, highlighting  its  robustness in heterogeneous  data distributions.
 {\color{blue}This is because the  clients exchange  prototypes in DFPL, which remain consistent across clients and are unaffected by data distribution.
 Consequently, DFPL exhibits strong  performance across  various data distributions and exhibits superior generalization capabilities across  datasets.}
 {\color{blue}In contrast, BLADE-FL shows inferior performance under these data distributions.
This is because different data distributions lead to distinct local optima, and the submitted model parameters in BLADE-FL are strongly affected by the data distribution.
MOON, FedIntR, and SWIM  outperform BLADE-FL in most cases, since they introduce additional constraints into the local optimization function.
However, their performance is still highly sensitive to data distributions.
In addition, FLPD adopts prototype-based aggregation by aligning local and global prototypes with cosine similarity, but it ignores the magnitude of prototypes.}
Although these methods improve the model performance of FL under heterogeneous data distributions,  their overall effectiveness is inferior to that of DFPL.
Additionally, the TAA of DFPL at communication round $R=6$ is almost identical to that at $R=10$.
This result is attributed to setting the number  in the local iteration to 20, which enables DFPL to converge by communication round $R=6$.
Therefore,  increasing the number of communication rounds   does not significantly improve  the performance of federated learning.

\begin{table}[t]
\tabcolsep= 0.075cm
\centering
\caption{The number of parameters transmitted by clients per Round.}
\begin{tabular}{l|c|c|c|c|c}
\hline\hline
\multirow{2}{*}{Datasets}&\multicolumn{5}{c}{Communication parameters}  \\
 &BLADE-FL&MOON&FedIntR&SWIM&DFPL \\\hline
MNIST&$4.37\times 10^5$&$4.37\times 10^5$&$4.37\times 10^5$&$4.37\times 10^5$&$1.00\times 10^4$ \\\hline
FMNIST&$4.37\times 10^5$&$4.37\times 10^5$&$4.37\times 10^5$&$4.37\times 10^5$&$1.00\times 10^4$ \\\hline
CIFAR10&$2.35\times 10^8$&$2.35\times 10^8$&$2.35\times 10^8$&$2.35\times 10^8$&$4.00\times 10^4$ \\\hline
SVHN&$2.35\times 10^8$&$2.35\times 10^8$&$2.35\times 10^8$&$2.35\times 10^8$&$4.00\times 10^4$ \\\hline\hline
\end{tabular} \label{11261948}
\end{table}

\subsection{Communication efficiency}

Due to inter-client communication,  communication costs have always been a major challenge for DFL, especially given the limitations of existing communication channels.
Table \ref{11261948} reports the number of raw    parameters transmitted by each client per round  under different  schemes.
 {\color{blue}Notably, the raw parameter count does not include the parameters related to signatures for authentication in the blockchain.}
Specifically, the communication cost of BLADE-FL, MOON, FedIntR, and SWIM is determined by the size of the model architecture.
 {\color{blue}In other words, the number of parameters required by architectures such as CNN or ResNet18 directly corresponds to the amount of parameters transmitted by clients in these schemes.
Thus, for the same model architecture, these baselines transmit the same number of parameters.
In contrast, the communication cost of DFPL depends only on the dimensionality of the feature extractor's output rather than the total number of model parameters.}
Notably, compared to these baselines, DFPL transmits fewer parameters per round   while still achieving better performance under heterogeneous data distributions.
This observation indicates  that transmitting more parameters does  not  {\color{blue}necessarily enhance the performance of federated training in heterogeneous environments.}

\subsection{Impact of Mining and Training on DFPL}

\begin{figure*}[t]
  \centering
  \subfloat[MNIST, \textit{Avg}=3, \textit{Std}=1]
  {
      \label{110916MNIST}  \includegraphics[width=0.18\linewidth]{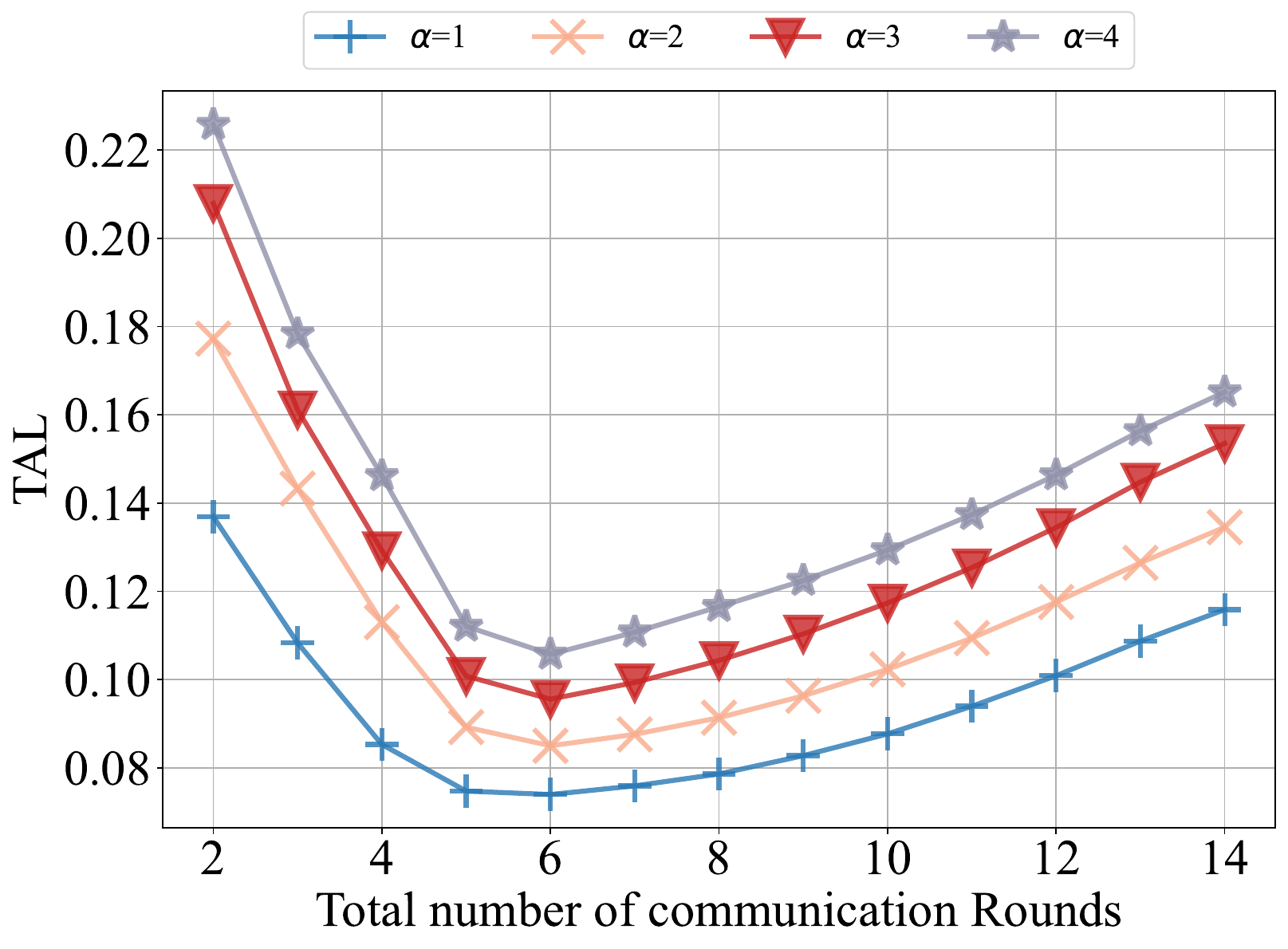}
  }
  \subfloat[MNIST, \textit{Avg}=3, \textit{Std}=2]
  {
      \label{110912MNIST}  \includegraphics[width=0.18\linewidth]{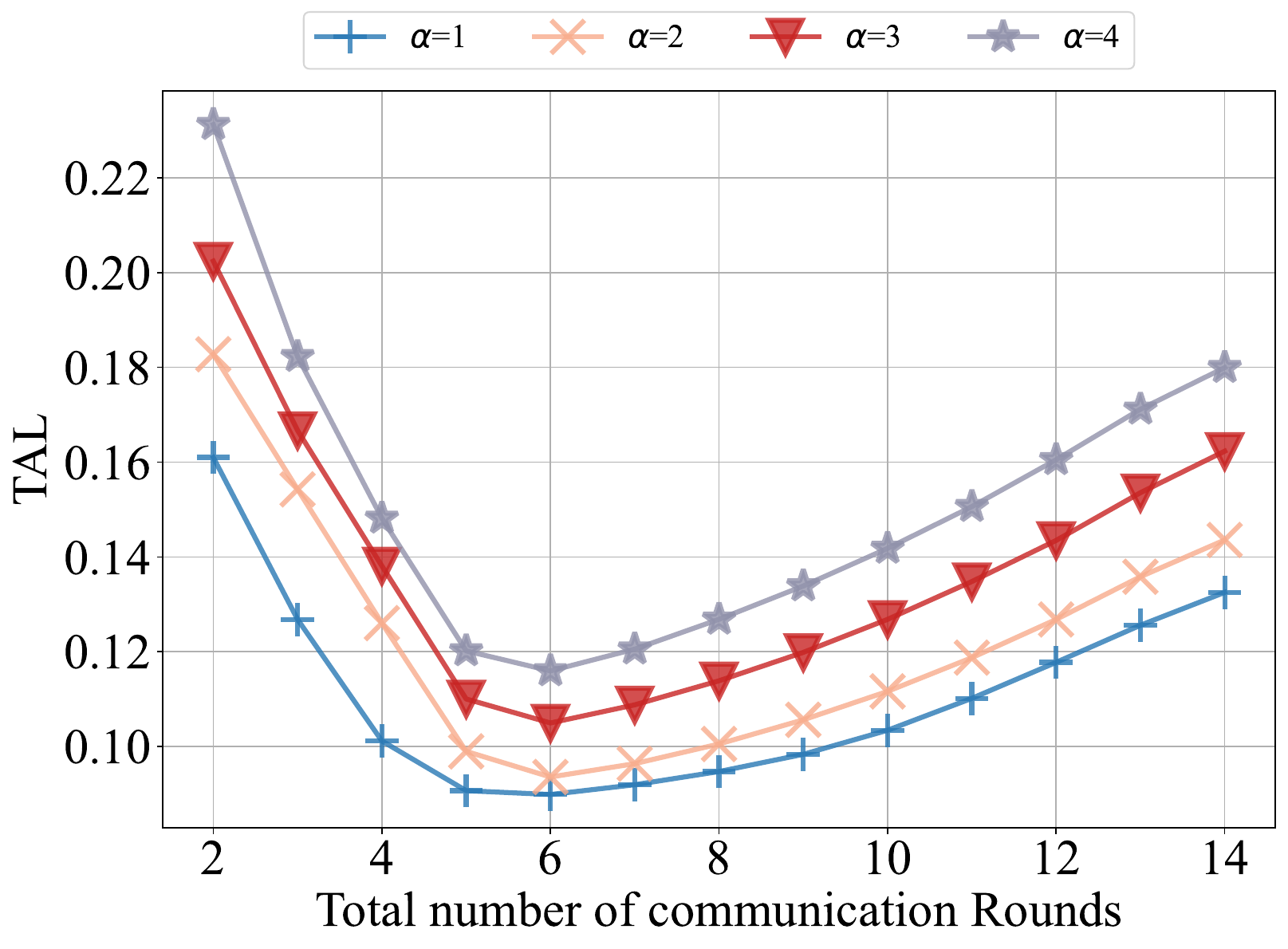}
  }
  \subfloat[MNIST, \textit{Avg}=4, \textit{Std}=1]
  {
      \label{110914MNIST}  \includegraphics[width=0.18\linewidth]{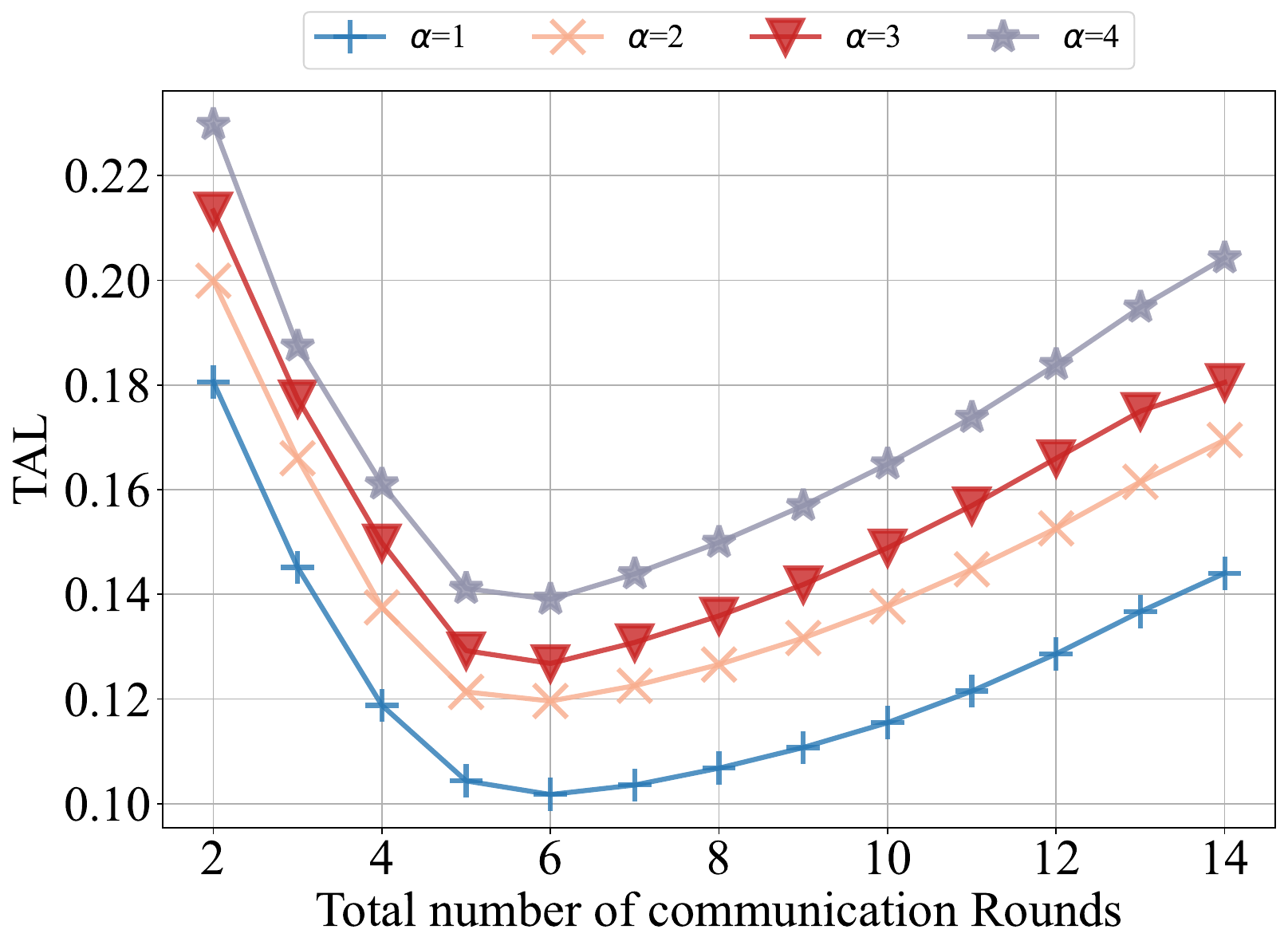}
  }
  \subfloat[FMNIST, \textit{Avg}=3, \textit{Std}=1]
  {
      \label{110916FashionMNIST}  \includegraphics[width=0.18\linewidth]{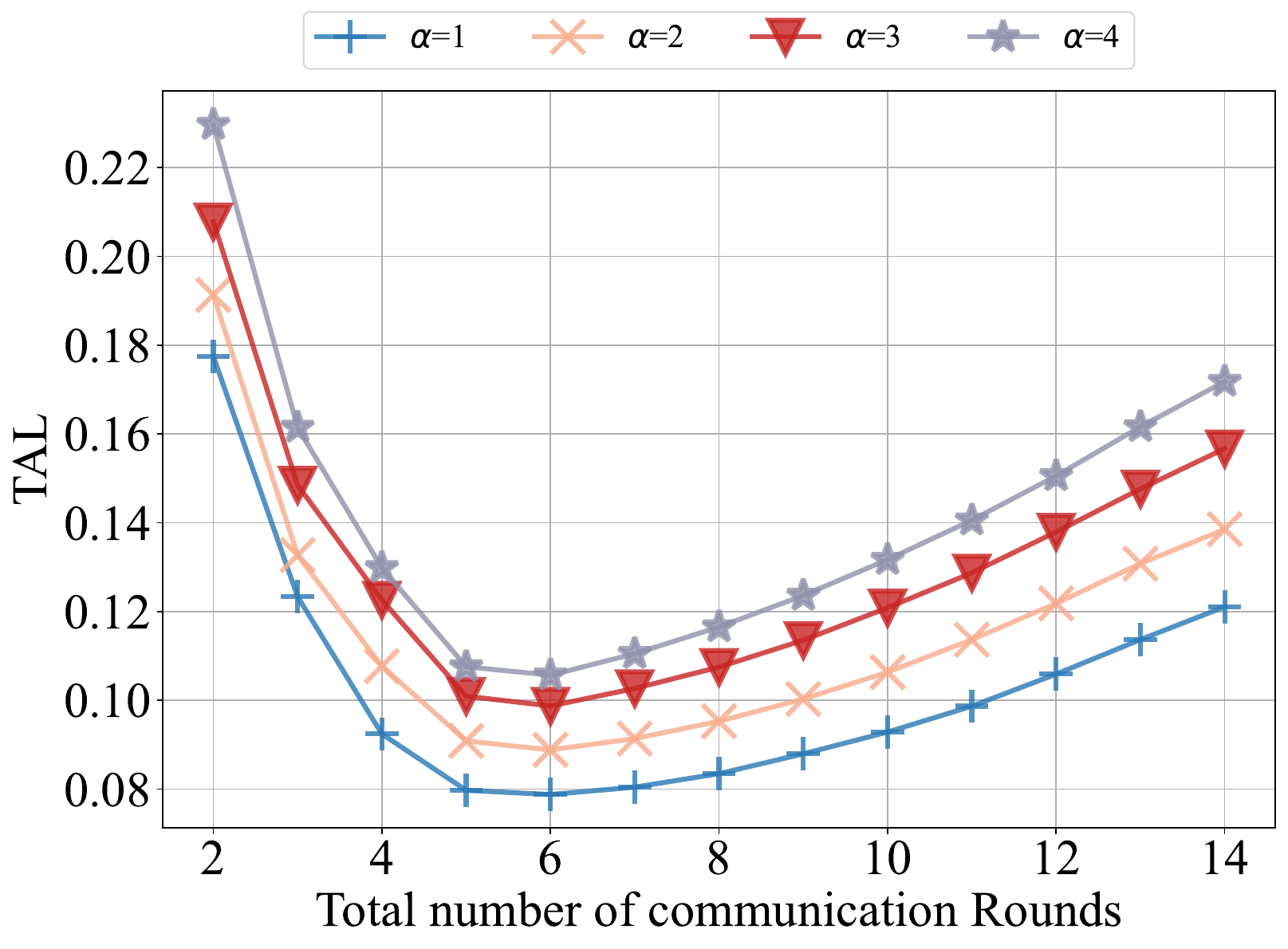}
  }
  \subfloat[FMNIST, \textit{Avg}=3, \textit{Std}=2]
  {
      \label{1109121FashionMNIST}  \includegraphics[width=0.18\linewidth]{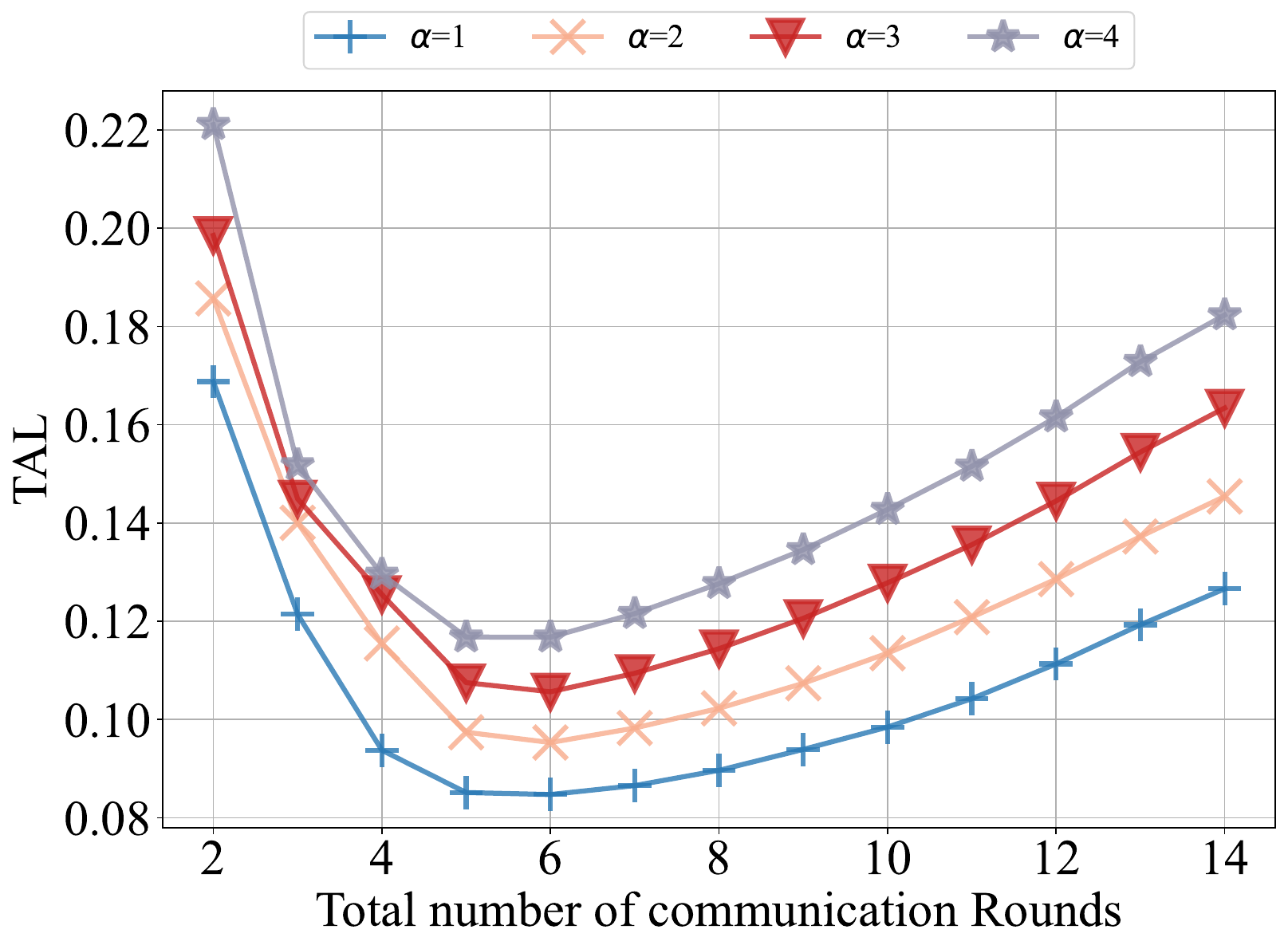}
  }

  \subfloat[FMNIST, \textit{Avg}=4, \textit{Std}=1]
  {
      \label{110914FashionMNIST}  \includegraphics[width=0.18\linewidth]{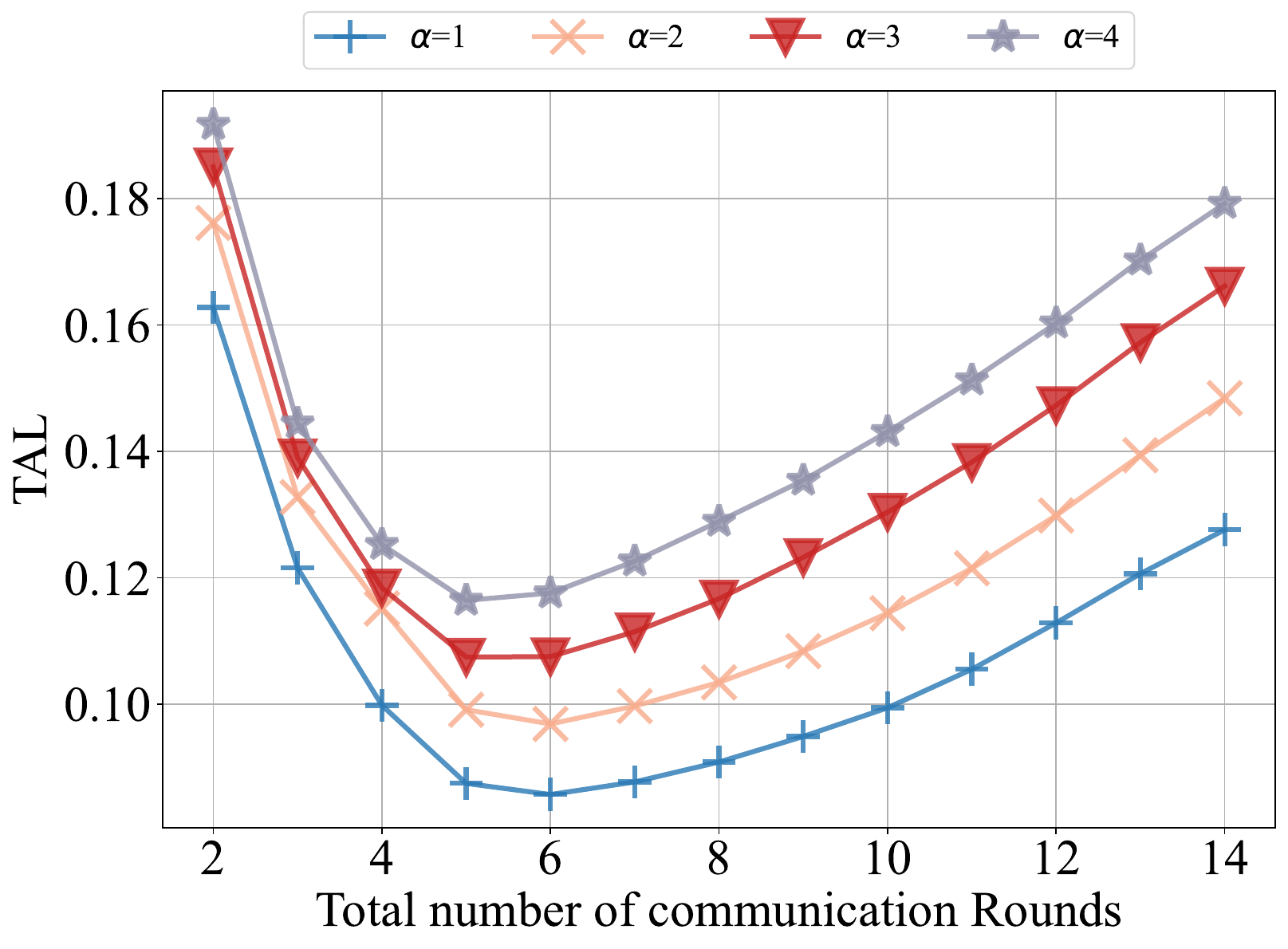}
  }
  \subfloat[CIFAR10, \textit{Avg}=3, \textit{Std}=1]
  { \includegraphics[width=0.18\linewidth]{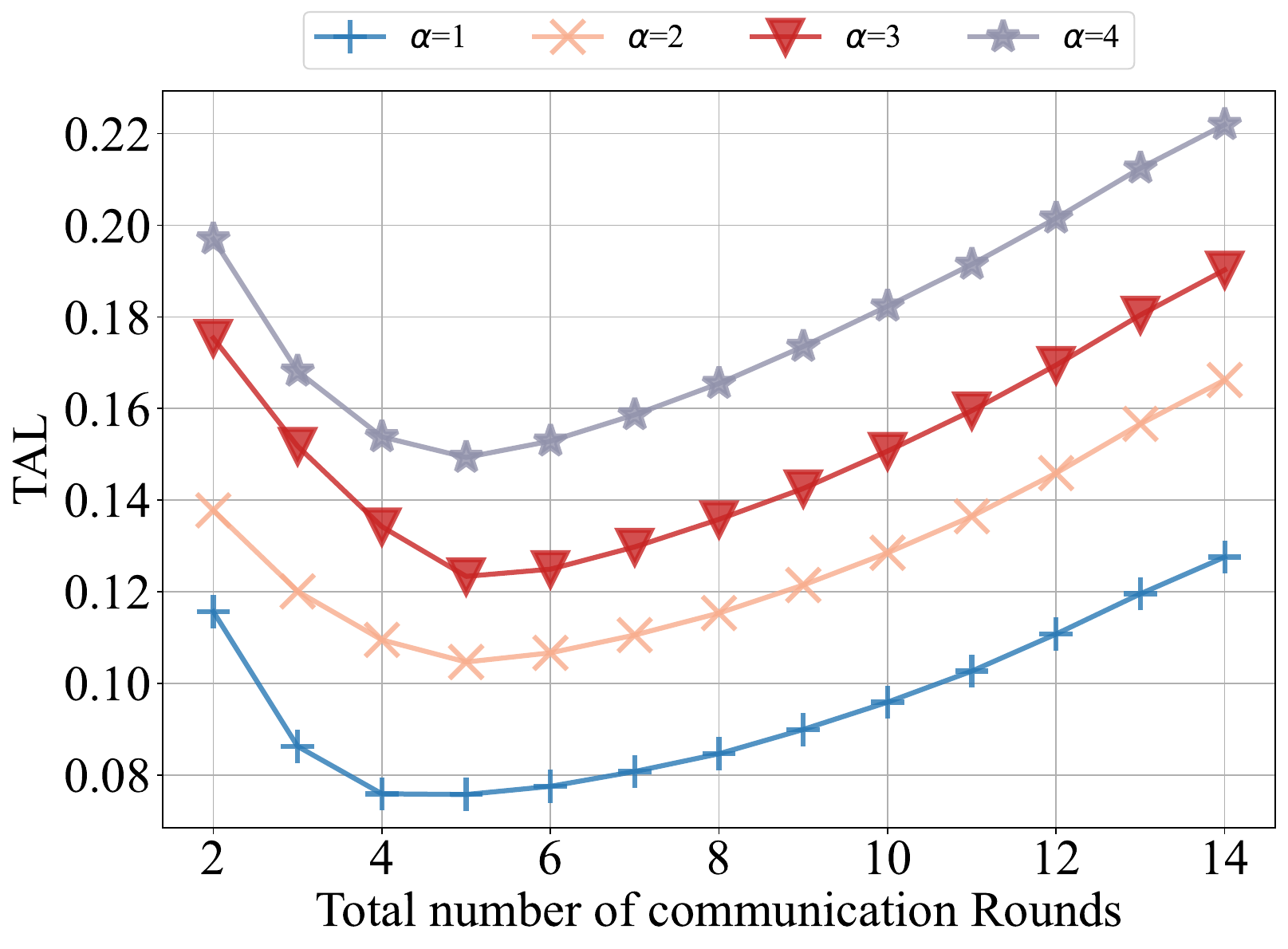}
  }
  \subfloat[CIFAR10, \textit{Avg}=3, \textit{Std}=2]
  { \includegraphics[width=0.18\linewidth]{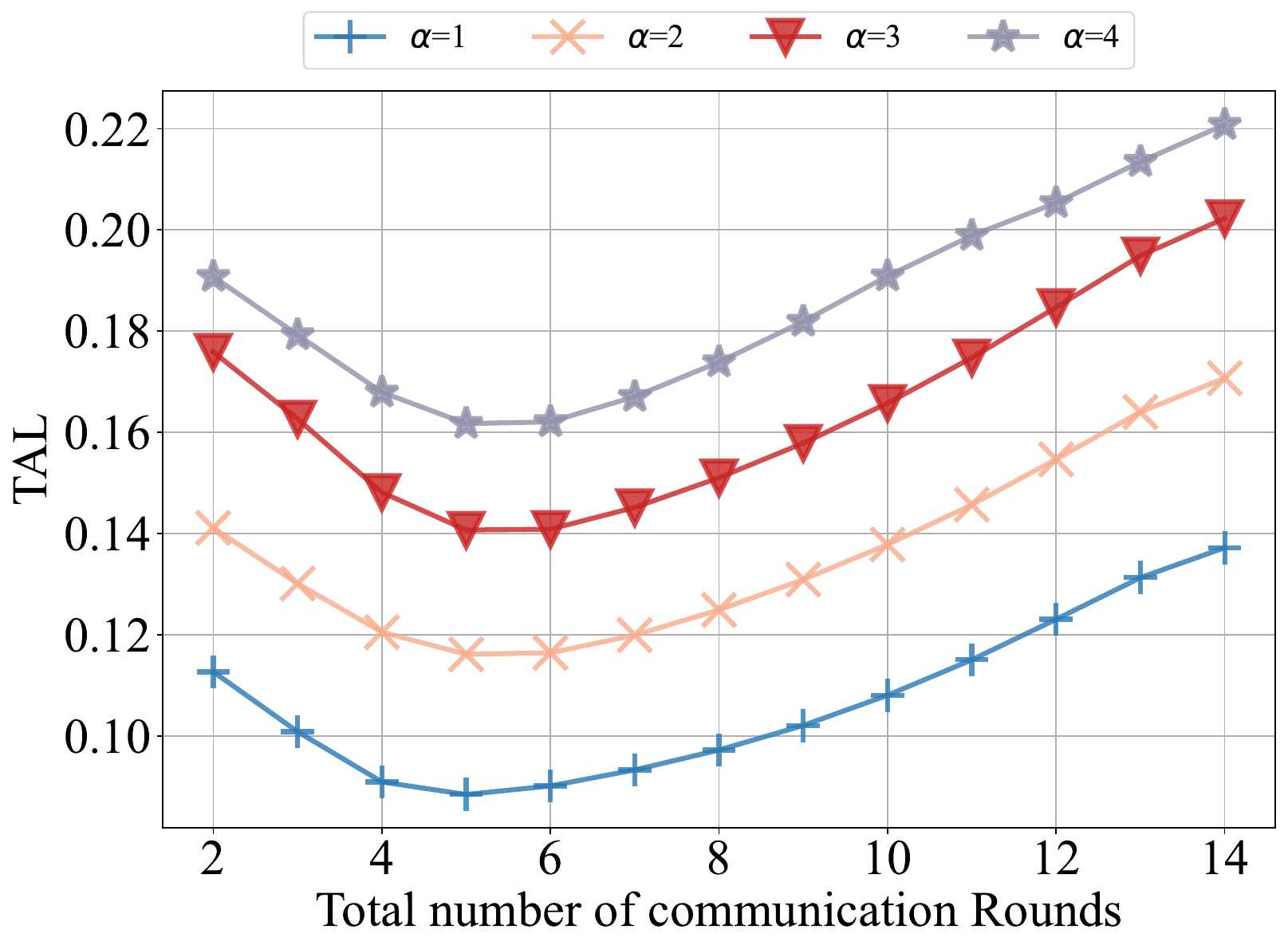}
  }
  \subfloat[CIFAR10, \textit{Avg}=4, \textit{Std}=1]
  { \includegraphics[width=0.18\linewidth]{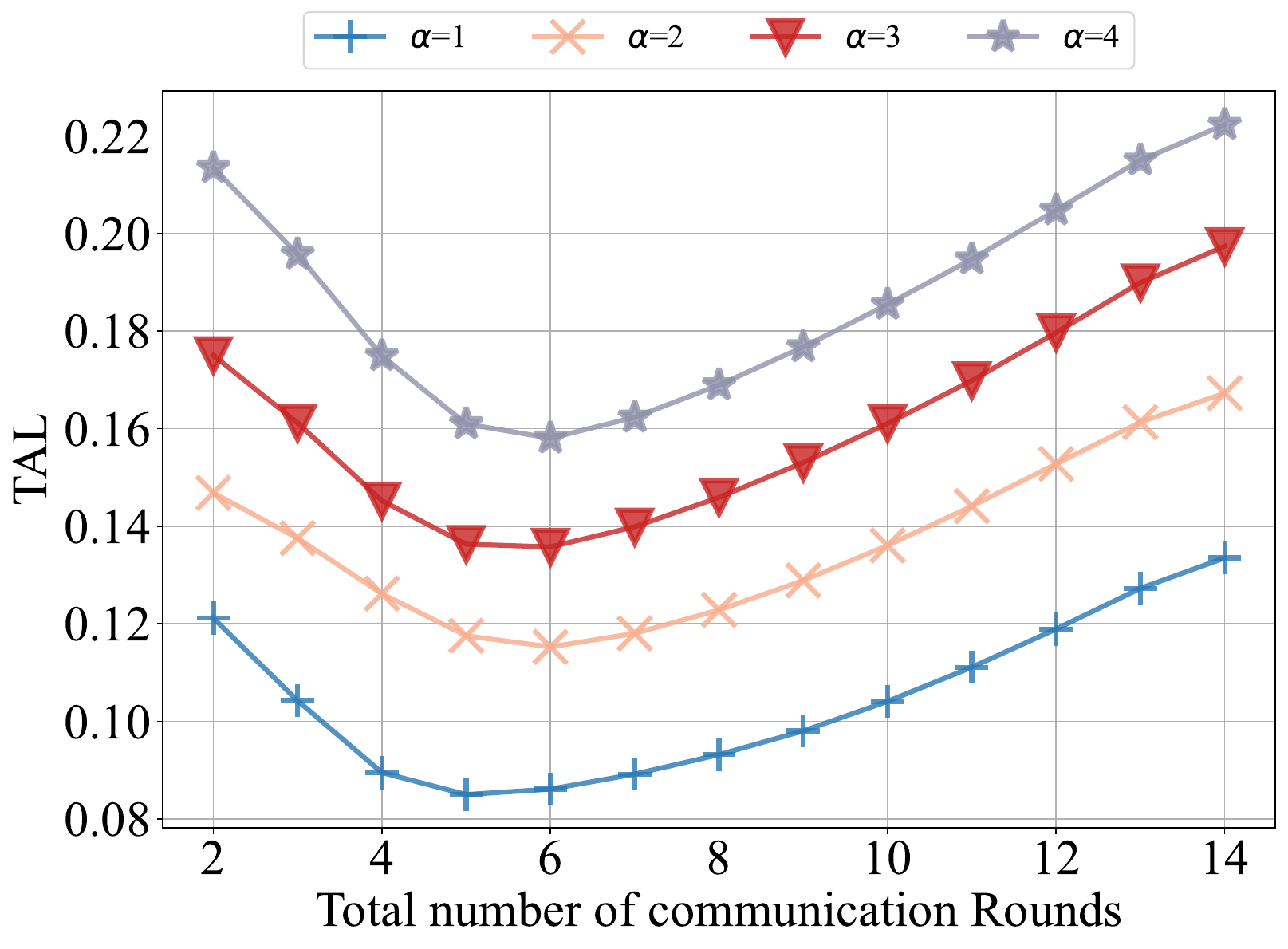}
  }
\caption{ TAL  versus $R$ for different $\alpha$ values  for MNIST,  FMNIST, and CIFAR10  with different data distributions.}
\label{11091}
\end{figure*}

\begin{figure*}[!t]
  \centering
  \subfloat[MNIST, \textit{Avg}=3, \textit{Std}=1]
  {
      \label{63105122MNIST}  \includegraphics[width=0.18\linewidth]{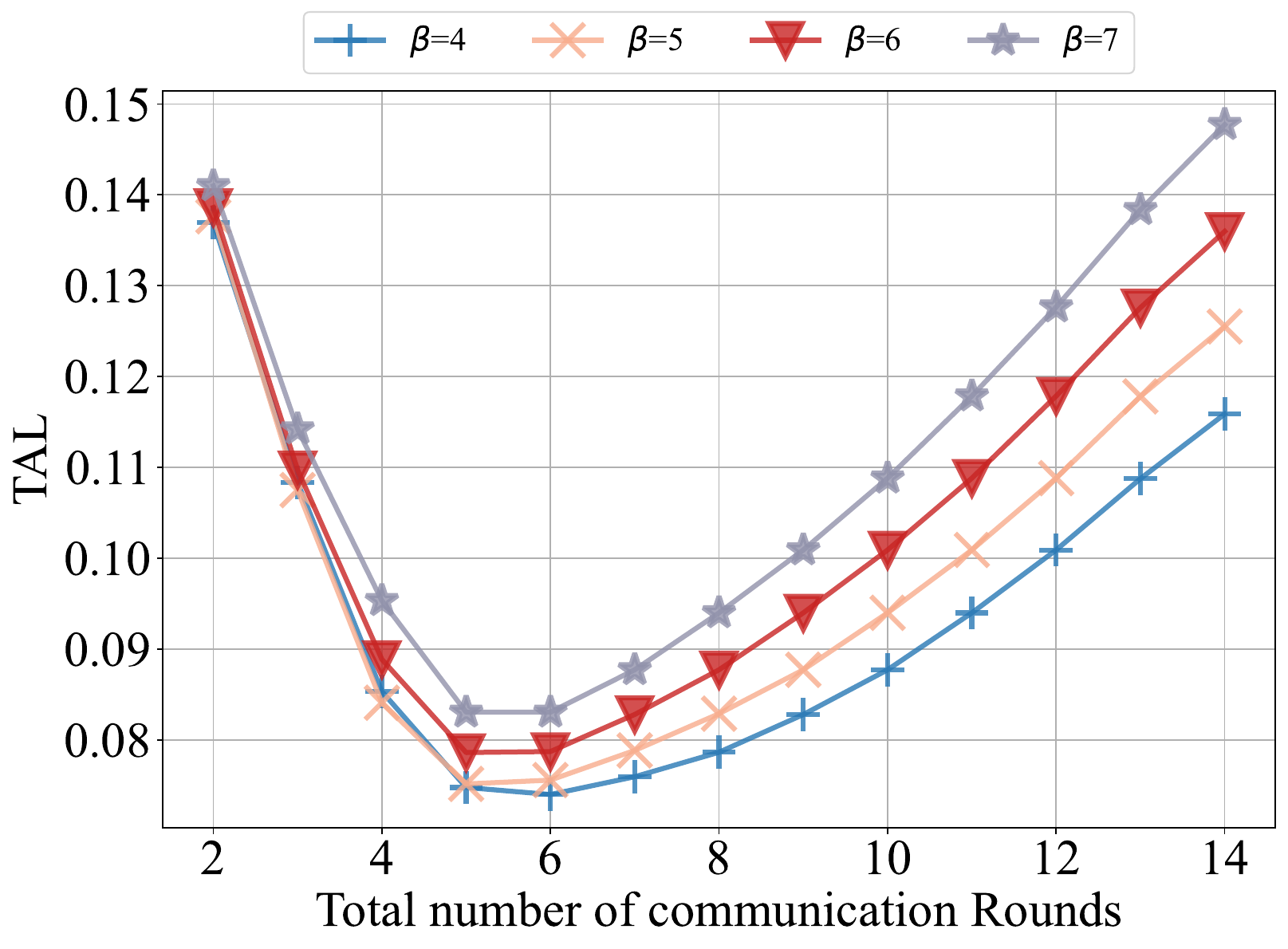}
  }
  \subfloat[MNIST, \textit{Avg}=3, \textit{Std}=2]
  {
      \label{6310514MNIST}  \includegraphics[width=0.18\linewidth]{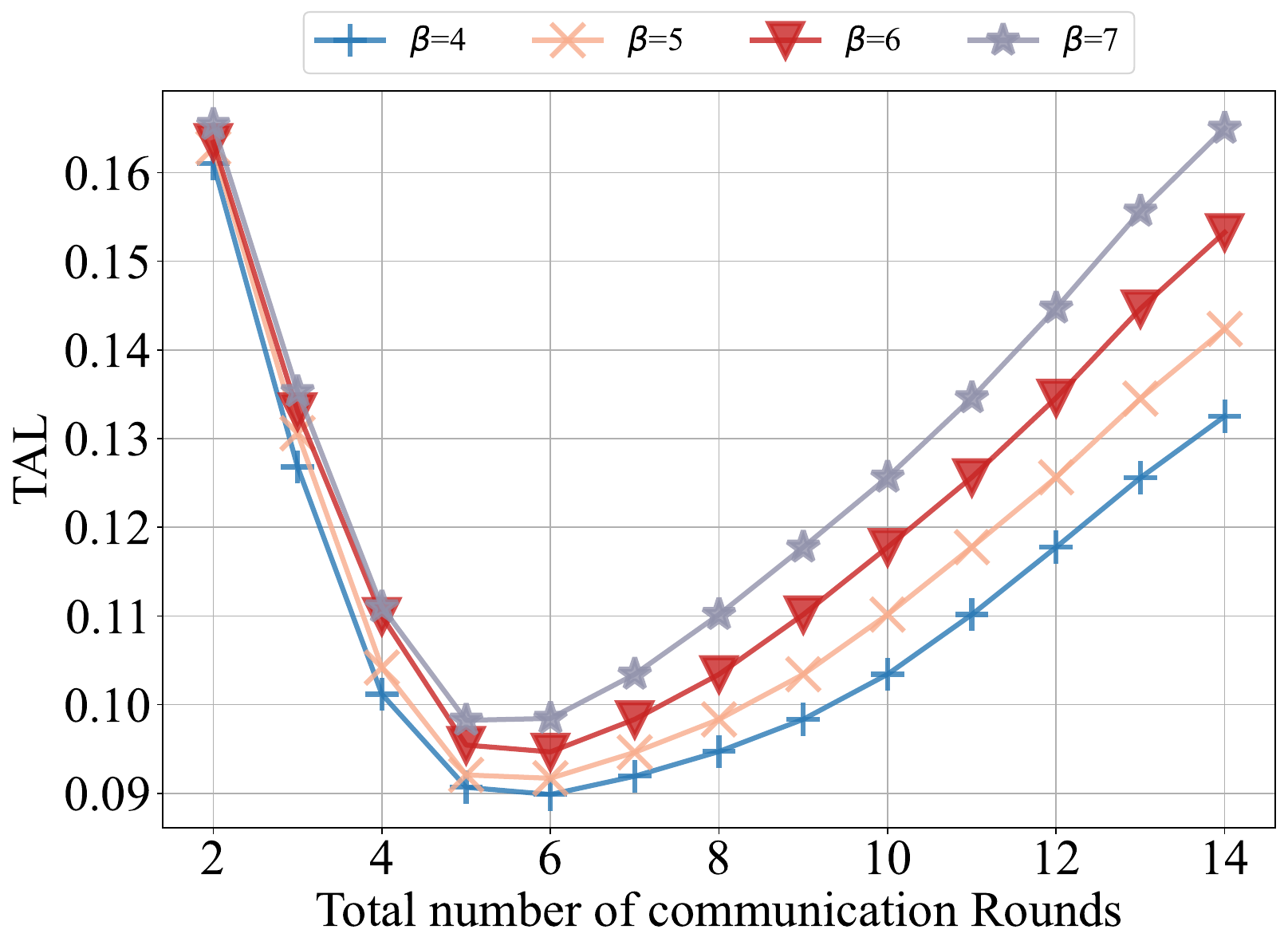}
  }
  \subfloat[MNIST, \textit{Avg}=4, \textit{Std}=1]
  {
      \label{6310516MNIST}  \includegraphics[width=0.18\linewidth]{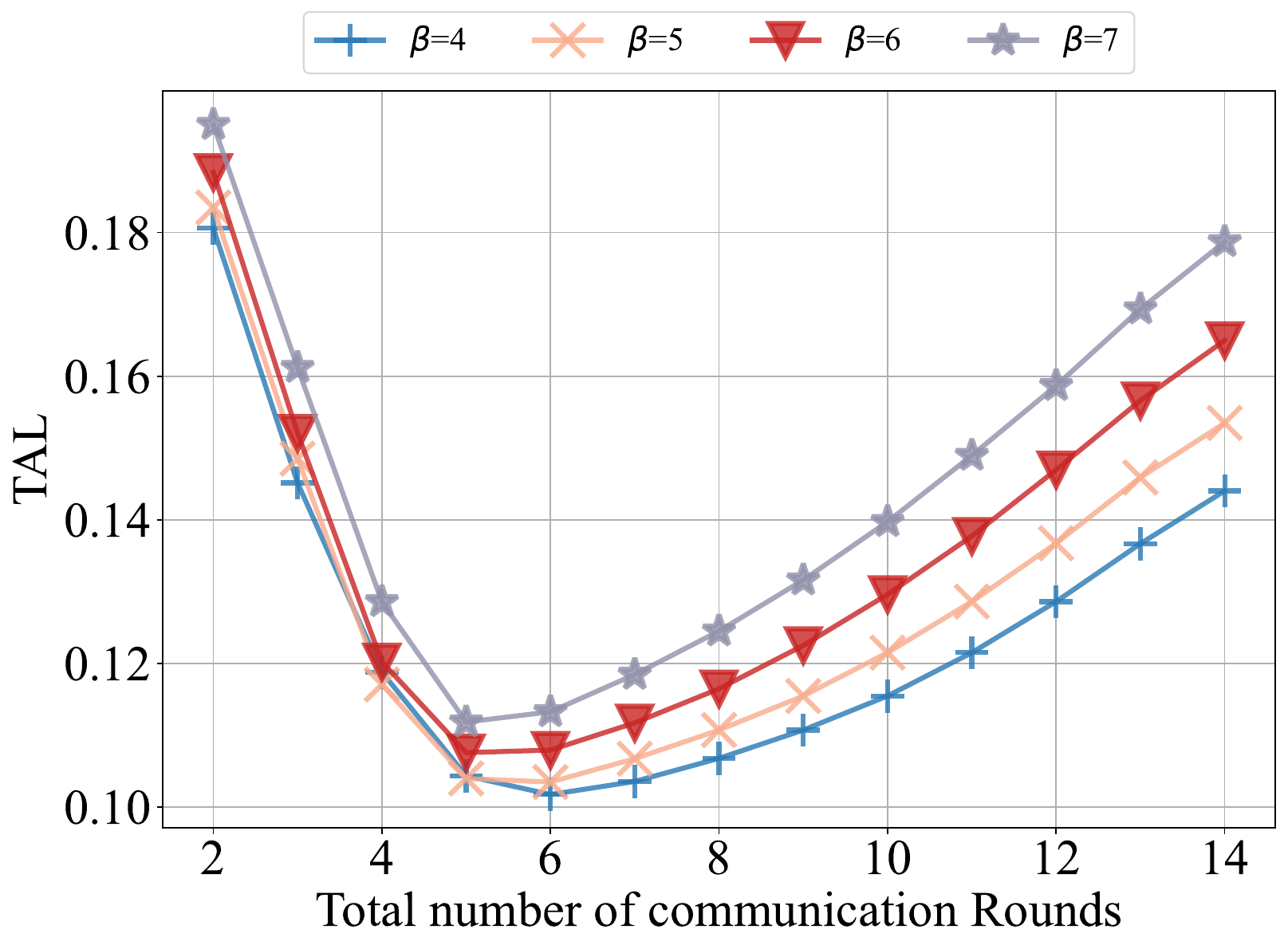}
  }
  \subfloat[FMNIST, \textit{Avg}=3, \textit{Std}=1]
  {
      \label{1109161FashionMNIST}  \includegraphics[width=0.18\linewidth]{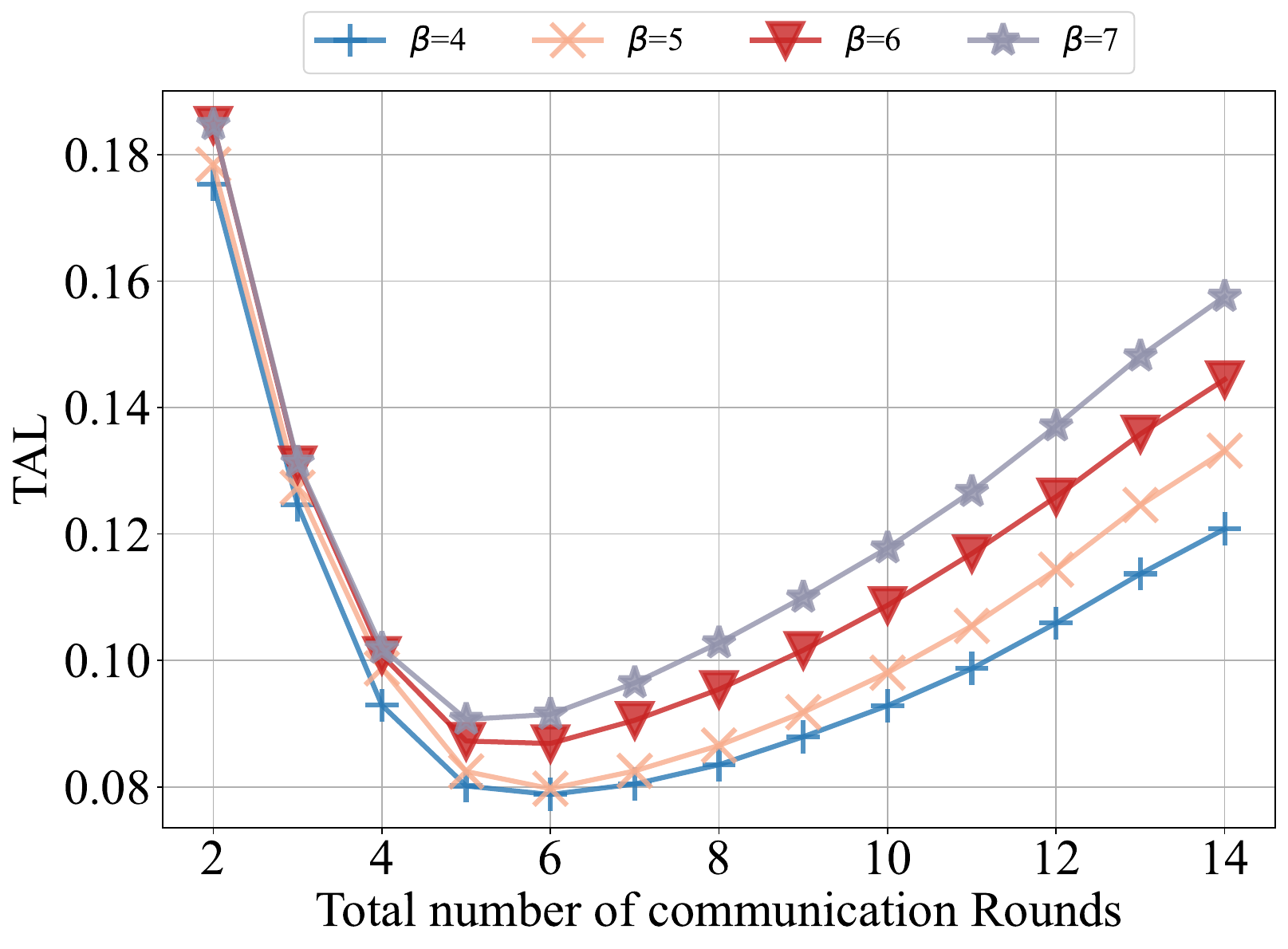}
  }
  \subfloat[FMNIST, \textit{Avg}=3, \textit{Std}=2]
  {
      \label{11091211FashionMNIST}  \includegraphics[width=0.18\linewidth]{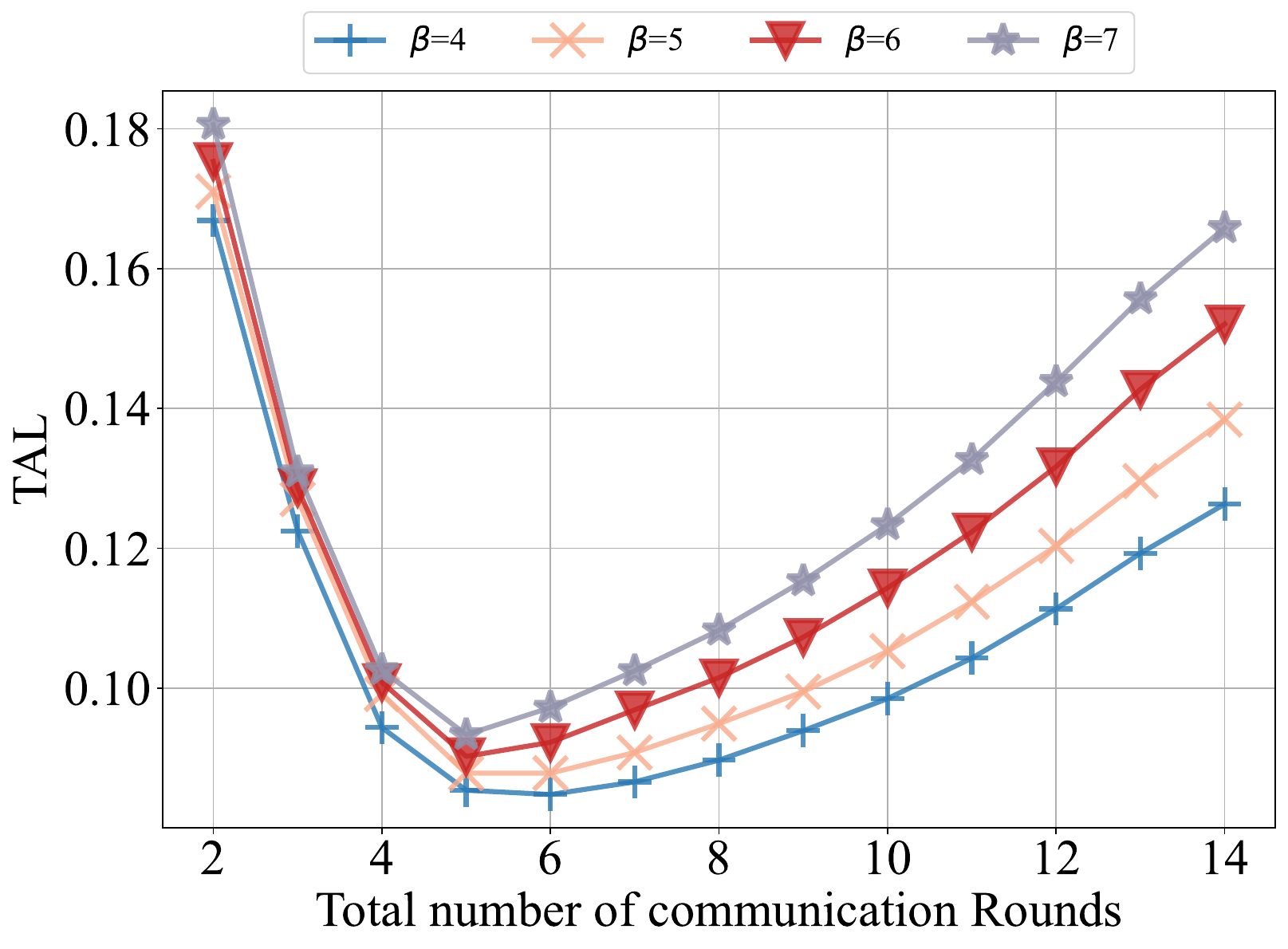}
  }

  \subfloat[FMNIST, \textit{Avg}=4, \textit{Std}=1]
  {
      \label{1109141FashionMNIST}  \includegraphics[width=0.18\linewidth]{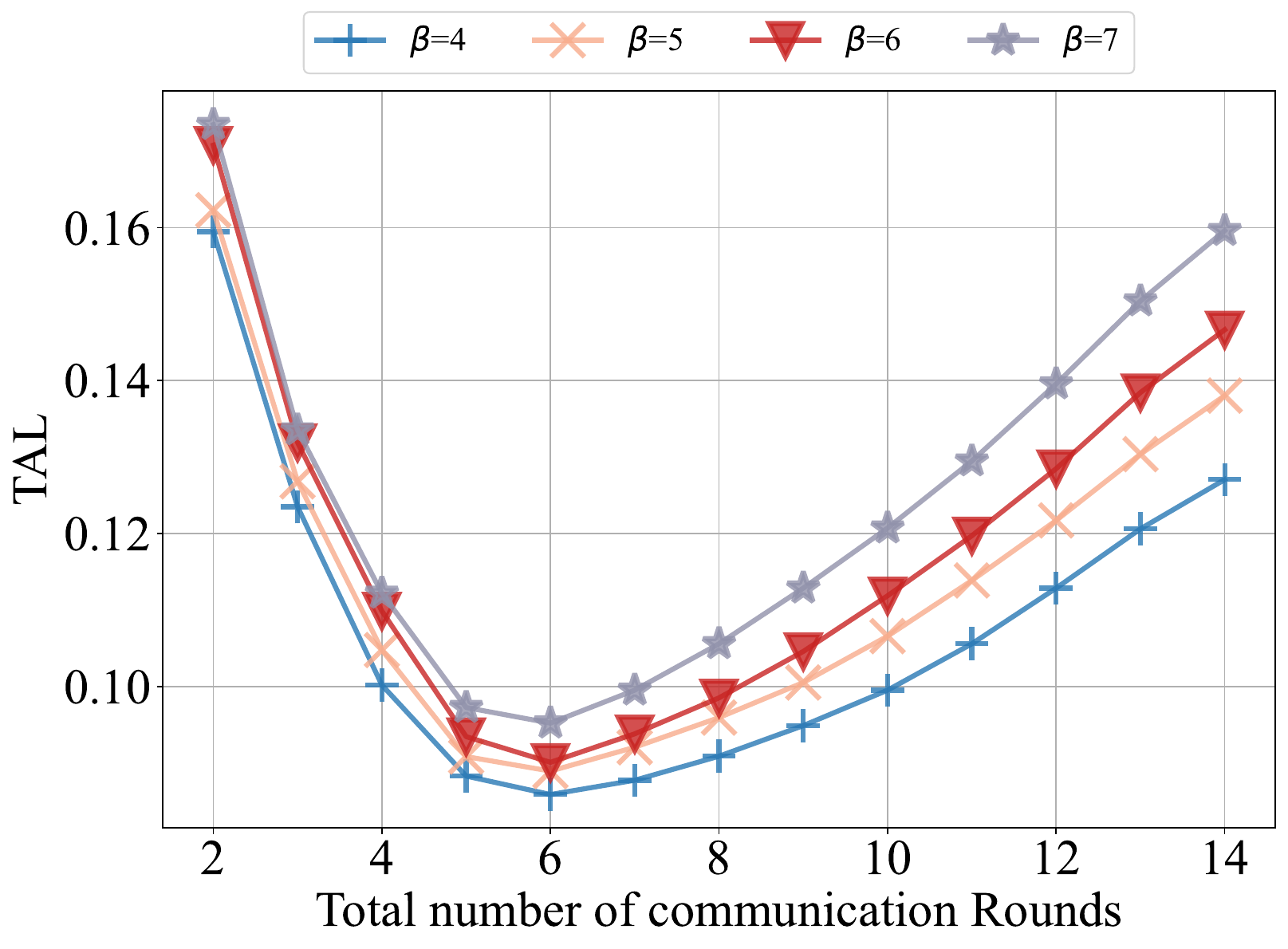}
  }
  \subfloat[CIFAR10, \textit{Avg}=3, \textit{Std}=1]
  {
      \label{110916CIFAR10}  \includegraphics[width=0.18\linewidth]{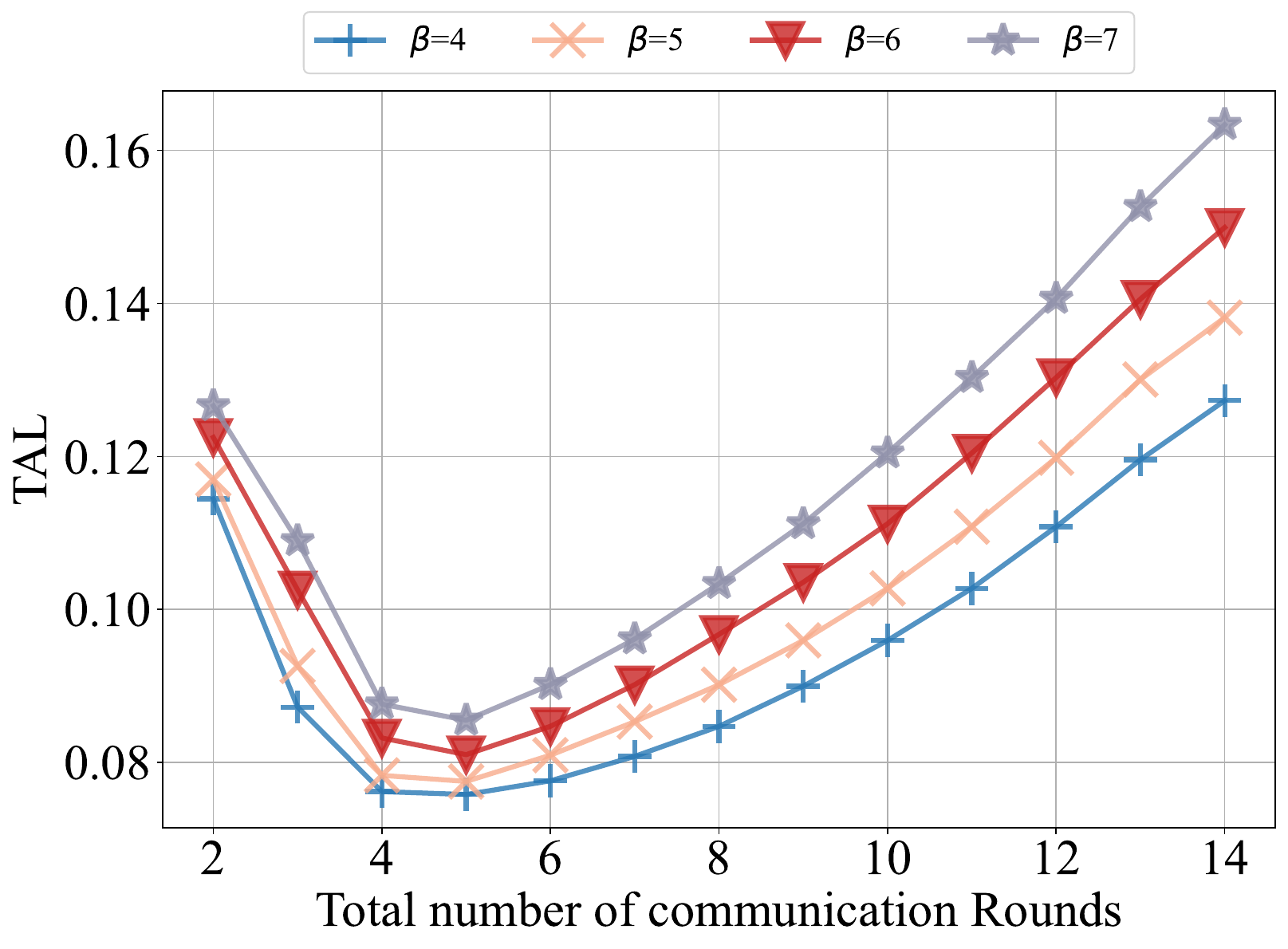}
  }
  \subfloat[CIFAR10, \textit{Avg}=3, \textit{Std}=2]
  {
      \label{110912CIFAR10}  \includegraphics[width=0.18\linewidth]{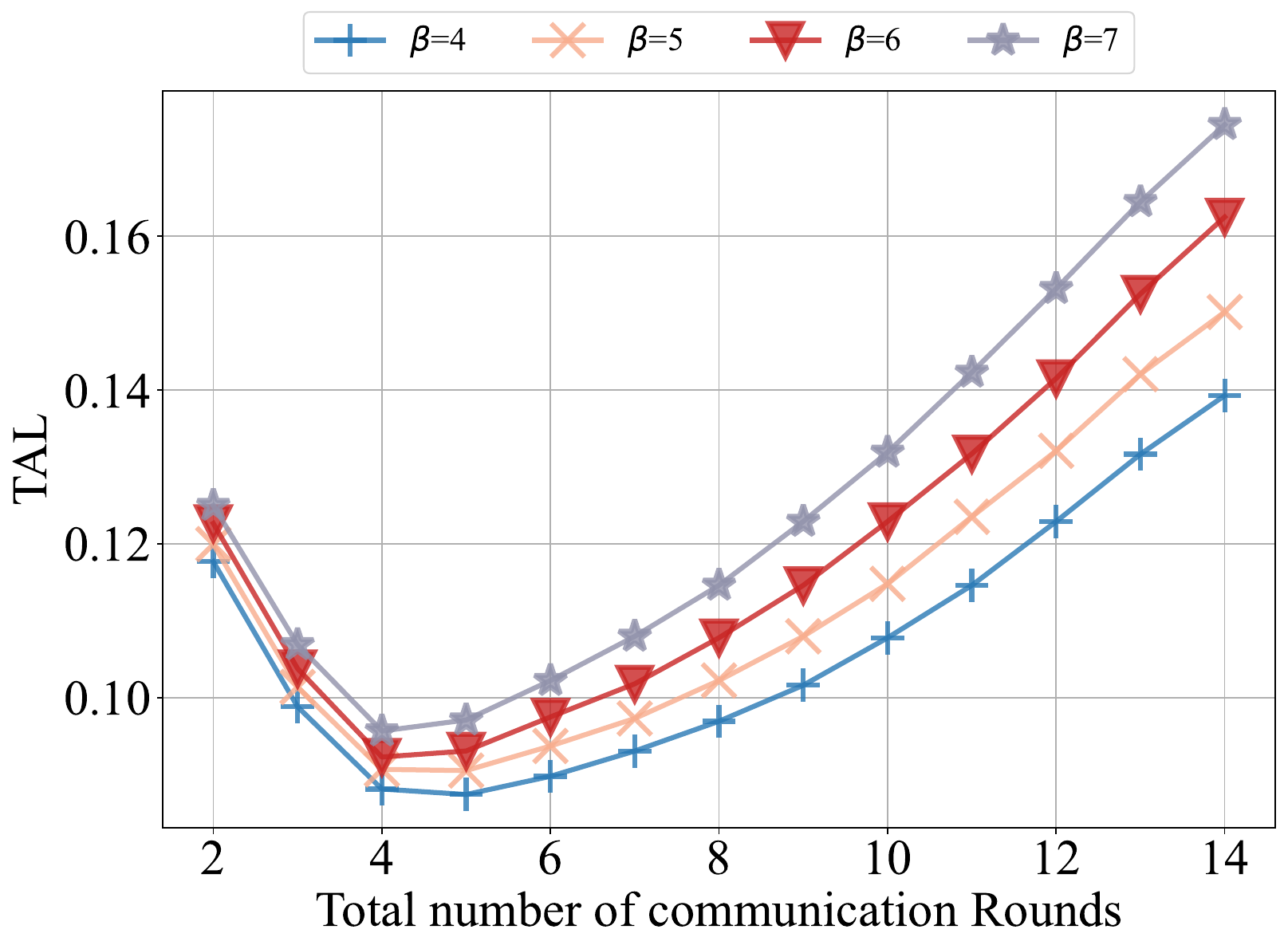}
  }
  \subfloat[CIFAR10, \textit{Avg}=4, \textit{Std}=1]
  {
      \label{110914CIFAR10}  \includegraphics[width=0.18\linewidth]{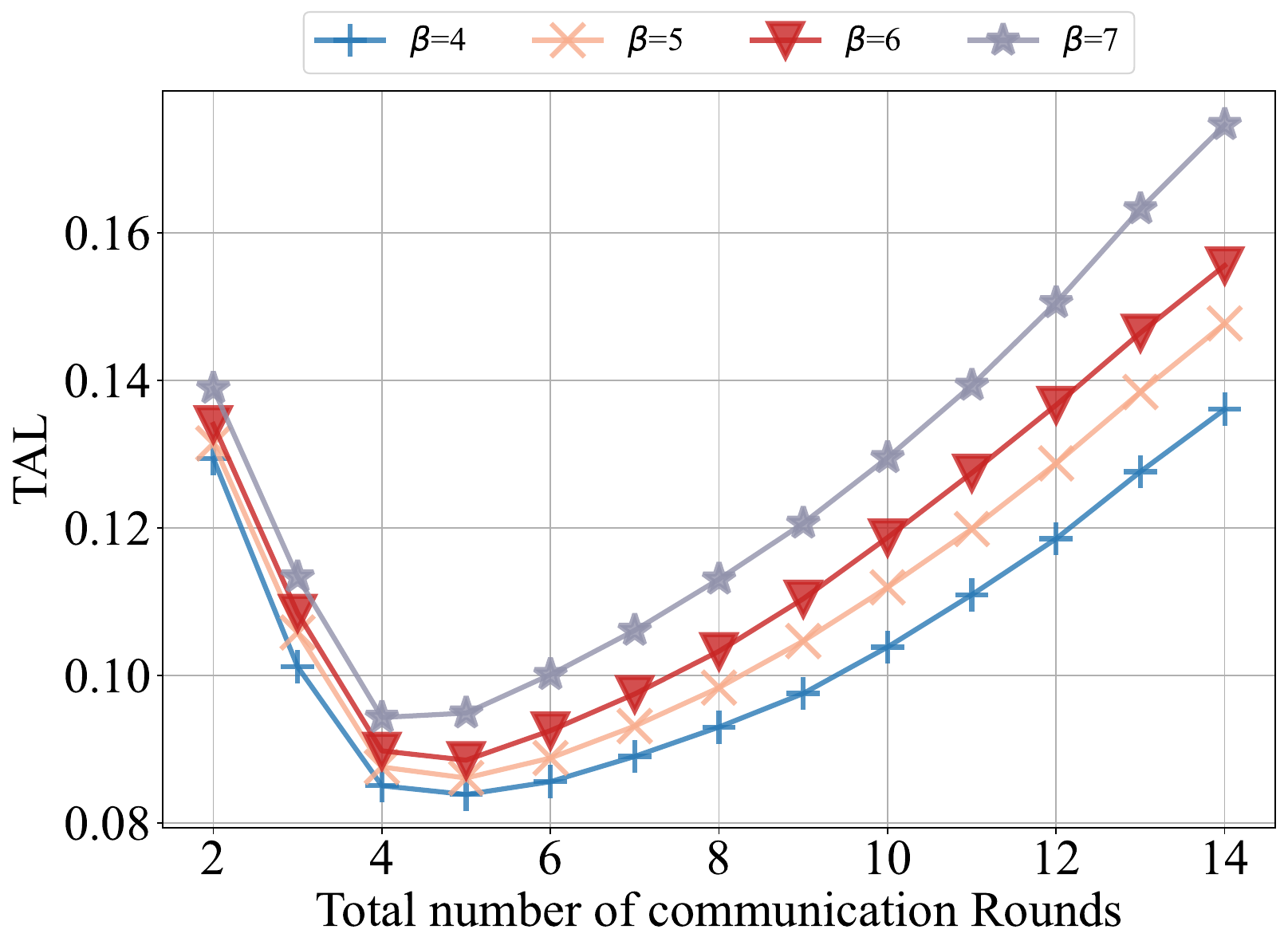}
  }
\caption{ TAL  versus $R$ for different $\beta$ values  for MNIST,  FMNIST, and CIFAR10  with different data distributions.}
\label{631051}
\end{figure*}

To study the impact of training time $\alpha$ and mining time  $\beta$ on  the performance of DFPL,  we evaluate   DFPL under different values of $\alpha$ and $\beta$, while fixing the total computation time  at $t_{sum}=100$.
For performance evaluation,  we employ three data distribution settings: $(\textit{Avg}=3, \textit{Std}=1)$, $(\textit{Avg}=3, \textit{Std}=2)$ and $(\textit{Avg}=4, \textit{Std}=1)$ on MNIST, FMNIST, and CIFAR10, respectively.
The  experimental results are  shown in Fig. \ref{11091} and Fig. \ref{631051}.
Notably,  within the same computation time (i.e., $t_{sum}$ = 100),
the number of executed communication rounds $R$ varies.
For example, the total computation time remains the same for $R = 2$, $4$, $6$, or $8$, while the number of local iterations per round differs.
The number of local training iterations is calculated as  $E = \lfloor\frac{1}{\alpha}({\frac{t_{sum}}{R}- \beta})\rfloor$.
Additionally, Tables \ref{01121} to \ref{01122} report the total  training time $E\alpha R$  required to achieve  optimal  performance under different data distributions, along with the corresponding number of communication rounds  $R$,  represented as tuples of the form  $(E\alpha R,R)$.

\begin{table}[t]
\centering
\caption{The optimal training time ($E\alpha R$) for different $\alpha$ values.  }
\tabcolsep= 0.175cm
\begin{tabular}{c|c|ccc}
\hline\hline
\multirow{2}{*}{ \makecell[c]{Data Distribution }} & \multirow{2}{*}{ \makecell[c]{Training Time \\Per Iteration}} & \multicolumn{3}{c}{$(E\alpha R,R)$}  \\
                       &                                                                                        & MNIST  & FMNIST   & CIFAR10       \\ \hline
\multirow{4}{*}{ \makecell[c]{\textit{avg}=3\\\textit{std} =1}}                     &              $\alpha $ = 1                                                                         &     (76, 6)   &        (76, 6)     &(80, 5)            \\
                       &                                                 $ \alpha$ = 2                                     &    (76, 6)   &      (76, 6)         &(80, 5)    \\
                       &                                                 $ \alpha$   = 3                                   &     (76, 6)     &         (76, 6)        & (80, 5)   \\
                       &                                           $\alpha$        = 4                                     &       (76, 6)     &        (76, 6)        &  (80, 5)     \\ \hline
 \multirow{4}{*}{ \makecell[c]{\textit{avg}=3\\\textit{std} =2}}                         &              $\alpha $ = 1                                                                            &       (76, 6)   &       (76, 6)     &(80, 5)        \\
                       &                                                       $ \alpha$  = 2                                   &   (76, 6) &         (76, 6)     &  (80, 5)   \\
                       &                                                 $ \alpha$          = 3                             &  (76, 6)    &            (76, 6)    &(76, 6)    \\
                       &                                                        $ \alpha$     = 4                               &    (76, 6)   &        (80, 5)     & (80, 5)   \\ \hline
 \multirow{4}{*}{ \makecell[c]{\textit{avg}=4\\\textit{std} =1}}                         &        $\alpha $ = 1                                                                                  &      (76, 6)   &         (76, 6)   & (80, 5)    \\
                       &                                               $ \alpha$    = 2                                       &     (76, 6)      &       (76, 6)     & (76, 6)    \\
                       &                                                 $ \alpha$       = 3                               &   (76, 6)   &          (80, 5)      & (76, 6)   \\
                       &                                                  $ \alpha$        = 4                                &   (76, 6)    &        (80, 5)       & (76, 6)       \\ \hline\hline
\end{tabular}
\label{01121}
\end{table}
\subsubsection{The Impact of Parameter $\alpha$ in  DFPL}
To evaluate the impact of $\alpha$ for  DFPL, we  set $\alpha \in \{1,2,3,4\}$  for evaluation.
In addition, we set the mining time $\beta$ fixed at 4.
The experimental results are shown in Fig. \ref{11091}.
Furthermore, Table \ref{01121} reports the training time and the number of communication rounds required to achieve minimum loss under  various data distributions,  where the majority of optimal results is achieved around 6 communication rounds with a total training time of approximately 76.
 {\color{blue}Moreover, we observe that, regardless of $\alpha$, TAL generally decreases at first  and then increases  with the number of communication rounds grows, reaching  the minimum loss at  the 6-th communication round.}
 {\color{blue}This indicates that, under the same computational budget,  the allocation of the number of communication rounds and local iterations has a significant impact on the model performance. 
When the number of communication rounds is too small, even prolonged local training cannot make up for the lack of information exchange among clients, resulting in higher loss.}
 {\color{blue}In contrast, when the number of communication rounds is too large, frequent information exchange shortens the time available for local training, preventing clients from fully leveraging their  data for local optimization, thereby degrading performance.
Thus, this experiment reflects the inherent trade-off between local computation and global communication, where insufficient communication weakens collaboration, while excessive communication reduces the efficiency of local optimization.}
Furthermore, we observe that as $\alpha$ decreases, TAA increases while TAL decreases.
This suggests that under the same computation time, the performance of federated learning improves as the CPU cycles $f$ per second  increase for   each client.
This is because a higher number of CPU cycles $f$ per second enables each client to perform more local iterations within the limited computation time, thereby enhancing the model's training effectiveness.

\begin{table}[t]
\centering
\caption{The optimal training time ($E\alpha R$) for different $\beta$ values.}
\tabcolsep= 0.175cm
\begin{tabular}{c|c|ccc}
\hline\hline
\multirow{2}{*}{ \makecell[c]{Data distribution}} & \multirow{2}{*}{ \makecell[c]{Mining time \\per round}} & \multicolumn{3}{c}{$(E\alpha R,R)$}  \\
                       &                                                                                        & MNIST  & FMNIST  &CIFAR10  \\ \hline
\multirow{4}{*}{ \makecell[c]{\textit{avg}=3\\\textit{std} =1}}                     &              $\beta $ = 4                                                                         &     (76, 6)   &     (76, 6)     & (80, 5)  \\
                       &                                                 $ \beta$ = 5                                     &     (75, 5)  &      (70, 6)      & (75, 5)  \\
                       &                                                 $ \beta$   = 6                                   &   (70, 5)   &        (64, 6)    & (70, 5)     \\
                       &                                           $\beta$        = 7                                     &    (58, 6)   &        (65, 5)      & (65, 5)    \\ \hline
 \multirow{4}{*}{ \makecell[c]{\textit{avg}=3\\\textit{std} =2}}                         &              $\beta $ = 4                                                                            &    (76, 6)    &  (76, 6) & (80, 5)            \\
                       &                                                       $ \beta$  = 5                                   &   (70, 6) &         (75, 5)    &(75, 5)      \\
                       &                                                 $ \beta$          = 6                             &   (64, 6)    &           (70, 5)    &(76, 4)      \\
                       &                                                        $ \beta$     = 7                               &     (65, 5)   &        (65, 5)   &(72, 4)       \\ \hline
 \multirow{4}{*}{ \makecell[c]{\textit{avg}=4\\\textit{std} =1}}                         &        $\beta $ = 4                                                                                  &       (76, 6)   &          (76, 6)  &(80, 5)      \\
                       &                                               $ \beta$    = 5                                       &     (75, 5)       &        (70, 6)       &(75, 5)     \\
                       &                                                 $ \beta$       = 6                               &  (70, 5)    &          (64, 6)      &  (70, 5)    \\
                       &                                                  $ \beta$        = 7                                &   (65, 5)     &        (58, 6)       & (72, 4)     \\ \hline\hline
\end{tabular}
\label{01122}
\end{table}
\subsubsection{The Impact of Parameter $\beta$ in DFPL}

Fig. \ref{631051} shows the  TAL performance of DFPL for different $\beta$ values under heterogeneous  data distributions with  $\alpha=1$. 
Table \ref{01122} reports the corresponding optimal results, which are consistent with those presented in Table \ref{01121}.
It is observed that TAL decreases with increasing $\beta$, indicating that the performance of DFPL deteriorates with larger $\beta$ values.
 {\color{blue}This is because higher $\beta$ leads to longer total mining time,} thereby reducing the available training time and causing underfitting of the local models.
Moreover, we observe that regardless of the data distribution, TAL initially decreases and then increases as the number of communication rounds increases.
This suggests that DFPL's performance depends on a proper balance between communication rounds and local training time.
Table \ref{01122} shows that most optimal results are achieved around 5 or 6 communication rounds.
This number of rounds ensures a balance between sufficient local training time and effective information exchange among clients.

\subsection{DFPL under varying $\lambda$ and $\eta$}
\begin{figure}[t]
  \centering
  \subfloat[MNIST ]
  {
      \label{112711411}  \includegraphics[width=0.48\linewidth]{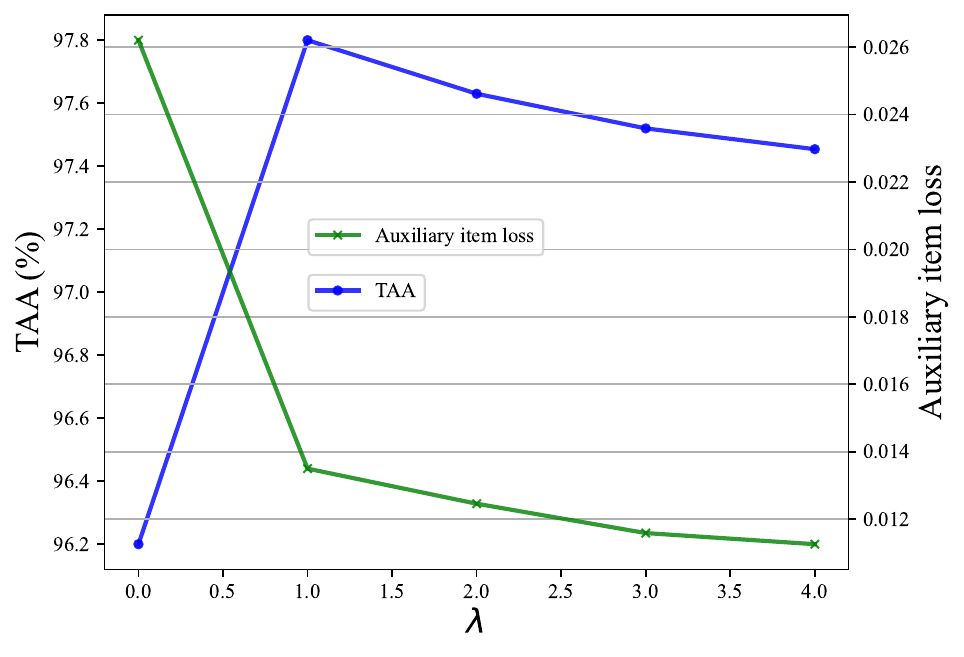}
  }
  \subfloat[FMNIST ]
  {
      \label{112711412}  \includegraphics[width=0.48\linewidth]{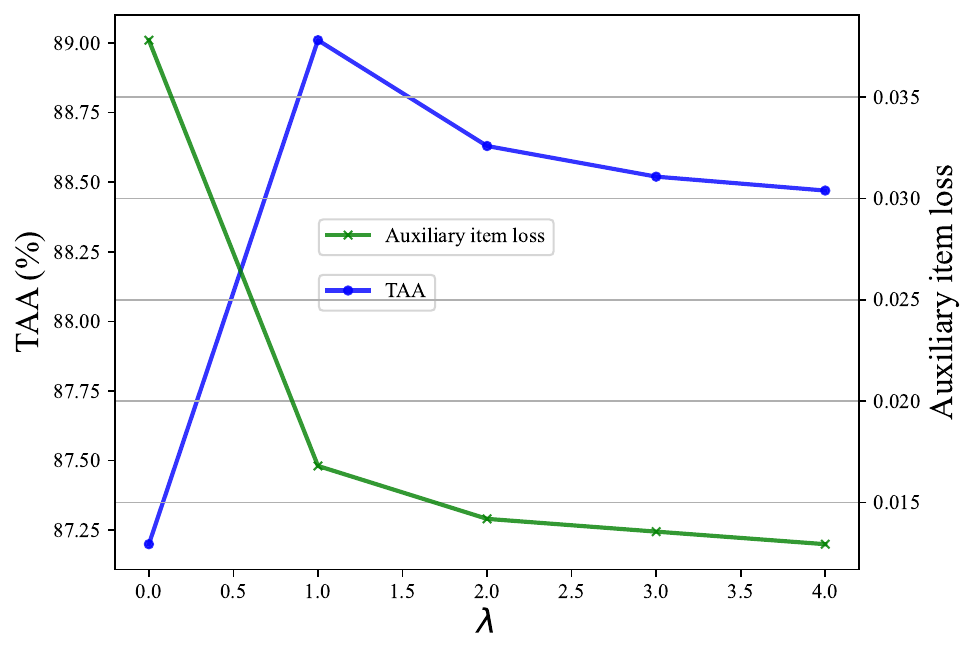}
  }

  \subfloat[CIFAR10 ]
  {
      \label{112711413}  \includegraphics[width=0.48\linewidth]{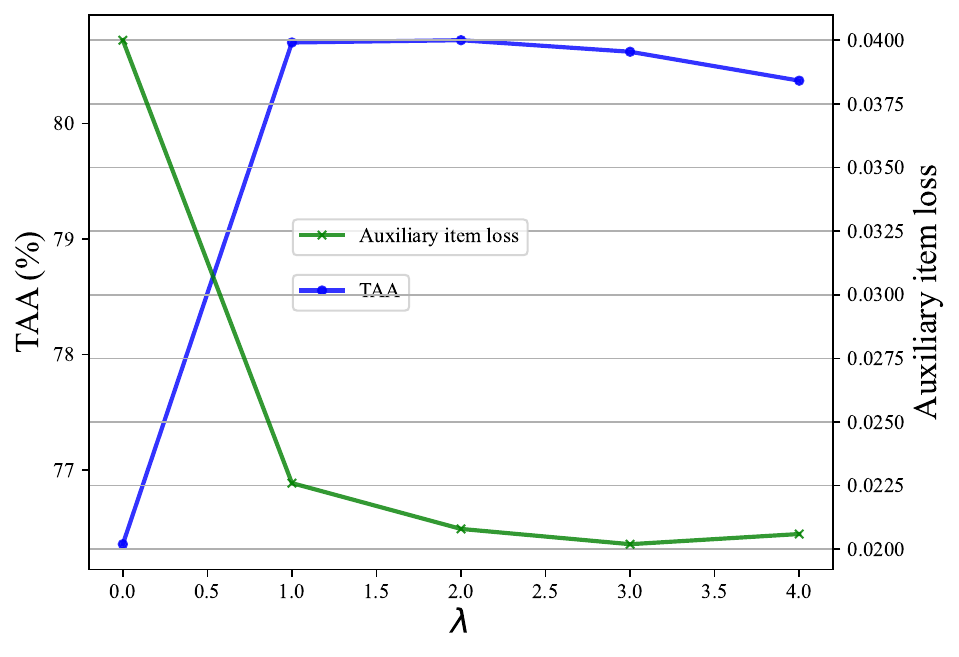}
  }
    \subfloat[SVHN ]
  {
      \label{112711413}  \includegraphics[width=0.48\linewidth]{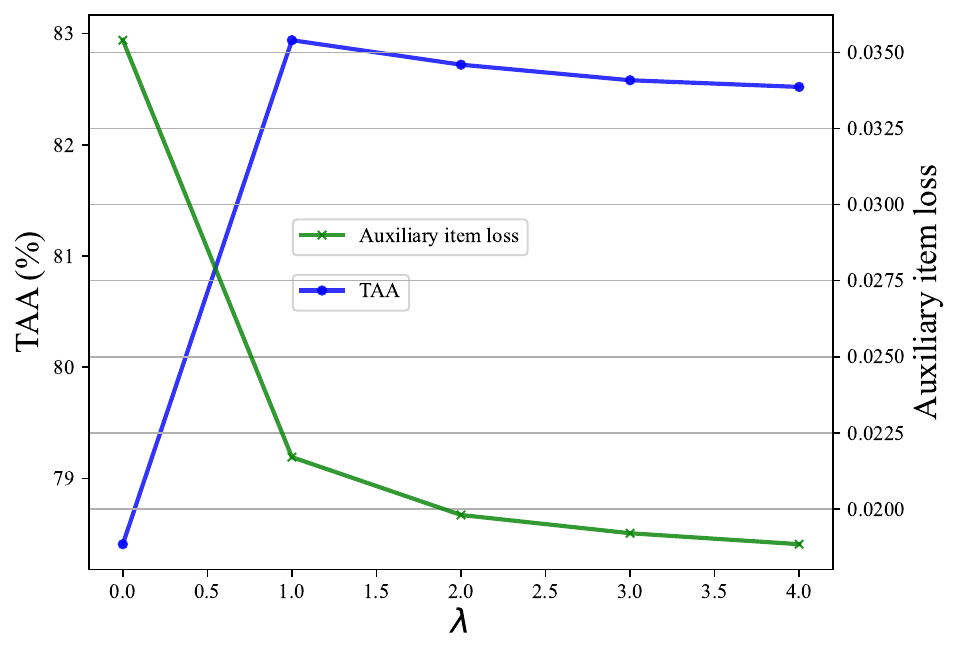}
  }
\caption{
DFPL's performance on the MNIST, FMNIST, CIFAR10, and SVHN  under varying importance weight $\lambda$.}
\label{11271141}
\end{figure}

To evaluate the impact of different importance weights $\lambda$ on the performance of DFPL, we conduct  experiments under the data distribution setting of $(\textit{Avg} = 3, \textit{Std} = 2)$.
Fig. \ref{11271141} illustrates   the varying performance of DFPL under different importance weights $\lambda$  within the optimization function. We select  a range of values from $[0, 4]$ and evaluate  the TAA and auxiliary term loss on the MNIST, FMNIST, CIFAR10, and SVHN datasets.
The experimental results show that DFPL achieves  optimal performance when $\lambda = 1$. Specifically, as $\lambda$ increases from 0 to 1, TAA rises significantly, indicating that  prototype aggregation can effectively improve the FL performance.
Meanwhile, the auxiliary loss decreases significantly as $\lambda$ increases from 0 to 1, indicating that clients learn more compact and consistent representations in the prototype space, thereby enhancing feature alignment among clients.
However, when $\lambda > 1$, although the auxiliary loss continues to decrease, the TAA value remains unchanged or may even decline.
This indicates that overemphasizing the auxiliary loss may lead to over-fitting of the prototype representations, reducing the performance of federated learning under  heterogeneous data distributions.

\begin{figure}[t]
\centering
    \subfloat[FMNIST]
      {\label{6411251}
  \centering \includegraphics[width=0.48\linewidth]{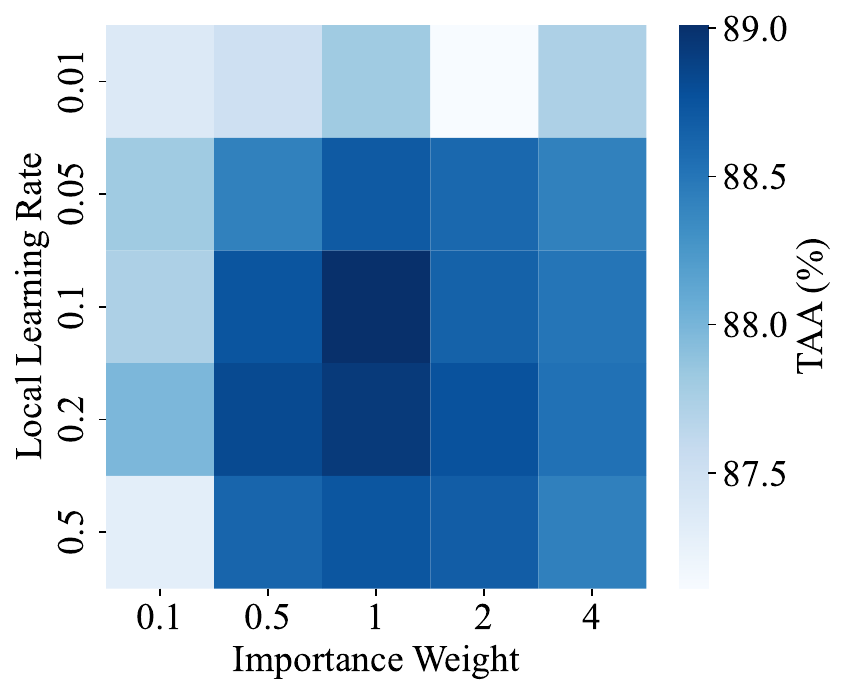}
  }
      \subfloat[CIFAR10]
      {\label{6411252}
  \centering \includegraphics[width=0.48\linewidth]{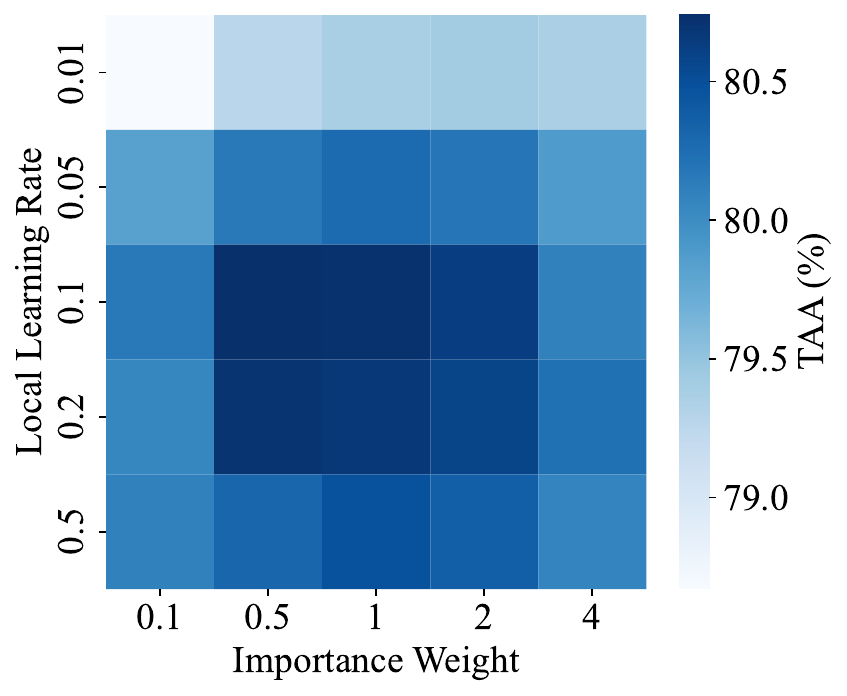}
  }
\caption{DFPL's TAA  in diverse learning rates  $\eta$ and importance weights $\lambda$.
}
\label{641125}
\end{figure}

Different local learning rates can greatly affect the FL performance.
To evaluate the robustness of DFPL under different local learning rate settings, we conduct  experiments on the FMNIST and CIFAR10 datasets under the data distribution setting of $(\textit{Avg} = 3, \textit{Std} = 2)$.
Specifically, we select   five different local learning rates $\eta$: 0.01, 0.05, 0.1, 0.2, and 0.5, and set five importance weights $\lambda$: 0.1, 0.5, 1, 2, and 4. The experimental results   are shown in Fig. \ref{6411251} and Fig. \ref{6411252}, with darker blues denoting higher TAA.
Experimental results show that different learning rate settings cause approximately $0\%\text{-}2\%$ variations in TAA.
The main reason for these fluctuations is that smaller local learning rates tend to slow down the convergence process during training, while larger local learning rates may introduce excessive oscillations in model updating, thereby affecting the stability of federated learning.

\subsection{Scalability of DFPL}

\begin{figure}[t]
  \centering
  \subfloat[\textit{Avg}=3, \textit{Std}=2]
  {
      \label{50701251}  \includegraphics[width=0.48\linewidth]{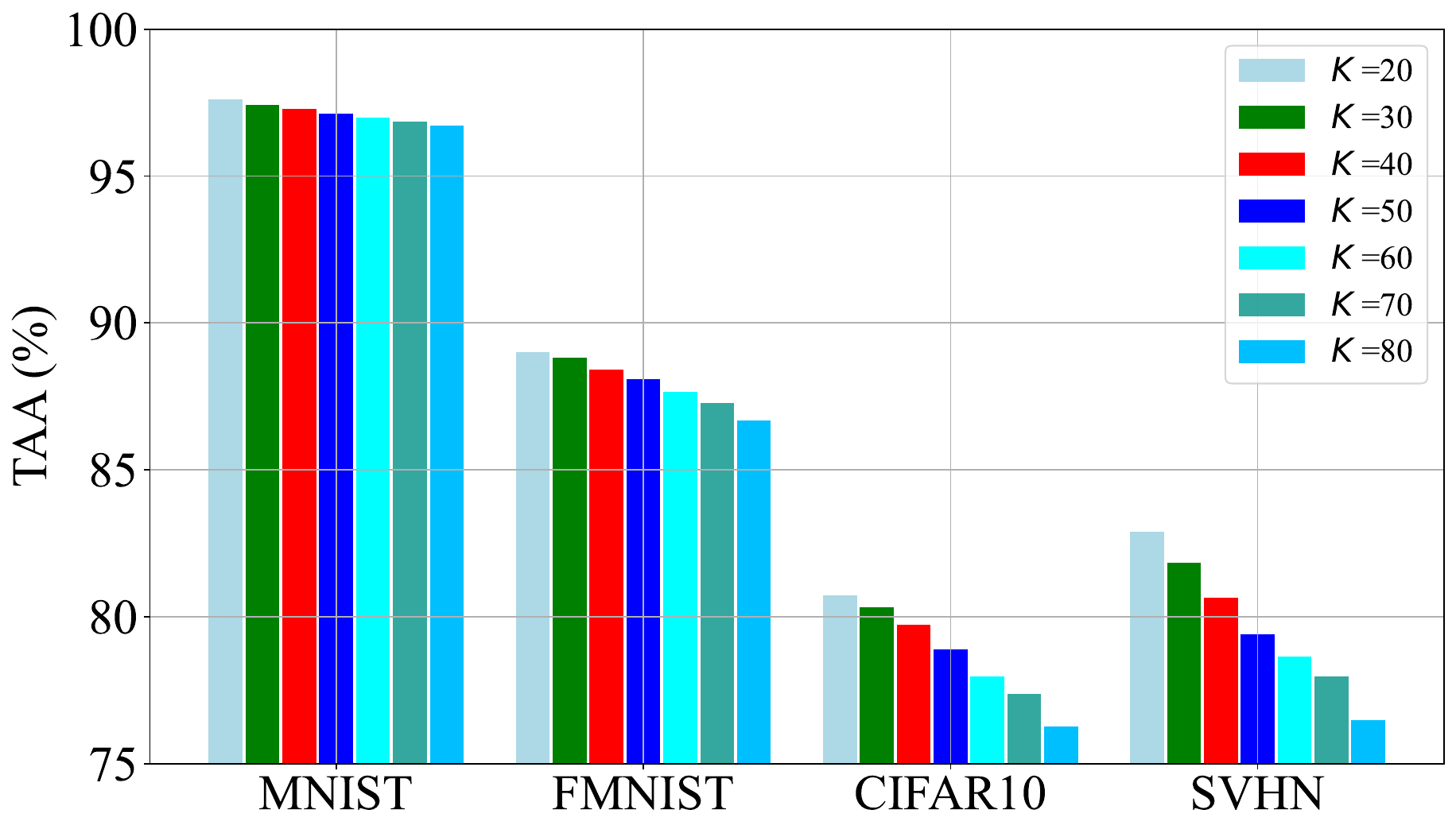}
  }
  \subfloat[\textit{Avg}=4, \textit{Std}=2]
  {
      \label{50701252}  \includegraphics[width=0.48\linewidth]{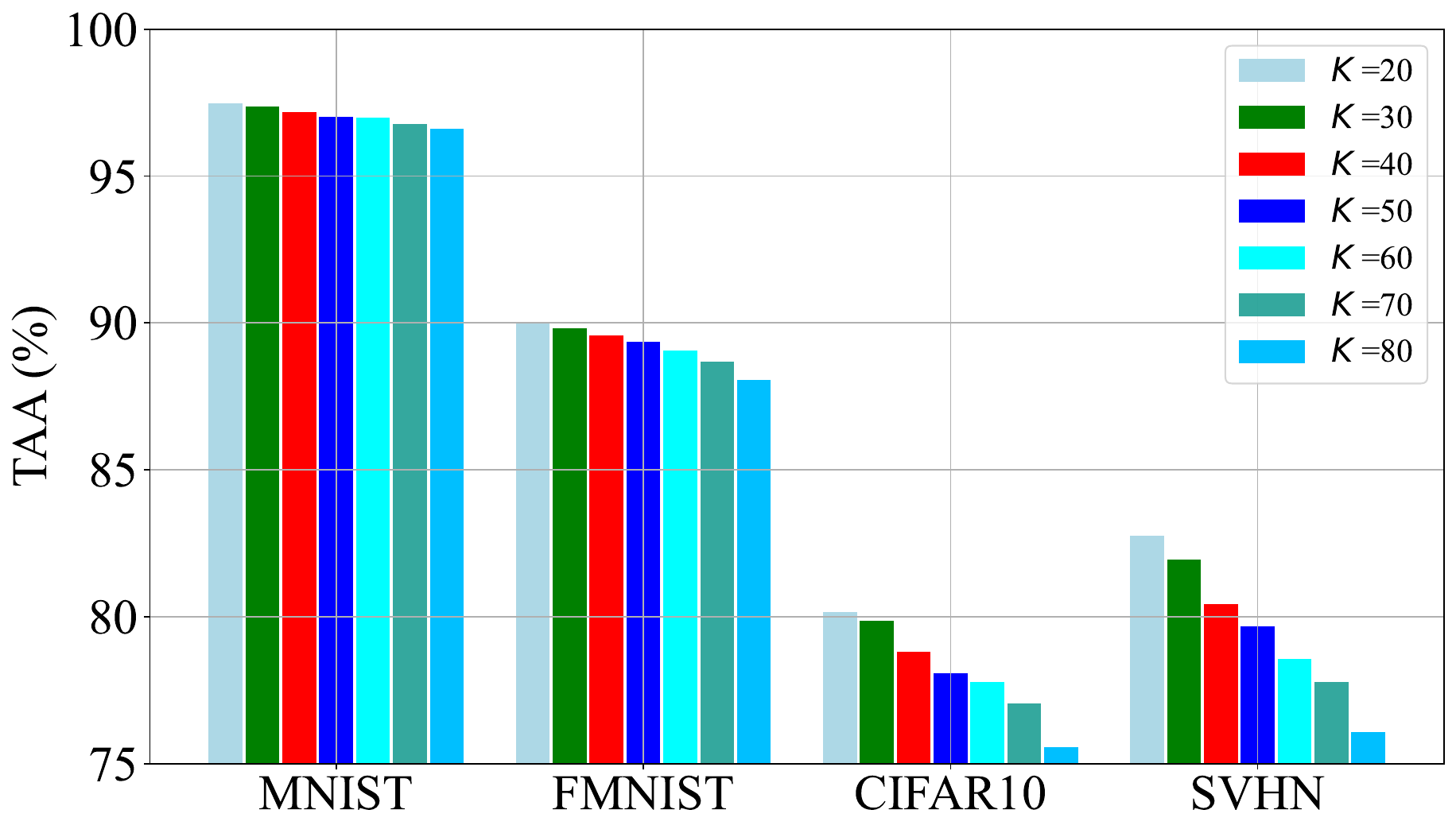}
  }
\caption{Scalability of DFPL on the MNIST, FMNIST, CIFAR10, and SVHN.}
\label{5070125}
\end{figure}

To evaluate the scalability of DFPL, we test  its performance under different numbers of clients.
Specifically, we set the total number of clients in federated learning to  {\color{blue}$K \in \{20, 30, 40, 50, 60, 70, 80\}$}, and experiments  are conducted under two data distribution conditions.
The experimental results are shown in Fig. \ref{5070125}.
It can be observed that as the number of clients gradually increases, the test average accuracy (TAA) of DFPL decreases slightly on the MNIST and FMNIST datasets.
This is because enlarging the system scale makes the data distribution across clients more heterogeneous, leading to minor performance degradation.
Notably, DFPL still maintains a relatively stable TAA with very limited degradation, demonstrating its adaptability to larger scales.
 {\color{blue}In contrast, on the CIFAR10 and SVHN datasets, the performance degradation becomes more pronounced as the number of clients grows.
The reason is that more complex datasets require larger local sample sizes to support effective local  training, while increasing the number of clients reduces the samples available to each client, thereby weakening local training.
This is a reasonable  phenomenon.
Overall, the experimental results verify that DFPL has good scalability to a certain extent.}

\subsection{Stability of DFPL}
\begin{figure}[t]
  \centering
  \subfloat[\textit{Avg}=3, \textit{Std}=2]
  {
      \label{41722251}  \includegraphics[width=0.48\linewidth]{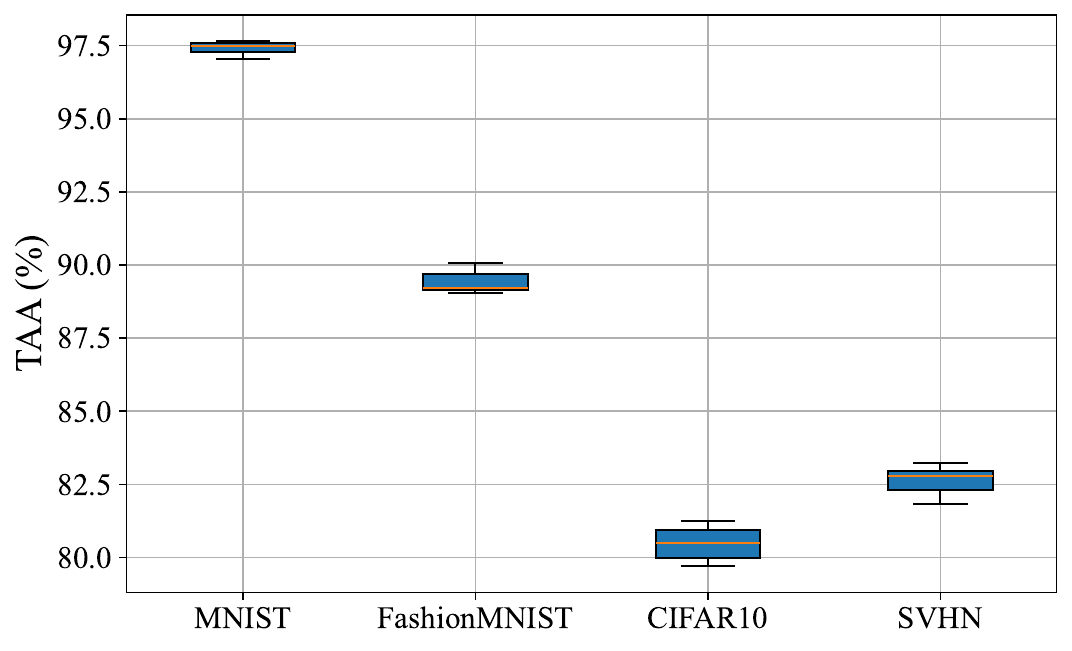}
  }
  \subfloat[\textit{Avg}=4, \textit{Std}=2]
  {
      \label{41722252}  \includegraphics[width=0.48\linewidth]{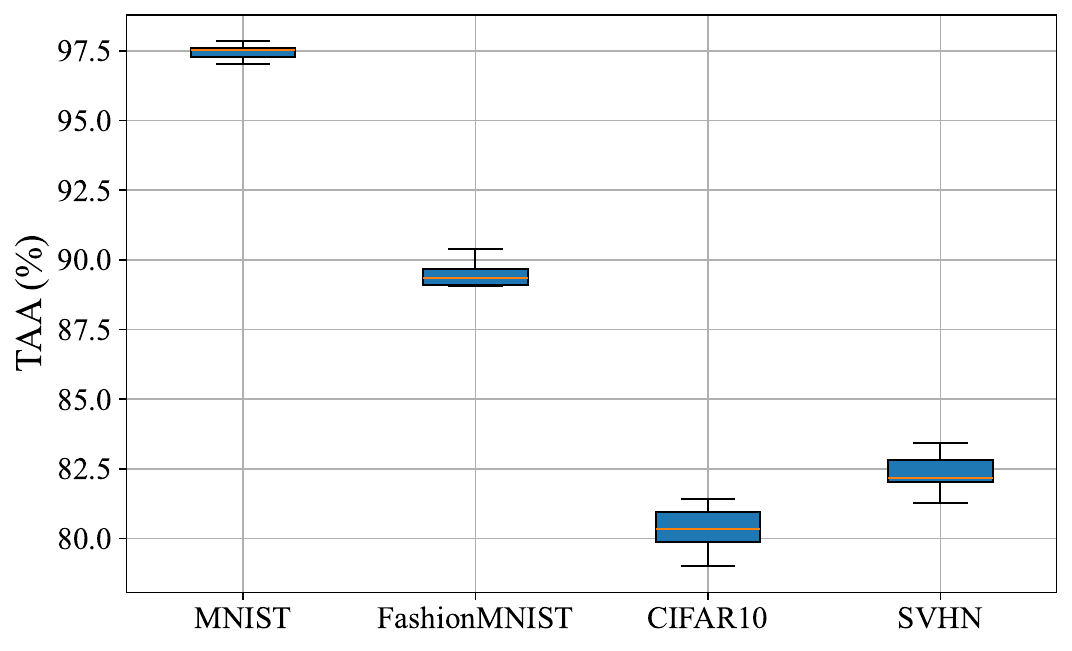}
  }
\caption{Stability of DFPL on the MNIST, FMNIST, CIFAR10, and SVHN.}
\label{4172225}
\end{figure}

To evaluate the robustness of DFPL, its stability across  repeated random experiments is crucial.
To verify DFPL's stability under heterogeneous data distribution, we conduct  experiments on four datasets: MNIST, FMNIST, CIFAR10, and SVHN.
Specifically, for each experimental setting, we repeated the experiments  10 times to evaluate the stability of DFPL under different random data samplings.
The experimental results are shown in the box plot Fig. \ref{4172225}.
It can be observed that on the CIFAR10 and SVHN datasets, the fluctuation range of DFPL's TAA is controlled within 2\%, while on the MNIST and FMNIST datasets, the fluctuation range is even smaller, showing higher stability.
This shows that DFPL can maintain consistent results in different data  scenarios.
Moreover,  the box plots show  that  there are no outlier results for DFPL in these experiments, indicating that DFPL is robust to different random parameters and  different data distributions.
Therefore, this result  validates the stability of the DFPL.

\subsection{DFPL under Model Heterogeneity}
\textcolor{blue}{Beyond data heterogeneity, model heterogeneity also exists in practical scenarios, where different clients  may adopt diverse model architectures.
To investigate the robustness of DFPL, we evaluate    its performance under model heterogeneity.
Specifically, to simulate architectural differences across clients, we vary the stride of convolutional layers in ResNet18 across clients when training on the CIFAR10 and SVHN datasets.
The experimental results are presented in Table \ref{012111110211}, where ``w/o'' denotes   ``without.'' 
We observe that the performance of DFPL decreases when model heterogeneity is introduced, compared with the setting without model heterogeneity.
This phenomenon indicates that model heterogeneity can weaken the performance of DFPL to some extent.
This is because the prototypes extracted by  clients exhibit differences.
Nevertheless, the results  demonstrate that DFPL is still able to maintain relatively stable performance under model heterogeneity, validating its applicability in more complex environments.}

\begin{table}[t]
  \centering
  \caption{ {\color{blue}Test average accuracy of DFPL with and without model heterogeneity.}}
  \setlength{\tabcolsep}{0.32cm}
  \scalebox{0.70}{
  \begin{tabular}{cc|ccc|ccc}
    \hline\hline
    \multirow{3}{*}{Dataset} & \multirow{3}{*}{\textit{Std}} & \multicolumn{6}{c}{Test Average Accuracy (\%)} \\
        & & \multicolumn{3}{c}{Model Heterogeneity}  &  \multicolumn{3}{c}{w/o  Model Heterogeneity}  \\
    & & \textit{Avg = }3 & \textit{Avg = }4 & \textit{Avg = }5 & \textit{Avg = }3 & \textit{Avg = }4 & \textit{Avg = }5 \\
    \hline
 \multirow{2}{*}{CIFAR10}  & 1 & 81.66 & 78.49 & 74.12 & 83.18 & 77.07& 75.25 \\
      & 2 & 80.30 & 77.84 & 75.25 & 80.72 & 80.17 & 76.41 \\\hline
 \multirow{2}{*}{SVHN}     &1 & 84.07 &82.63 & 81.75 & 84.81 & 82.12 & 83.07 \\
       & 2 & 83.19 &80.97 & 79.26 & 82.88& 82.70& 81.54 \\
    \hline\hline
  \end{tabular}
  }
  \label{012111110211}
\end{table}

\subsection{Time Cost}
\begin{figure}[t]
  \centering
  \subfloat[Local model training time]
  {
      \label{50810281}  \includegraphics[width=0.5\linewidth]{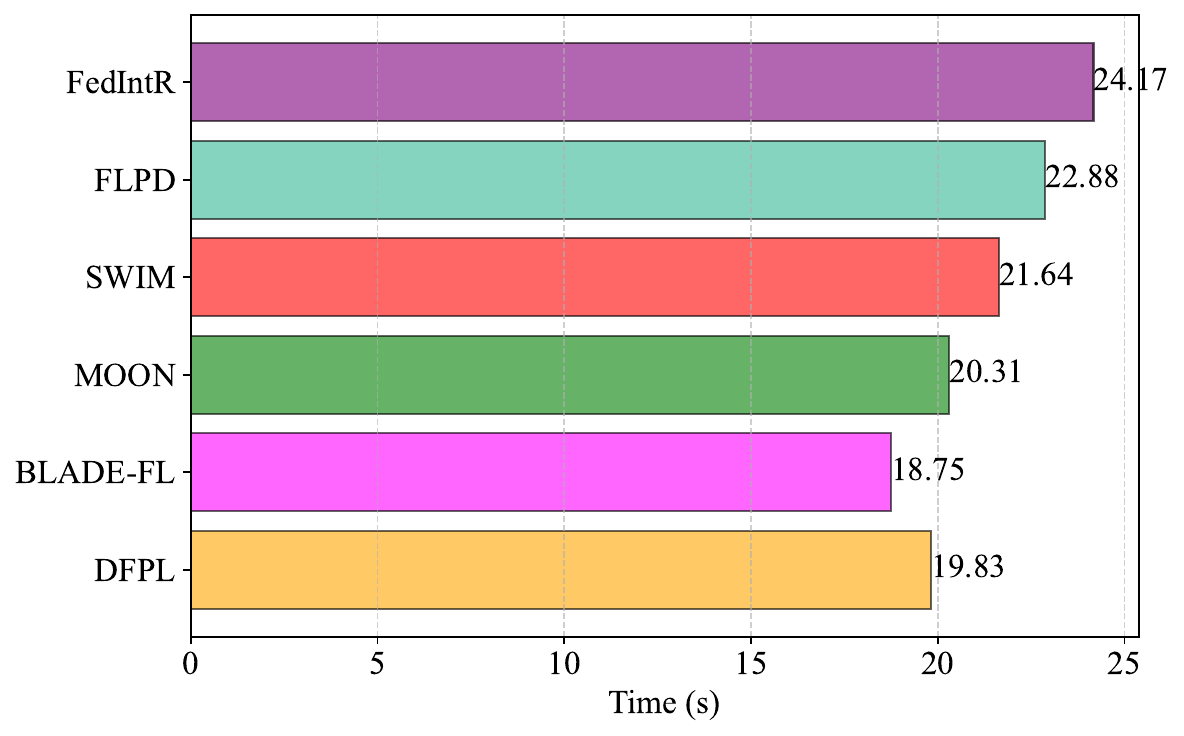}
  }
  \subfloat[Aggregation time]
  {
      \label{50810282}  \includegraphics[width=0.48\linewidth]{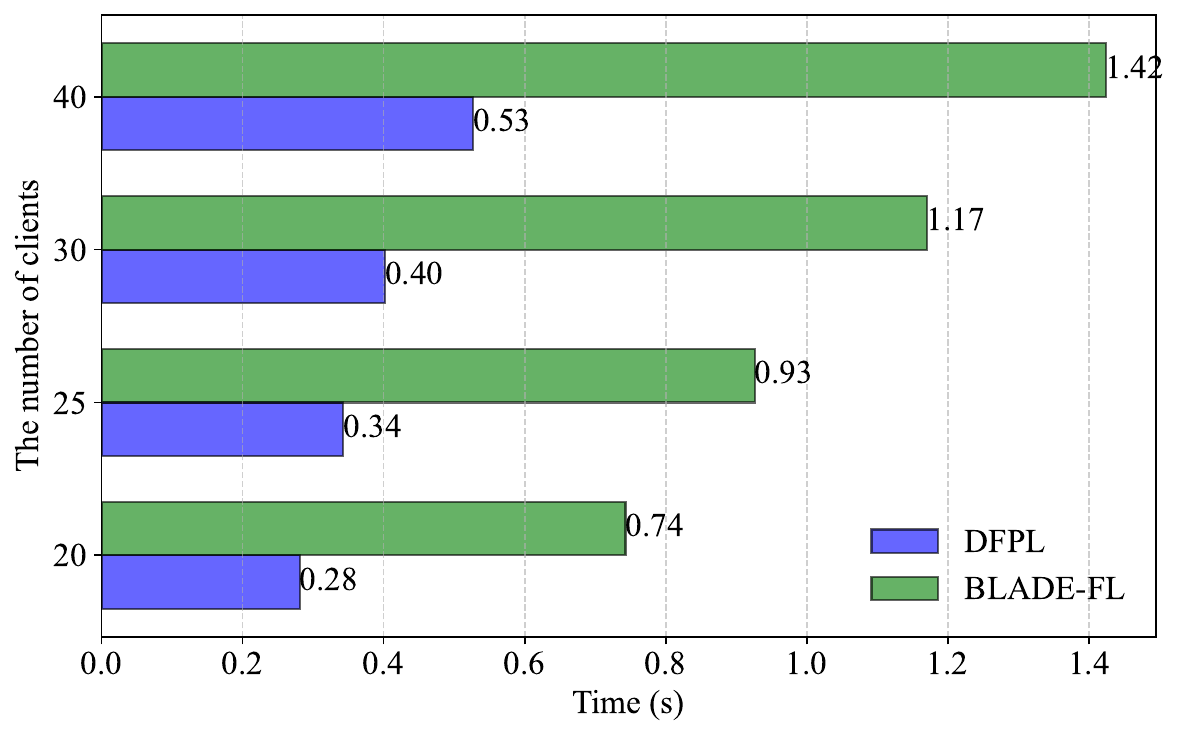}
  }
\caption{Time cost of different  schemes in CIFAR10 dataset.}
\label{5081028}
\end{figure}

To evaluate the time overhead of DFPL, we measure   its local model training time and aggregation time, respectively.
The time evaluation is  {\color{blue}conducted} on PC equipped with Intel(R) Xeon(R) Silver 4210R CPU, 64 GB RAM, and NVIDIA GTX 3090 GPU.
Specifically, we measure the time cost of local model training in different schemes on the CIFAR10 dataset with 20 local iterations.
The results are shown in Fig. \ref{50810281}.
The experimental results show that, except for BLADE-FL, the training time of all schemes  is higher than that of DFPL.
BLADE-FL {\color{blue}exhibits}    the lowest time spent since its optimization objective does not involve additional mechanisms for handling Non-IID data.
In contrast, FedIntR incurs the highest local training time {\color{blue}because it uses}   layer-wise model representation comparisons in the local optimization function to  enhance the consistency of model updates across clients, thereby introducing substantial computational overhead.
The DFPL leverages the output of the feature extractor as a constraint during local training, thereby avoiding the high computational cost associated with layer-wise comparisons.
Similarly, MOON incorporates a contrastive term based on the output of the feature extractor into its optimization function.
However, it introduces a Softmax operation for smoothing, which results in slightly higher training time than DFPL.
In summary, DFPL guarantees performance while minimizing computational overhead under heterogeneous data distributions.

In addition, we compare the aggregation time of DFPL and BLADE-FL in   a single communication round.
The measured aggregation time refers to the time required to aggregate all model updates, excluding the client's local mining process.
The experimental results are shown in Fig. \ref{50810282}.
The results show that the aggregation time of DFPL is significantly lower than that of BLADE-FL.
This advantage mainly comes from the fact that clients in DFPL only need to upload prototype parameters, whereas BLADE-FL requires uploading the local model parameters.
As shown in Table \ref{11261948}, the number of parameters transmitted in DFPL is significantly reduced.
Therefore, DFPL can effectively reduce the aggregation time without sacrificing performance.

\section{CONCLUSION}\label{sec7}
In this paper, we propose DFPL to improve the performance of distributed learning under heterogeneous data distributions and effectively reduce the parameters transmitted  between clients.
In addition, we theoretically analyze the convergence of DFPL in combination  with the resource allocation between training and mining of clients.
 Extensive experiments demonstrate the superior performance of DFPL under various heterogeneous data distributions.
 DFPL can be effectively applied in healthcare to provide intelligent services by achieving highly accurate models through distributed training while ensuring the privacy of sensitive data.
 
  {\color{blue}However, beyond data heterogeneity, model heterogeneity  also exists in practical scenarios.
 Since DFPL exchanges prototypes rather than  model parameters, it remains applicable under heterogeneous model settings.
 Nevertheless, our work assumes that all local models have the same architecture but different data distributions, and does not provide theoretical analysis  under model heterogeneity.
 In future work, we will focus on two directions: (\textit{i}) addressing the challenges of  federated learning when both model and data heterogeneity, and (\textit{ii}) exploring privacy-preserving decentralized federated learning to ensure the protection of client privacy.
}

\bibliographystyle{IEEEtran}
\bibliography{reference}

\begin{IEEEbiography}[{\includegraphics[width=1in,height=1.25in,clip,keepaspectratio]{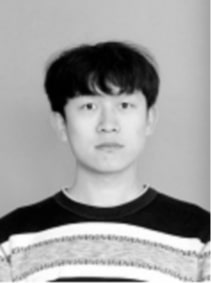}}]{Hongliang Zhang} received the B.S. degree and  M.S. degree in Computer Science from Qilu University of Technology in 2019 and 2022 respectively.
He is currently studying for a Ph.D. in Computer Science at Qilu University of Technology. His  research interests include federated learning and blockchain.
\end{IEEEbiography}
\vspace{-1.5cm} %
\begin{IEEEbiography}[{\includegraphics[width=1in,height=1.25in,clip,keepaspectratio]{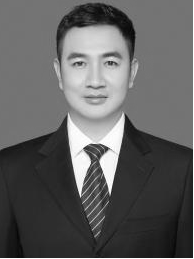}}]{Fenghua Xu}
received M.E degree from Peking University in 2018. He is currently studying for a Ph.D. in cyber security at University of Science and Technology of China. His research interests include wireless security and AI security.
\end{IEEEbiography}
\vspace{-1.5cm} %
\begin{IEEEbiography}[{\includegraphics[width=1in,height=1.25in,clip,keepaspectratio]{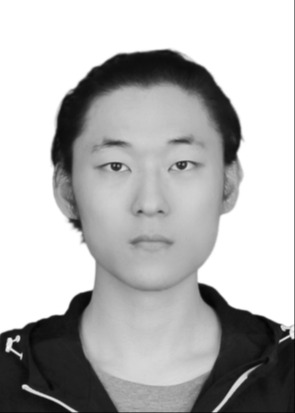}}]{Zhongyuan Yu}
received B.S. degree from School of Information Science and Engineering at Lanzhou University in 2024.
He is currently studying for M.E degree at China University of Petroleum.
His research interests include distributed computing, machine learning and cybersecurity.
\end{IEEEbiography}
\vspace{-1.5cm} %
\begin{IEEEbiography}[{\includegraphics[width=1in,height=1.25in,clip,keepaspectratio]{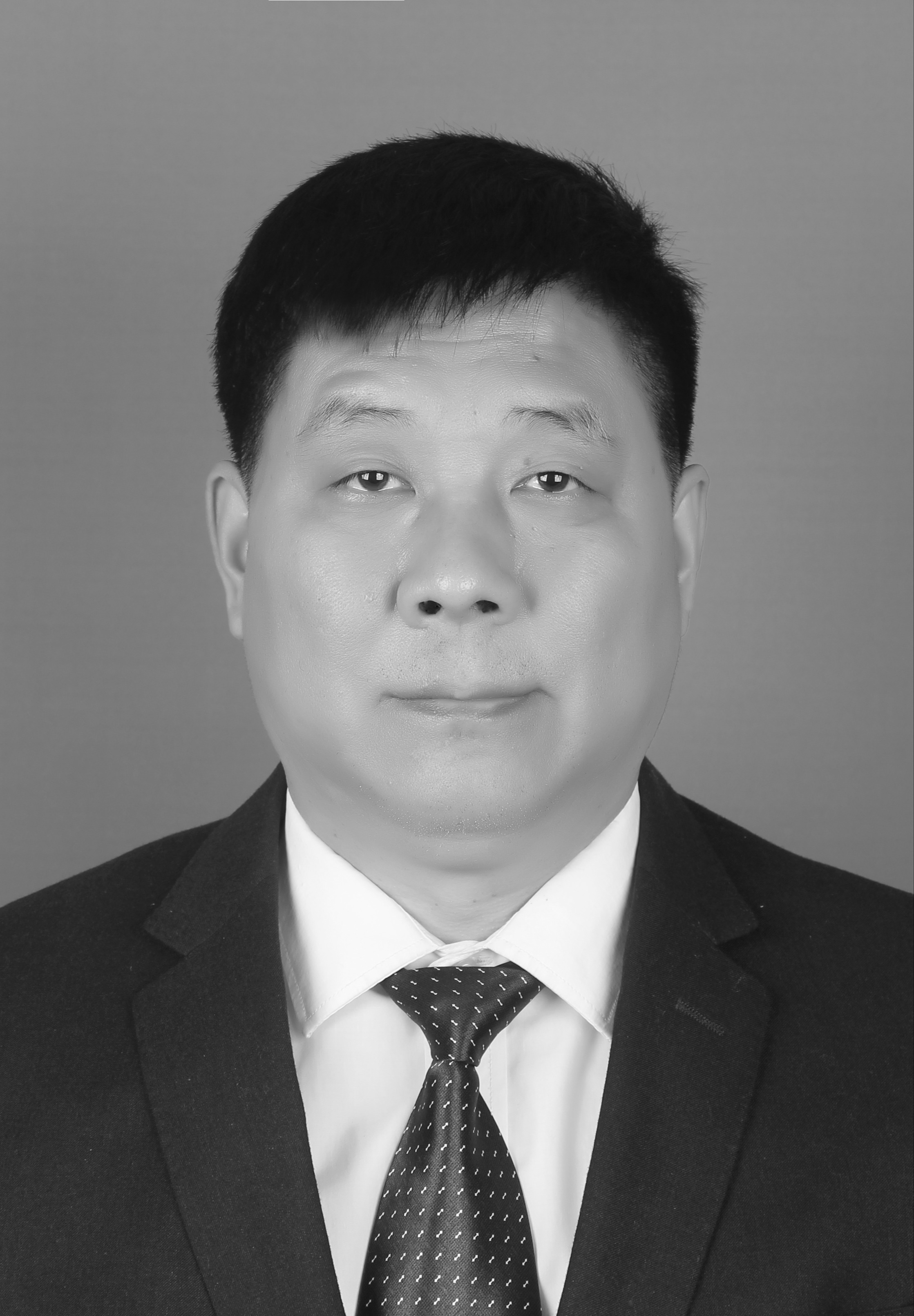}}]{Shanchen Pang} received the graduation degree from the Tongji University of Computer Software and Theory, Shanghai, China, in 2008. He is a Professor from the China University of Petroleum, Qingdao, China. His current research interests include the theory and application of Petri Net, service computing, trusted computing, and edge computing.
\end{IEEEbiography}
\vspace{-1.5cm} %
\begin{IEEEbiography}[{\includegraphics[width=1in,height=1.25in,clip,keepaspectratio]{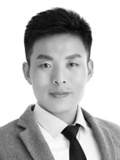}}]{Chunqiang Hu} received the B.S. degree in computer science and technology from Southwest University, Chongqing, China, in 2006, the M.S. and Ph.D. degrees in computer science and technology from Chongqing University, Chongqing, China, in 2009 and 2013, respectively, and the Ph.D. degree in computer science from George Washington University, Washington, DC, USA, in 2016. He is currently a Faculty Member with the School of Software Engineering, Chongqing University. His current research interests include privacy-aware computing, big data security and privacy, wireless and mobile security, applied cryptography, and algorithm design and analysis. Dr. Hu was a recipient of the Hundred-Talent Program by Chongqing University. He is a member of ACM.
\end{IEEEbiography}
\vspace{-1.5cm} 
\begin{IEEEbiography}[{\includegraphics[width=1in,height=1.25in,clip,keepaspectratio]{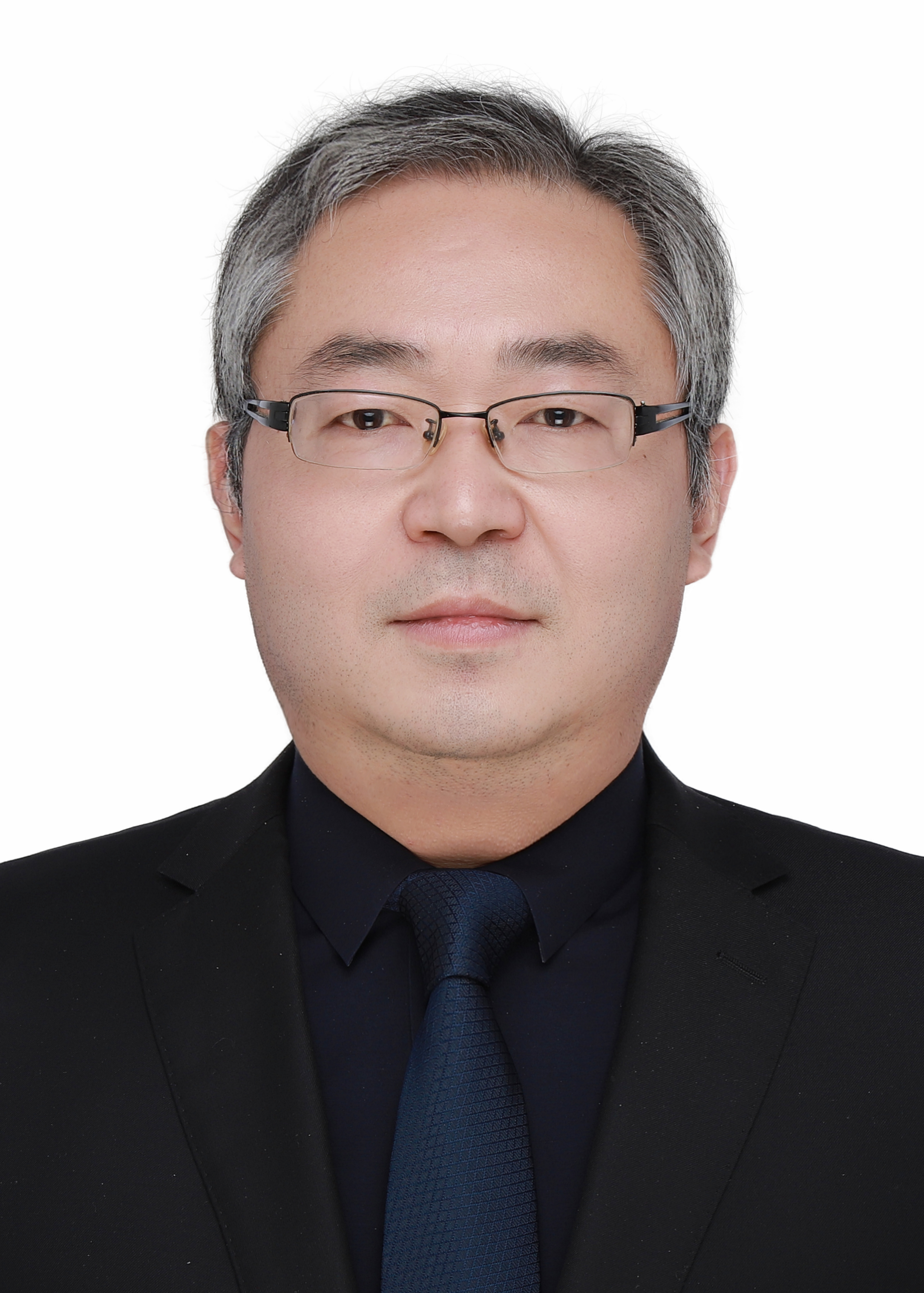}}]{Jiguo Yu} received the Ph.D. degree from the School of mathematics, Shandong University in 2004. In 2007, he was a full professor with the School of Computer Science, Qufu Normal University, Shandong, China. He is currently a full professor with University of Electronic Science and Technology of China, Chengdu, and a joint professor at Qilu University of Technology. His research interests include privacy-aware computing, wireless networking, distributed computing, blockchain and AI Security. He is a Fellow of IEEE, a member of ACM and a senior member of the China Computer Federation (CCF).
\end{IEEEbiography}

\end{document}